\documentclass[a4paper,11pt]{article}
\usepackage[utf8]{inputenc}
\usepackage{amssymb}
\usepackage[T1]{fontenc}
\usepackage[utf8]{inputenc}
\usepackage[english]{babel}
\usepackage{graphicx}
\usepackage{amsmath}
\usepackage{mathtools}
\usepackage[dvipsnames]{xcolor}
\usepackage{amssymb}
\usepackage{amsthm}
\usepackage{amsfonts}
\usepackage{listings}
\usepackage{bm}
\usepackage{pgfplots}
\usepackage{booktabs}
\usepackage{multirow}
\usepackage{eurosym}
\usepackage{longtable}

\usepackage{indentfirst}

\usepackage{graphicx}
\usepackage{fancyhdr}
\usepackage{enumerate}
\usepackage{caption}
\usepackage{subcaption}
\usepackage{float}
\usepackage{booktabs}
\usepackage{siunitx}
\usepackage{verbatim}
\addto\captionsitalian{}

\usepackage{titlesec}

\usepackage{subcaption}

\usepackage{scalerel}
\usepackage{mathtools}
\newcommand\myabs[1]{%
  \setbox1\hbox{$#1$}%
  \stretchrel{\lvert}{\usebox1}\stretchrel*{\lvert}{\usebox1}%
}

\titleformat{\paragraph}
{\normalfont\normalsize\bfseries}{\theparagraph}{1em}{}
\titlespacing*{\paragraph}
{0pt}{3.25ex plus 1ex minus .2ex}{1.5ex plus .2ex}

\usepackage{natbib}
\begin{document}

\title{\textbf{{A Public-Private Insurance Model for Natural Risk Management: an Application to Seismic and Flood Risks on Residential Buildings in Italy.}}\thanks{The idea for this project was inspired by author's collaboration in \cite{CasaItalia}, and for this reason we would like to thank Giovanni Azzone and his research team from Politecnico di Milano. We would also like to thank Giuseppe Di Capua (INGV), Andrea Flori (Politecnico di Milano), Ilan Noy (Victoria University of Wellington), Valentina Tortolini (IMT School for Advanced Studies Lucca), and Francesco Turino (Fundamentos del An\'alisis Econ\'omico, Universidad de Alicante) for their valuable comments and advice.}
\author{{Selene Perazzini$^{a}$, Giorgio Stefano Gnecco$^{a}$, and Fabio Pammolli$^{b}$}
\\ \\
\small{$^{a}$IMT School for Advanced Studies Lucca}\\
\small{$^{b}$Politecnico di Milano} }}
\date{\small{\today}}
\maketitle
\thispagestyle{empty}

\begin{abstract}
This paper proposes a public-private insurance scheme for earthquakes and floods in Italy in which property-owners, the insurer and the government co-operate in risk financing. Our model departs from the existing literature by describing a public-private insurance intended to relieve the financial burden that natural events place on governments, while at the same time assisting individuals and protecting the insurance business. Hence, the business is aiming at maximizing social welfare rather than profits.\\
Given the limited amount of data available on natural risks, expected losses per individual have been estimated through risk-modeling. In order to evaluate the insurer's loss profile, spatial correlation among insured assets has been evaluated by means of the Hoeffding bound for r-dependent random variables. Though earthquakes generate expected losses that are almost six times greater than floods, we found that the amount of public funds needed to manage the two perils is almost the same. We argue that this result is determined by a combination of the risk aversion of individuals and the shape of the loss distribution. Lastly, since earthquakes and floods are uncorrelated, we tested whether jointly managing the two perils can counteract the negative impact of spatial correlation. Some benefit from risk diversification emerged, though the probability of the government having to inject further capital might be considerable.\\
Our findings suggest that, when not supported by the government, private insurance might either financially over-expose the insurer or set premiums so high that individuals would fail to purchase policies.\\ 

\end{abstract}

\section{Introduction}\label{sec1}

Natural risks pose a broad range of social, financial and economic issues, with potentially long-lasting effects. Historically, governments have mostly addressed the financial effects of natural events on an ad-hoc basis, but countries are now increasingly focusing on proactive planning before a disaster strikes \citep{WB}. Among others, OECD, G20 \citep{OECD1}, the World Bank and GFDRR \citep{WB} claim that governments should guide citizens towards recovery by implementing both risk reduction and financial protection. In particular, the \cite{WB} argues that \textit{``financial protection complements risk reduction by helping a government address residual risk, which is either not feasible or not cost effective to mitigate. Absent a sustainable risk financing strategy, [...], a country with an otherwise robust disaster risk management approach can remain highly exposed to financial shocks, either to the government budget or to groups throughout society''}.

While guaranteeing social assistance, governments should at the same time encourage private initiatives in prevention and financial protection. As emphasized by the \cite{OECD2}, improving public awareness reduces the human-induced factors that make a major contribution to the cost of disasters and alleviates losses on public finances. In particular, since private insurance is the main risk financing tool for businesses and households, the \cite{OECD1} recommends that governments \textit{``assess their availability, adequacy and efficiency to the population and within the economy, as well as their costs and benefits relative to other types of possible risk reduction measures''}.

A series of challenges hinder the development of the insurance business in protection from natural disasters. First of all, \cite{Kousky} shows that spatial correlation creates the potential for enormous losses at the aggregate level, and insurers therefore need to access a large amount of capital in order to offer the cover and meet solvency constraints. As a consequence, they are often forced to drive up premiums, which could become so high that it would not be rational for individuals to purchase the policy. Large insurers can significantly reduce the probability of insolvency by pooling risks from more independent regions or by transferring a portion of their portfolio through reinsurance. However, while lowering premiums for regions with a higher risk, this solution might raise those of those with a lower risk and, especially in a competitive market, low risk-individuals might fail to purchase, therefore leaving the company with an extremely risky pool. As shown by \cite{Charpentier}, the free market does not necessarily provide an efficient level of natural-catastrophe insurance, but government-supported insurance allows losses from disasters to be spread equally among policyholders thanks to the government's easier access to credit.

Climate change also exacerbates these issues: the \cite{GenevaAssociation} warns that return periods and correlation among claims for several high-loss extreme events are \textit{``ambiguous rather than simply uncertain''}, and raises concerns about the future sustainability of insurance business on natural risks. Social assistance policies may also hinder the development of private markets and increase the financial burden of natural disasters on public finances due to charity hazard \citep{WB}.

Against this background, a number of economies have established various forms of public-private co-operation to support the insurance business, and several countries have decided to enter the market by establishing a public-private company entirely devoted to insuring citizens' properties against natural disasters at a discounted price (e.g. Spain, France, Australia, Turkey, New Zealand, Taiwan, USA, etc..) \citep{Consorcio}.

This work proposes a public-private insurance scheme for Italy. Italy is highly exposed to natural risks, especially earthquakes and floods, but there is currently no well-defined loss allocation mechanism at national level. A few people insure their properties \citep{Maccaferri} and expect social assistance from the government instead. Each natural event is evaluated by public authorities when it occurs, social assistance depends on the decisions of the parties in charge and is therefore commensurate with the financial resources available at the time. In recent years public debate has increasingly shifted towards natural risk management and planning, although at the moment no initiative has been undertaken.

In defining the public-private insurance scheme for Italy, our work addresses three issues:
\begin{itemize}
    \item \textbf{Loss estimation and lack of data on past losses}.\\
    Insurance companies need big loss database for premium rating, but there is currently no source that collects information on natural impacts in Italy at national level. Lack of data on the impacts of natural disasters is a widespread issue and in order to overcome this problem, the world's biggest insurance companies have developed sophisticated models for loss estimation based on engineering and geology studies. In this paper, one of these models has been applied to estimate earthquake losses, while an alternative approach is proposed for floods.
    \item \textbf{Public-private insurance model}.\\
    Once losses have been estimated, we define a public-private insurance model. Our model departs from the existing literature by addressing a public-private partnership, which therefore modifies the fundamental hypotheses of traditional insurance. Our contribution to the literature can be summarized in three aspects. First, the purpose of the business is social assistance, and premium collection serves solely to risk management and to guarantee quick compensation to the damaged population. Therefore, rates do not include any profit load and are commensurate to citizens' demand. Second, we introduce the government as a social guarantor that contributes to reserves and provides public funds in case reserves are not sufficient for claim compensation. Finally, our model includes spatial correlation by applying the Hoeffding bound for r-dependent random variables.
   \item \textbf{Multi-hazard management.}\\
    As well known in finance, merging portfolios is beneficial only if risks are uncorrelated, as floods and earthquakes are likely to be. It remains to be seen whether the benefits from risk diversification counteract the negative impact of spatial correlation. The last part of this work extends the public-private insurance scheme to multi-hazard management.
\end{itemize}

The paper presents and discusses results for each of these three aspects. We found that seismic risk produces the highest expected losses at national level, but floods may generate the highest losses per square metre. The two perils differ in geographic extent: while the seismic risk involves almost all the nation, floods concern approximately two thirds of the territory. Though the seismic risk generates expected losses that are almost six times greater than floods, we found that the amount of public funds needed to manage them is almost the same. Our analysis shows that the public-private insurer can benefit from risk differentiation by jointly managing earthquake and flood risks through a multi-hazard policy: the amount of public capital needed is lower than would be necessary if the two risks were managed separately. Another desirable feature emerges: rates for multi-hazard policies are more geographically homogeneous, and therefore promote fairness perception among the population. However, it emerged that under no circumstances does the maximum premium that individuals are willing to pay match the insurer capital constraints. Without the government as a guarantor, it would therefore be impossible for the company to offer policies throughout the territory.

The paper is organized as follows: Section \ref{Risk} describes risk assessment models for flooding and seismic hazards and concludes presenting expected losses in Italy; Section \ref{Premium} defines the insurance model for single hazard policies and applies it to the two risks; Section \ref{ch:MH} extends the model to the analysis of multi-hazard policies; Section \ref{ch:conclusion} concludes.

\section{Risk assessment}
\label{Risk}
Expected losses are traditionally estimated from records of past events, but when data are too scarce or not available, alternative techniques are needed. During the last decades, a new family of models inferring losses from the characteristics of soil and structures has emerged \citep{Grossi}. According to this branch of literature, risk can be reconstructed as a combination of four components:\\
\begin{itemize}
\item	Hazard ($H$) provides a phenomenon description based on physical measurements, usually frequency, severity and location.
\item	Exposure ($E$) identifies the object at risk.
\item	Vulnerability ($V$) defines the relationship between hazard and exposure, quantifying the impact of the catastrophic event on the property under analysis.
\item	Loss ($L$) converts physical damages into monetary values.
\end{itemize}
Each component is defined on a series of geophysical, engineering or financial variables and relations, and equally contribute to the overall estimate of risk \citep{Mitchell-Wallace}. Through a proper definition and combination of these components, a risk model should describe the geological or environmental features of the peril in analysis and should also capture differences in impacts on the relevant structural typologies.

Although this line of research is growing fast, not many models are currently available and not any peril has been satisfactorily described. Moreover, these models, while not requiring data on losses, need a large amount of information on soil, weather, and housing. In addition to the difficulty of finding this data, models strongly depend on geographical and urban features of the area they have been defined on, and therefore can hardly be adapted to other territories \citep{Hufschmidt,Scorzini}.

As far as Italy concerns, current literature offers some analysis that allow to appreciate seismic risk on the whole territory, while little is still known about floods. We therefore refer to the existing literature for seismic risk, and develop a new model for flood assessment. Our methodology is similar to \cite{IVASS}.

After a brief presentation of the database, the following two subsections present earthquake and flood risk assessment respectively. Although the two model strongly differ, they both combine the four risk components as:
\begin{equation}\label{eq:risk_assessment}
\mbox{Expected monetary damage} = L \times E \times \int V(H) \mbox{d}(H).
\end{equation}
After a general description of the model, each of these subsection discuss the components separately. The section concludes presenting estimated expected losses. Our analysis considers residential housing only, furniture not included. Multi-hazard risk assessment is postponed to the Section \ref{ch:MH}.

\subsection{Data}
\label{Data}
There is currently no database collecting records on impacts from natural disasters in Italy, but some information on national riskiness is available, thought data quality is sometimes questionable. In particular, our models require data about hazard and exposure.
\begin{itemize}
    \item \textbf{Hazard}\\
    While seismic hazard is well documented, flood data are strongly affected by the lack of a single body responsible for physical detection.\\
     Seismic movements are in fact regularly monitored by the National Institute of Geophysics and Volcanology (INGV), that freely provides daily-updated databases both on past events and about several seismic indicators. Records are georeferenced and cover almost all the national territory, indicators are presented for different probability scenarios and associated to an accuracy index. Data for the analysis of earthquake has been drawn from INGV's maps of seismic riskiness.
    On the other side, flood monitoring is demanded to a number of regional authorities - named ``basins’ authorities'' - that independently choose collection methods and indicators. These differences in data collection often leads to inconsistencies and poor comparability among regions \citep{Molinari2012}. The main database on hydrological risk in Italy is the AVI (``Aree Vulnerate Italiane'' - ``Italian Vulnerable Areas'') archive managed by National Research Council \citep{Guzzetti}. The archive collects historical information on flood events in Italy (mainly from 1900 to 2002). However, records are mostly gathered from local journals and, unfortunately, are rarely suitable to scientific analysis: information are provided in a narrative form, georeferencing is poor, physical phenomena description is not uniform and data quality depends on the original source \citep{Molinari2014}. Despite these limitations, the archive is currently among the best representation of the flood hazard, and has therefore been used here.
    Information from the archive have been integrated with data from ``Italian Flood Risk Maps'' (EU Directive 2007/60/CE) indicating the perimeter of geographic areas that could be affected by floods according to three probability scenarios \citep{DecretoLegislativo}: extreme events with time to return 500 years (P1); events with time to return of 100-200 years (P2); events with time to return between 20-50 years (P3).\\
    \item \textbf{Exposure}\\
    As far as exposure concerns, we refer to the ``Mappa dei Rischi dei Comuni Italiani'' (``Riskiness Map of Italian Municipalities'' - MRCI). This database has been created during a recent institutional project - ``Casa Italia'' - to the aim of providing the best representation of major natural risks in Italy (volcanic, seismic, hydrological, geological). Among several risk indicators, the database presents a fairly rich representation of Italian real estate. Additional information on regional average house's squared metres and the average dwelling value are estimates by the Revenue Agency \citep{AgenziaEntrate}.
    
\end{itemize}

\subsection{Earthquake}
Earthquakes and land movements are among the most studied risks in the literature, but most of the analysis focus on vulnerability and explore the relationship that links hazard intensity and damage to buildings. As far as Italy concerns, a few analysis investigate the number of deaths, missing persons and/or injured people \citep{Cascini,Salvati,MarzocchiGarcia}, while, to our knowledge, risk assessment on residential risk is presented in \cite{Asprone} only. The latter model follows the structure specified in eq. (\ref{eq:risk_assessment}) and has been tested on the L'Aquila earthquake, therefore we are referring to it for seismic loss estimates. Some slight modification of the model has been introduced in order to update the analysis with latest released data on hazard and to consider a wider range of potential loss scenario. Moreover, our real-estate database provides a more detailed representation of residential housing, thus allowing for higher accuracy of the estimates.

Damages have been estimated per municipality relating the peak ground acceleration (PGA) and its exceedance probability $\lambda(PGA)$ with the existing residential building stock by means of fragility curves. Given a certain set of ``limit states'' ($LS$) representing subsequent level of damage (usually from ``no damage'' to ``collapse''), a fragility curve describes the probability of reaching a given limit state as a consequence of the observed PGA, $P \left(LS \vert PGA \right)$. Expected loss can be estimated by comparing fragility curves of each $LS$. Damages are then monetarily quantified by means of a function $RC(LS)$ linking the property's value to the level of damage.

Literature offers many fragility curves' models, and we rely on \cite{Asprone} selection for Italy (Table \ref{tab:fragility models}).
Each model $k$ applies to a number of specific building structures and is defined on $N_{LS_k}$ limit states chosen by the authors to describe the impact of earthquakes on the $j$-th structure. Since many models may address the same $j$-th structure, losses are estimated by averaging results from the $K_j$ models describing $j$.

Municipal residential housing stock is divided into five relevant structural typologies - thus fixing $j=1, \dots, 5$ - and seismic losses per square metre $l^s$ are computed for each $j$ and each municipality $c$.

Given the probability $P_k \left(LS+1 \vert PGA \right)$ of the structural typology $j$ of suffering a damage level $LS$ given a certain PGA, expected losses are estimated as: 
\begin{multline}\label{eq_model}
l_{j,c}^{s} =
\frac{1}{K_j} \sum_{k=1}^{K_j} \sum_{LS=1}^{N_{LS_k}} RC(LS) \int_0^{\infty} \left[ P_k \left(LS \vert PGA \right)- P_j \left(LS+1 \vert PGA \right)\right]   \mbox{d} F_c \left( PGA \right) =\\
=\frac{1}{K_j} \sum_{k=1}^{K_j} \sum_{LS=1}^{N_{LS_k}} RC(LS) \cdot \int_0^{\infty} \left[ P_k \left(LS \vert PGA \right)- P_k \left(LS+1 \vert PGA \right)\right] \myabs{\frac{ \mbox{d} \lambda_c \left( PGA \right) }{\mbox{d}(PGA)}} \mbox{d}(PGA).
\end{multline}
where $F_c(PGA) = 1-\lambda_c(PGA)$ is the cumulative density function of PGA for the $c$-th municipality. According to \cite{Asprone}, we assume $P_k \left(N_{LS_k}+1 \vert PGA \right) = 0$. Model (\ref{eq_model}) combines a probability distribution with domain $[0, \infty)$ and a damage function increasing with $PGA$. $PGA$ is traditionally expressed in gravity acceleration units $g$ and \cite{Asprone} bounds the integration variable $PGA$ to $\left[0,2g\right]$. Since we wanted to include as many scenarios as possible, we extended the domain to include even most unlikely events, and therefore the considered domain is $\left[0,\infty \right)$.

The five municipal losses estimates have been multiplied by municipal exposure and then aggregated into municipal total seismic losses $L_c^{s}$.
\begin{equation}
    L_c^{s} = \sum_{j=1}^5 l_{j,c}^{s} \cdot E_{j,c}^{s}.
\end{equation}

\begin{figure}
\centering
\caption{PGA exceedance probability.}\label{fig:ag}
\includegraphics[width=0.5\textwidth, trim=10mm 5mm 10mm 20mm, clip]{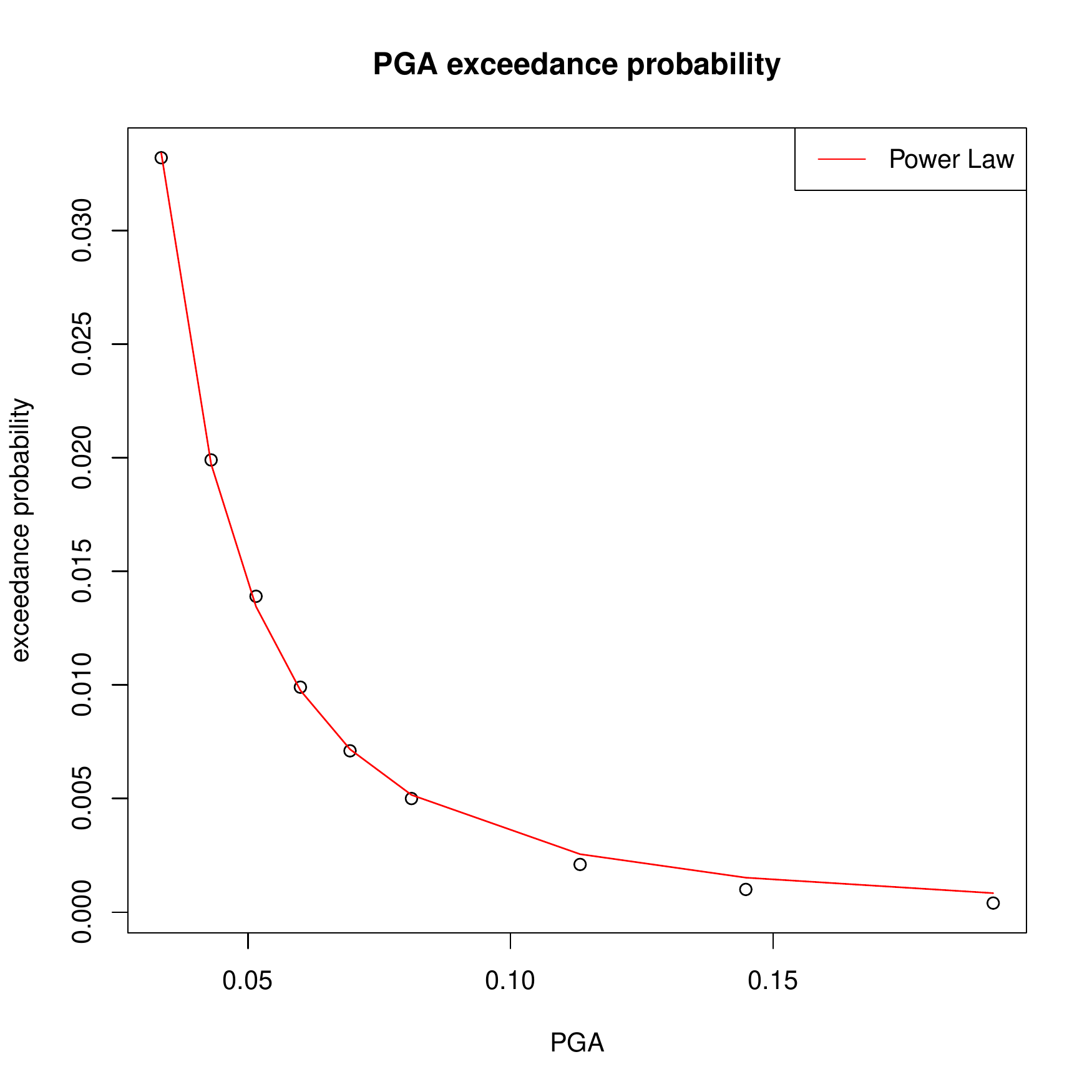}
\caption*{\textbf{Note}: the plot shows the PGA distribution of a random municipality. The nine points are data by INGV, and the red line represent fitting with the power law distribution.}\notag
\end{figure}

\subsubsection{Hazard}\label{seismic_hazard}

Seismic hazard is represented by PGA and its annual probability of exceedance, which are both available on the INGV website \citep{GruppoMPS} for most of Italy\footnote{Sardinia, Alicudi, Filicudi, Panarea, Pantelleria, Pelagie Islands, Stromboli, Ustica not included.}.

INGV released seismic maps for 9 probabilities of exceedance in 50 years \citep{Meletti}. Those $PGA$ measurements are presented for points in a 0.05 degree grid drawn on the Italian map. Grid points are defined by longitude and latitude, and can be associated to a municipality by means of reverse geocoding, that led to the definition of a PGA distribution for over 4600 municipalities. Sometimes more points referred to the same municipality, hence their average value has been considered. In order to capture the widest possible representation of the territory, missing municipalities have then been approximated by averaging the neighbours' PGA values. However, we failed to represent the whole national territory since Sardinia and many other small islands cannot be captured by neighborhood (missing municipalities can be seen in Figure \ref{fig:seismic_loss_map}). Our database is thus composed of 7685 municipalities.

The 9 INGV measurements describe the tail of $\lambda(PGA)$ for each grid point (a grid point's $PGA$ curve example is plotted in Figure \ref{fig:ag}). \cite{Asprone} assumed uniform seismicity in each municipality, but the known curve's sections in Figure \ref{fig:ag} do not seem to reflect this hypothesis. Moreover, since the left-side of the curve is missing, classical fitting methodologies led to unsatisfactory results, often overestimating tails. Therefore, parameters of the distribution have been estimated by regression. Best fitting results have been obtained by the power law distribution.  

In order for the hazard curves to reflect the soil category at the building foundation, \cite{OPCM} and \cite{DM14012008} state that PGA values at the bedrock should be multiplied by the stratigraphic $S_S$ and topographic $S_T$ amplification factors. These factors have been computed by \cite{Colombi} for all the Italian municipalities and kindly provided by INGV.

\begin{table}[ht]
\centering
\caption{Number of buildings per seismic structural typology.}\label{Table:seismic_exposure}
\begin{tabular}{|c|ccc|}
\hline
      & Material            & Building Code & Buildings (u=1000) \\
      \hline
RC gl & Reinforced concrete & Gravity Load  & 2853.96  \\
RC sl & Reinforced concrete & Seismic Load  & 636.92\\
M     & Masonry             & Gravity Load  & 6975.98\\
A gl  & Other Structures    & Gravity Load  & 1406.21\\
A sl  & Other Structures    & Seismic Load  &  260.88\\
\hline
\end{tabular}
\end{table}

\subsubsection{Exposure}
As seismic events differently affect buildings, relevant structural typologies have been identified on the basis of the information available.

First, the MRCI database divides municipal housing stock into: masonry, reinforced concrete, and other; \cite{Asprone} argue that buildings of type ``other''  contain both components of reinforced concrete and masonry structures, so we assumed this category to be a mixture of these two.

These structures may then have been built in compliance with modern anti-seismic requirements or not. Since the database does not include this information, we refer to the construction year and building laws in force. In fact, from 1974 a series of subsequent laws \citep{Legge64} led to the progressive re-classification of risk-prone areas, where more restrictive anti-seismic construction requirements entered into force, thus substantially modifying buildings’ structures. The process ended in 2003 when anti-seismic laws \citep{OPCM} were extended to the whole Italian territory. Thus, we define reinforced concrete and other structures as seismic loaded if built after these laws entered into force, or gravity loaded otherwise\footnote{As far as the year of construction concerns, ISTAT does not specify the exact year in which the building has been built, but a time interval which is approximately ten-years long. We assumed that the number of buildings constructed in any year of the interval is constant.}. According to \cite{Asprone}, we assumed masonry as seismic loaded only. Therefore, we refer to 5 structural typologies (see Table \ref{Table:seismic_exposure}) : masonry ($M$), and gravity or seismic loaded reinforced concrete ($RC.gl$ and $RC.sl$), gravity or seismic loaded other-type structures ($A.gl$ and $A.sl$).

Since $l_{j,c}^{s}$ is the expected seismic loss of the structure type $j$ in the municipality $c$ per square metres, $E^s_{j,c}$ is obtained by multiplying the number of buildings $B_{j,c}$ by the average apartment's surface $\bar{s}_c$ \citep{AgenziaEntrate} and the average number of apartments per building $\bar{A}_c$ (ISTAT, census 2015):
\begin{equation}
E^s_{j,c}= \bar{s}_c \cdot B_{j,c} \cdot \bar{A}_c.
\end{equation}

\begin{table}[t!]
\centering
\small
\caption{Fragility curves for seismic risk assessment.}\label{tab:fragility models}
\begin{tabular}{|c|c|c|cc|cc|}
 \toprule[.1em]
\multirow{2}{*}{Structure}&\multirow{2}{*}{Model ($k$)}& \multirow{2}{*}{$N_{LS_k}$} & \multicolumn{2}{c|}{gravity load} & \multicolumn{2}{c|}{seismic load}\\
 & & & 	$\mu$ &	$\sigma$				& $\mu$ &	$\sigma$\\
 \midrule[.1em]
\multirow{15}{*}{Masonry} & \multirow{3}{*}{\cite{Rota2}} & \multirow{3}{*}{3} & & & -2.03 & 0.36\\
	&								&					& & & -1.65 & 0.27\\
	&								&					& & & -1.35 & 0.22\\
\cline{2-7}
& \multirow{4}{*}{ \cite{Ahmad}} & \multirow{4}{*}{4} & & & -1.13 & 0.35\\
	&								&					& & &-1.03 &	0.35\\
	&								&					& & &-0.85 &	0.26\\
	&								&					& & &-0.77 &	0.23\\
\cline{2-7}
& \multirow{2}{*}{\cite{Erberik}} & \multirow{2}{*}{2}	& & & -0.47 &	0.35\\
				&					&			& & & -0.33 &	0.35\\
\cline{2-7}
& \multirow{2}{*}{\cite{Lagomarsino}}	& \multirow{3}{*}{3} & & &	-1 &	0.41\\
			&						&			& & & -0.75 &	0.34\\
				&					&			& & & -0.61 &	0.37\\
\cline{2-7}
& \multirow{2}{*}{\cite{Rota1}} &  \multirow{3}{*}{3}  & & &	-0.85 &	0.24\\
		&							&			& & &	-0.7 &	0.18\\
			&						&			& & &	-0.58 &	0.14\\
\hline
\multirow{30}{*}{Reinforced Concrete} & \multirow{4}{*}{\cite{Kappos2006}} & \multirow{4}{*}{4} & -1.78 & 1.14 & -1.32 & 0.29\\
&													&		& -1.12 & 0.8 &	-0.95 &	0.27\\
&													&		& -0.7 & 0.63 &	-0.57 &	0.27\\
&													&		& -0.59 & 0.57 & -0.24 & 0.28\\
\cline{2-7}
&\multirow{4}{*}{\cite{Spence}} & \multirow{4}{*}{4} 	&	-1.01 &	0.32 & -0.87 &	0.29\\
&												&	&	-0.55 &	0.32 &	-0.46 &	0.28\\
&												&	&	-0.28 &	0.31 & -0.02 & 0.29\\
&												&	& -0.09 & 0.32 & 0.15 &	0.27\\
\cline{2-7}
&\multirow{2}{*}{\cite{Crowley}} & \multirow{2}{*}{2} & -0.77 & 0.24 & -0.8 &	0.18\\
&													&		& -0.62 &	0.26 &	-0.61 &	0.22\\
\cline{2-7}
&\multirow{3}{*}{\cite{Ahmad}} & \multirow{3}{*}{3} & -1.07 & 0.22 & -1.07 & 0.22\\
&													&	& -0.91 &	0.29 & -0.91 &	0.29\\
&													&	& -0.59 &	0.26 &	-0.44 &	0.26\\
\cline{2-7}
&\multirow{2}{*}{\cite{Borzi1}} & \multirow{2}{*}{2} &  -0.74 &	0.32 &	-0.56 &	0.32\\
&													&	& -0.46	& 0.34 & -0.37	& 0.33 \\
\cline{2-7}
&\multirow{2}{*}{\cite{Borzi2}} & \multirow{2}{*}{2} &-0.68	& 0.45 & -0.41	&0.35\\
&													&	&-0.41	& 0.36 & -0.31	&0.35\\
\cline{2-7}
&\multirow{3}{*}{\cite{Kostov}} & \multirow{3}{*}{3}	&-0.48	& 0.47 & -0.44	& 0.48\\
	&												&		&-0.34	& 0.48 & -0.28	& 0.49\\
	&												&		&-0.29	& 0.48 & -0.19	& 0.49\\
\cline{2-7}
&\multirow{2}{*}{\cite{Kwon}} & \multirow{2}{*}{2}	&-1.08	& 0.22	& &\\	
&													&	&-0.73	& 0.22	& &\\	
\cline{2-7}
&\multirow{2}{*}{\cite{Ozmen}} & \multirow{2}{*}{2}	&-0.37	& 0.35	&-0.36	&0.3\\
&													&		&-0.17	& 0.23	&-0.12	&0.15\\
\cline{2-7}
&\multirow{4}{*}{\cite{Kappos2003}} & \multirow{4}{*}{4}	&-1.57	& 0.44	&-1.14	&0.43\\
&													&		&-0.92	& 0.44	&-0.57	&0.43\\
&													&		&-0.67	& 0.44	&-0.18	&0.43\\
&													&		&-0.51	& 0.44	&0.1	&0.43\\
\cline{2-7}
&\multirow{2}{*}{\cite{Tsionis}} & \multirow{2}{*}{2}	&-0.67	& 0.27	&-0.64	&0.28\\
&													&		&-0.22	& 0.38	& 0.18	&0.79\\
\hline
\multirow{3}{*}{Other} & \multirow{3}{*}{\cite{Kostov}} & \multirow{3}{*}{3} &	-0.62 &	0.5 & -0.52 &	0.49\\
&											& &	-0.44 &	0.49 &	-0.34 &	0.49\\
&											& & -0.35 &	0.49 &	-0.24 &	0.49\\
\hline							
\end{tabular}
\caption*{\textbf{Note}: this Table reproduces the selection of seismic fragility curves per building structural typology by \cite{Asprone}.}
\end{table}

\subsubsection{Vulnerability}\label{seismic_vulnerability}
Seismic vulnerability is represented by fragility curves, that provide the probability of exceeding a certain damage state, given some hazard parameters. Several curves are offered by the seismic engineering literature, each referring to a specific building structural category. We rely on Asprone et al. (2013) selection of curves, that is reported in Table \ref{tab:fragility models}. The selection contains 5 models for masonry structures, 11 for reinforced concrete ones, and 1 for the other typology. Each model $k$ is defined on a different set of $N_{LS_k}$ limit states representing building's structural damage conditions (the last limit state always corresponds to collapse) and provides one fragility curve for each limit state. Our fragility curves are log-normally shaped and require PGA values as unique input.

\subsubsection{Loss}
\label{ch: seismic_loss}
The loss component is represented by the function $RC(LS)$ transforming structural damages into monetary losses. We assume that the property value equals its reconstruction cost - on average 1500 euro per square metre, constant among all the municipalities \citep{AgenziaEntrate} - and define $RC(LS)$ as a fraction of the total reconstruction cost $RC$ through a function $RC(LS)$:
\begin{equation}
RC(LS) = \left(\frac{LS}{N_{LS_k}}\right) ^\alpha RC.
\end{equation}
where each limit state is represented by a positive integer and $N_{LS_k}$ is the number of limit states of model $k$. According to \cite{Asprone}, we assume $\alpha =1$.

\subsection{Flood}
Hydraulic literature offers very little about flood damage in Italy because the lack of uniform data at national level hinders research in this field. A few studies concern small geographical areas (usually cities, sometimes sections of river basins) and focus on the estimation of damages in the immediate follow-up of an event. Most of the analysis study the relationships between some flood's physical measurements and expected losses, and the most common output are depth-percent damage curves. Machine learning techniques have been recently applied to the creation of river basins hazard maps \citep{Degiorgis, Gnecco}. However, these techniques still require quite accurate data on past loss. Few example of probabilistic risk assessment have been developed for other countries also, and, similarly to \cite{Apel}, we decided to extend the deterministic post-event models available in the literature to probabilistic assessment. In this respect, we estimated expected losses by means of depth-percent damage curves from the existing literature and additional information on hazard and exposure from our database. In particular, two functions characterize our model: depth damage curves $g(\cdot)$ and depth probability, that might be represented by the density $f_{\delta}(\delta)$, the cumulative distribution $F_{\delta}(\delta)$ and the exceedance probability $\lambda(\delta) = 1 - F_{\delta}(\delta)$.

Similarly to seismic fragility curves, depth-damage curves refer to structural typologies. In particular, we consider the buildings' number of storeys and classify the housing stock into 3 classes ($j$) - 1, 2 and 3 or more storeys . A sample of depth-damage curves $g_j(\delta)$ has been selected from the engineering literature per each structural typology $j$. Unlike seismic fragility curves, depth-damage curves do not specify the probability that a given level of depth might produce a certain damage, and return the most likely outcome only. Moreover, the selected curves are ``depth-percent damage'', and indicate damages as percentages of property's total value.

Given the building's reconstruction cost $RC$, expected flood loss per square metre $l^f_{j,c}$ on a $j$-type building in the municipality $c$ can be estimated as:
\begin{multline}\label{eq:modello_flood}
l^{f}_{j,c} = \frac{RC}{100} \int_{0}^{\infty} \left[ g_j(\delta) \myabs{\frac{ \mbox{d} \left( \lambda  \left( \delta \right) \right) }{\mbox{d}\delta}} \right] \mbox{d}\delta = \frac{RC}{100} \int_0^{\infty} \left[ g_j(\delta) \myabs{\frac{ \mbox{d} \left[ 1 - F_{\delta}(\delta) \right]}{\mbox{d}\delta}} \right] \mbox{d}\delta = \frac{RC}{100} \int_0^{\infty} g_j(\delta) f_{\delta}(\delta) \mbox{d}\delta .
\end{multline}
By construction, there is a value $\delta_{j,max}$ after which a $g_j(\delta)=100$. Thus, equation (\ref{eq:modello_flood}) can be split in two parts as:
\begin{equation}\label{eq:modello_flood2}
l^{f}_{j,c} = \frac{RC}{100} \cdot \Bigg[ \int_0^{\delta_{max}} g_j(\delta) f_{\delta}(\delta)  \mbox{d}\delta +  100 \cdot \int_{\delta_{max}}^{\infty} f_{\delta}(\delta) \mbox{d}\delta \Bigg] .
\end{equation}
Bayes' theorem allow us to express $f_{\delta}(\delta)$ as the product of the probability of $\delta$ conditional to the occurrence of at least a flood event $f_{\delta\vert N_F}(\delta \vert N_F \geq 1)$ and the probability that at least one flood event occurs in a year:
\begin{equation}\label{eq:f_prob}
f_{\delta}(\delta)=P(N_F \geq 1) f_{\delta\vert N_F}(\delta \vert N_F \geq 1).
\end{equation}
When estimating losses, we are considering $N_F \geq 1$ only, thus substituting eq. (\ref{eq:f_prob}) into eq. (\ref{eq:modello_flood2}) leads to:
\begin{equation}
\begin{split}
l^{f}_{j,c} = \frac{RC}{100} \cdot P(N_F \geq 1) \cdot \Bigg[ \int_0^{\delta_{max}} g_j(\delta) f_{\delta\vert N_F}(\delta \vert N_F \geq 1) \mbox{d}\delta +  100 \cdot \int_{\delta_{max}}^{\infty} f_{\delta\vert N_F}(\delta \vert N_F \geq 1) \mbox{d}\delta \Bigg].
\end{split}
\end{equation}
Since $\int_{\delta_{max}}^{\infty} f_{\delta\vert N_F}(\delta \vert N_F \geq 1) \mbox{d}\delta = 1 - F_{\delta \vert N_F}(\delta_{max} \vert N_F \geq 1) = \lambda_{\delta\vert N_F}(\delta_{max} \vert N_F \geq 1)$, the model becomes:
\begin{equation}\label{eq:final_flood_model}
l^{f}_{j,c} = \frac{RC}{100} \cdot P(N_F \geq 1) \cdot \Bigg[ \int_0^{\delta_{max}}  g_j(\delta) f_{\delta\vert N_F}(\delta \vert N_F \geq 1)  \mbox{d}\delta +  100 \cdot \lambda_{\delta\vert N_F}(\delta_{max} \vert N_F \geq 1)\Bigg].
\end{equation}
Loss estimates per square metre per municipality and structural typology are multiplied by municipal exposure and aggregated into municipal flood losses $L_c^{f}$\\
\begin{equation}
    L_c^{f} = \sum_{j=1}^3 l_{j,c}^{f} \cdot E_{j,c}^{f}.
\end{equation}

\begin{figure}[t]
    \centering
    \caption{Flood frequency distribution.} \label{fig:cluster}
    \includegraphics[width=0.65\textwidth, trim=10mm 5mm 10mm 20mm, clip]{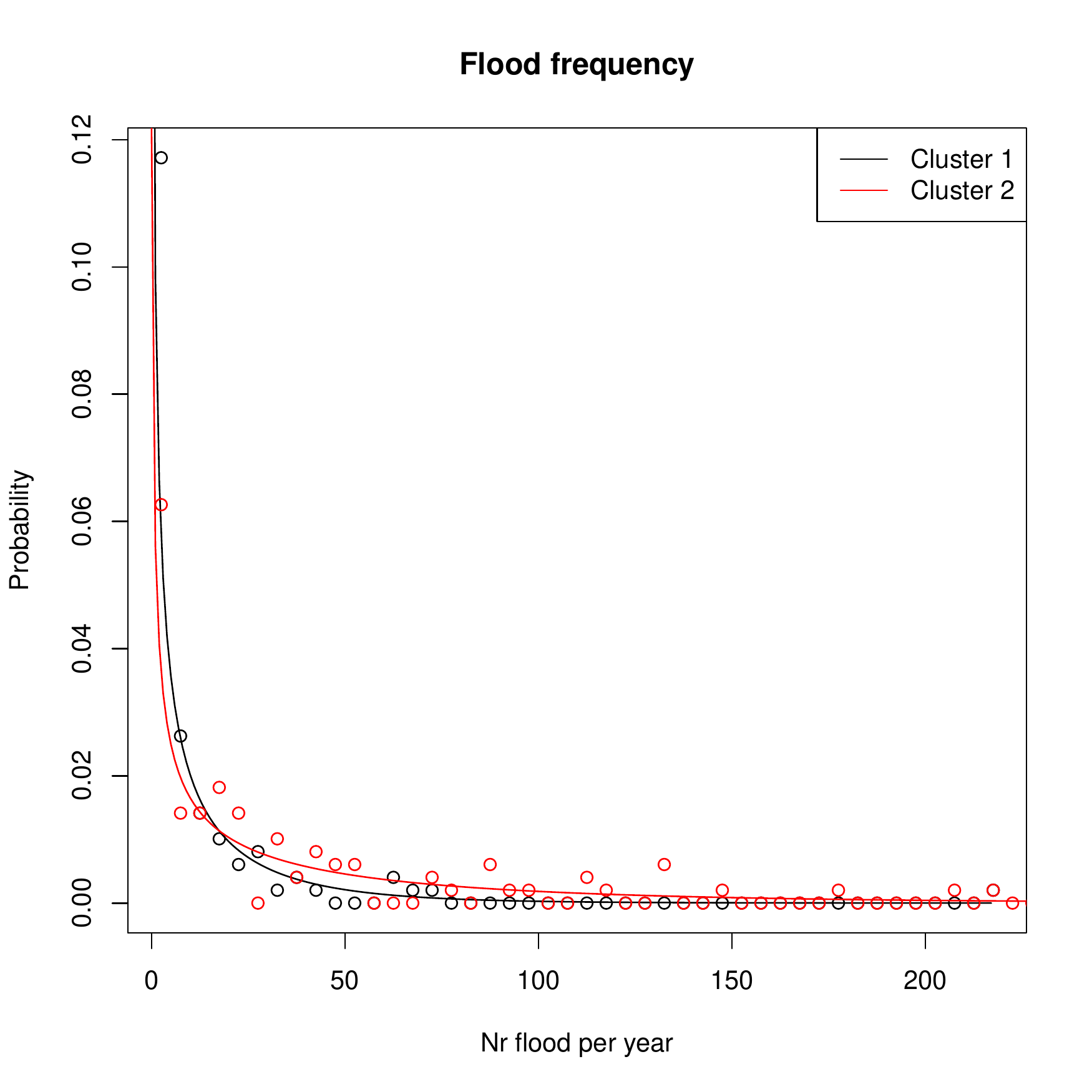}
    \caption*{\textbf{Note}: the plot divides observations (points) in two clusters: records from municipalities with $0<P2<0.5$ and $P2 \geq 0.5$. Both the clusters have been fitted with a negative binomial, as shown by the black and red lines.}
\end{figure}

\begin{figure}[t]
    \centering
    \caption{Depth probability distribution.}  \label{fig:depth frequency}
    \includegraphics[width=0.65\textwidth, trim=10mm 5mm 10mm 20mm, clip]{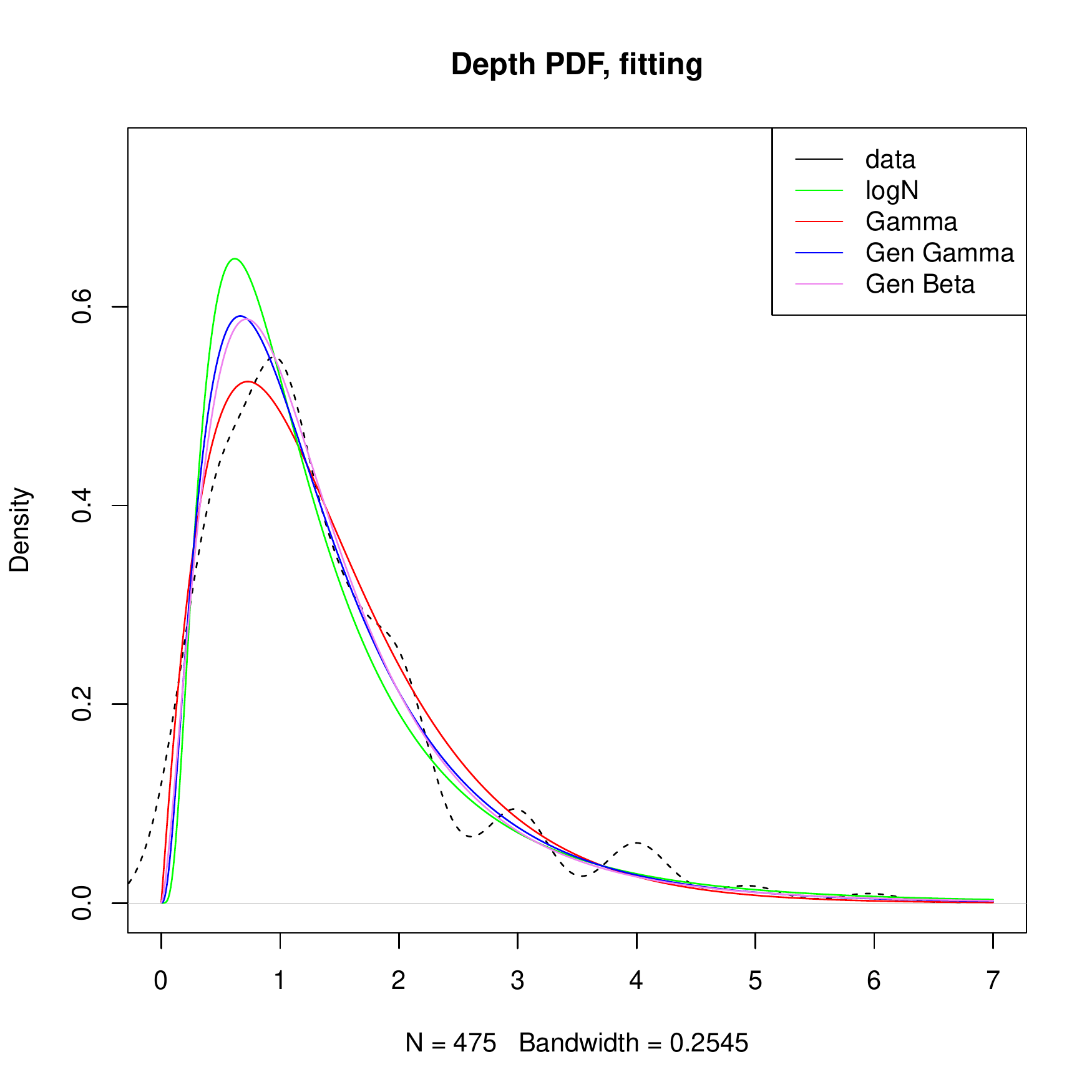}
    \caption*{\textbf{Note}: the dotted line is the empirical distribution $f_{\delta\vert N_F}(\delta \vert N_F \geq 1)$, and colored lines shows fitting.}
\end{figure}

\subsubsection{Hazard}
\label{ch:flood_hazard}
Flood hazard has been represented by frequency and depth probabilities. Both the distributions have been estimated from the AVI database and fitted by means of non-parametric techniques due to the lack of data. Since AVI gathers information from local press, it is likely that most remote events have not been captured. In particular, the number of floods listed after 1900 in the AVI archive is much higher than those recorded before and therefore we considered events occurred from that date onward only. Unfortunately, only $795$ events remain and they are too few to fit distributions at municipal level.

Frequency has been described by the probability density function of the number of floods in a year $f_{N_F}(N_F)$. In order to capture differences between the frequency of occurrence among the municipalities, data have been divided into two clusters - $A_{P_1}$ (120 obs.) and $A_{P_2}$ (620 obs.) - on the basis of the hydrological hazard index $P2$ from MRCI. Figure \ref{fig:cluster} shows that frequencies $f^{A_P}_{N_F}$ approximate negative binomial behaviour in both the two clusters. Despite the curves appear so close, they strongly differ in mean (the average number of floods per year is $11.95$ in $A_{P_1}$ and $42.58$ in $A_{P_2}$).

The probability of flood returns in each cluster is then adapted to fit the municipal and individual risk: since each flood involves a certain number of municipalities within the cluster $A_P$, the municipal probability of experiencing at least one flood in a year is estimated by multiplying $f^{A_P}_{N_F}$ times the average number $\bar{c}^{f}$ of municipalities flooded in $A_P$ over the number of municipalities $N^{A_P}_c$ in $A_P$:
\begin{equation}
    F^{c}_{N_F}(1) = \left( 1- f^{A_P}_{N_F}(0) \right) \frac{\bar{c}^f}{N^{AP}_c} \qquad c \in A_P.
\end{equation}
Floods usually strike several municipalities at the same time, but not all the properties in a flooded municipality will be hit by the flood. Therefore, the individual flood frequency does not coincide with the municipal one. We approximated the individual frequency probability by means of the $P3$ index in MRCI\footnote{Indicators P3 are not available for the entire Italian territory, since data are missing for part of Marche and Emilia-Romagna Regions.}, that indicates the percentage of municipal surface flooded in a 20-50 years probabilistic scenario. We indicate the index as $ext_{P3}$. Assuming homogeneously distributed buildings among the municipal area, the individual probability of flood returns is:
\begin{equation}\label{eq:flood_prob}
P(N_F\geq 1) = F^{c}_{N_F}(1) \cdot ext_{c,P3}.
\end{equation}
In addition to frequency, we estimate the probability of water to reach a certain depth during a flood. Depth information are missing for most of the events in the AVI database and sometimes are replaced by hydrometric heights measuring water depth from the river bed. We excluded hydrometric heights and assumed that depth levels reported in the database always correspond to the maximum reached in the area, which is a reasonable hypothesis since records in AVI are largely gathered from local press or compensation claims.

We found no significant difference in depth distributions between differently-exposed areas $A_P$ but this may be due to the low amount of available data, and therefore decided to estimate a unique function $f_{\delta\vert N_F}(\delta \vert N_F \geq 1)$ for the entire national territory. Since a flood usually hits more municipalities, a number of depth measurements are often reported for the same event but we represented each event with the maximum depth reported in the database. Hence, estimates have been computed on 475 observations.

The depth empirical distribution estimated from AVI data $f_{\delta\vert N_F}(\delta \vert N_F \geq 1)$ is shown in Figure \ref{fig:depth frequency}, where a graphical comparison between some distributions is presented too. Satisfactory fittings have been reached with the Generalized Beta (GB), the Generalized Gamma (GG) and the Gamma distributions. Table \ref{tab:flood.fit} shows that GG and GB's led to similar sum of squared errors and sum of absolute errors, while errors are much higher for the Gamma. The Chi squared goodness of fit test confirms the higher performance of GG and GB with respect to the Gamma, even though none of them reached a positive outcome. However, the likelihood ratio test shows weak evidence that the GG is more appropriate, therefore the Gamma has been chosen because of computational advantages.

\begin{table}[t!]
\centering
\caption{Flood depth distribution, goodness of fit.}\label{tab:flood.fit}
\begin{tabular}{c|cc}
  \hline
 & SSE & SAE \\ 
  \hline
Gamma & 0.02194857 & 0.2493763 \\ 
GG & 0.01328367 & 0.1951612 \\ 
GB & 0.01444061 & 0.2024778 \\ 
   \hline
\end{tabular}
\caption*{\textbf{Note}: this Table shows the sum of squared errors (SSE) and sum of absolute errors (SAE) obtained when fitting flood depth distribution with Gamma, Generalized Gamma (GG) and Generalized Beta (GB) distributions.}
\end{table}

\subsubsection{Exposure}
When evaluating structural vulnerability to floods, the number of storeys of the building is a fundamental feature to take into account. Therefore, buildings have been classified in three groups according to the number of storeys - one, two and three or more - in MRCI. Another element significantly affecting buildings resistance to floods is the presence/absence of a basement floor; since this information is not available, we assumed the two features to be equally distributed.

Given the number of buildings per structural typology within the municipality $B_{j,c}$, the average number of apartments per building $\bar{A}_c$ (ISTAT, census 2015) and the average apartment’s surface $\bar{s}_c$ \citep{AgenziaEntrate}, exposure has been estimated as:
\begin{equation}
E^f_{j,c}= \bar{s}_c \cdot B_{j,c} \cdot \bar{A}_c .
\end{equation}

\begin{table}[h]
\centering
\caption{Number of buildings per number of storeys.}\label{Table:flood_exposure}
\begin{tabular}{|c|c|}
\hline
   Number of Storeys &  Buildings (u=1000) \\
      \hline
 1 & 2083.39  \\
 2 & 5981.26 \\
 3 \mbox{ or more} & 4123.05 \\
\hline
\end{tabular}
\end{table}

\begin{figure}[t!]
\centering
\caption{Depth-percent damage curves for flood risk assessment.}  \label{fig:depth-percent damage curves}
\includegraphics[width=\textwidth, trim=0mm 0mm 0mm 10mm, clip]{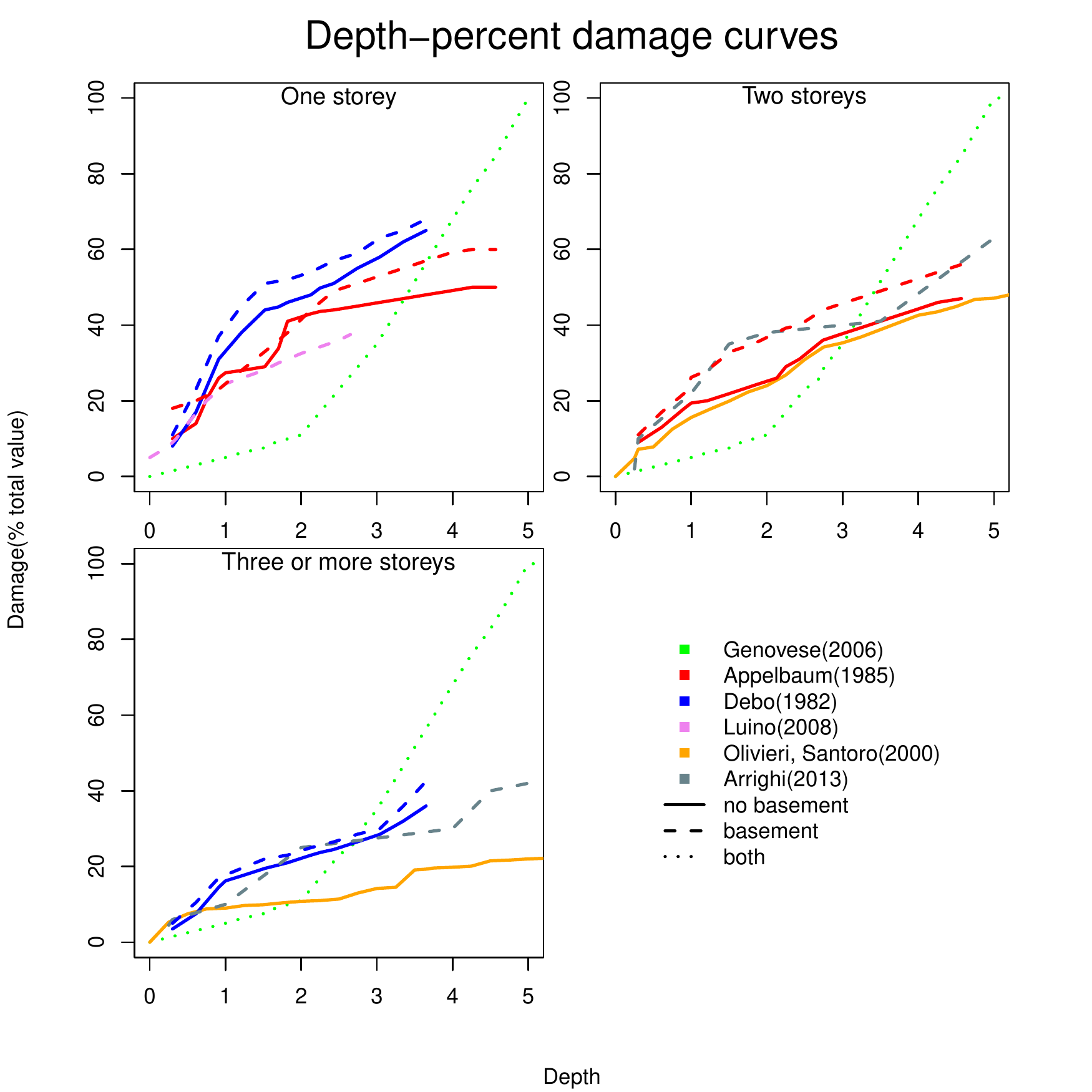}
\caption*{\textbf{Note}: selection of depth-percent damage curves for flood risk assessment. Curves are listed per buildings' number of storeys and can refer to dwellings with and/or without basement.}
\end{figure}

\begin{figure}[t]
\centering
\caption{Depth-percent damage curves.}
\includegraphics[width=0.65\textwidth, trim=10mm 5mm 10mm 20mm, clip]{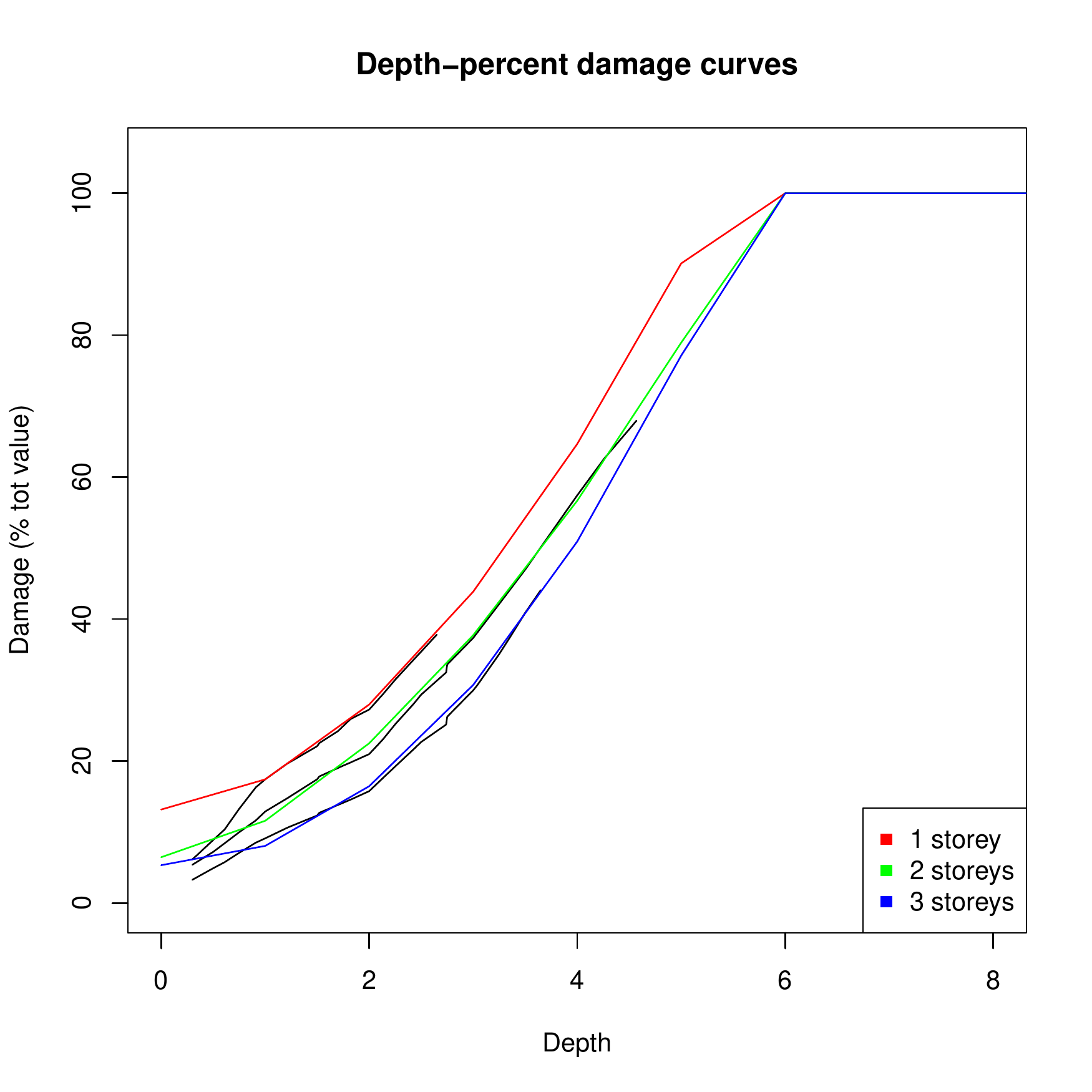}
\caption*{\textbf{Note}: black lines represent the average values of the curves selected per number of storeys. Red, green and blue lines show the functions fitted by polynomial regression.}
\label{fig:dd}
\end{figure}

\subsubsection{Vulnerability}\label{flood_vulnerability}
Flood's vulnerability is evaluated by depth damage curves defined on the building's number of storeys. The most widely adopted curves in hydraulic literature express damage as a percentage of building's total value and therefore called ``depth-damage curves''. Conversely to the curves expressing damages in absolute values, percentages curves are not affected by monetary volatility and are more reliable \citep{Appelbaum}.

Many studies have led to the definition of different depth-percent damage curves, that are strongly geographical-dependent \citep{Scorzini}: being derived from the analysis of historical data, they are in fact defined on the characteristics of the area under analysis and tend to lose accuracy when applied to contexts whose urban and territorial features differ too much from the original site.

We have selected depth-percent damage curves from six previous works \citep{Appelbaum,Arrighi,Debo,Genovese,Luino,OlivieriSantoro}, all either defined or tested on Italian data. The selection is reported in Figure \ref{fig:depth-percent damage curves}. Selected curves per structural typology have then been averaged into three new curves in order to guarantee higher reliability of results at the national level. Curves have been fitted by polynomial regressions, as shown in Figure \ref{fig:dd}.

\subsubsection{Loss}
\label{ch:flood_loss}
Structural damages have been converted into monetary terms by means of the function $\frac{RC}{100}$. Similar to the seismic model, we assume that the property value is equal to its reconstruction cost - on average 1500 euro per square metre, constant among all the municipalities \citep{AgenziaEntrate}.

\begin{center}
\begin{table}[ht]
\centering
\caption{Estimated seismic expected losses.}\label{tab:max.val_s1}
\begin{tabular}{rllllr}
  \hline
 & Structure & $\max(l_{j,c}^{s})$ (euro) & $\max(L^s_{j,c})$(Mln euro) & tot $L^s_j$ (Mln euro)\\ 
    \hline
\multirow{6}{*}{$\lambda(PGA) \sim PL$}  & RC.gl & 10.53 Castelbaldo (Padova) &  216.79 Roma  & 2223.61 \\ 
& RC.sl & 3.83 Castelbaldo (Padova) & 3.54 Roma & 130.70 \\ 
& A.gl & 4.03 Castelbaldo (Padova) & 7.16 Roma & 233.76 \\ 
& A.sl & 3.22 Castelbaldo (Padova) & 0.43 Roma & 30.73 \\ 
& M & 12.69 Castelbaldo (Padova) & 109.54 Roma & 3615.87 \\ 
& tot &  &  & 6234.661 \\ 
  \hline
\multirow{6}{*}{Asprone (2013)} & RC.gl & 17.04 Giarre (Catania) & 51.5 Roma & 1186.8 \\ 
 & RC.sl & 11.34 Navelli (L'Aquila) & 8.0 Reggio di Calabria & 489.9 \\ 
 & A.gl & 14.51 Giarre (Catania) & 25.1 Roma & 667.2 \\ 
  & A.sl & 11.71 Navelli (L'Aquila) & 2.4 Napoli & 174.0 \\ 
  & M  & 29.99 Giarre (Catania) & 196.4 Roma & 8661.8 \\ 
& tot & & & 11179.6 \\ 
   \hline
\end{tabular}
\caption*{\textbf{Note}: the table lists some descriptive statistics about estimated seismic expected losses per structural typology; in order: maximum expected loss per square metre $l_{j,c}^{f}$, maximum expected loss at the municipal level $L^s_{j,c}$ and the total expected loss $L^s_{j}$. The upper part describes current results, obtained by fitting PGA with a power law distribution; the lower side reports results by \cite{Asprone} for comparison.}
\end{table}
\end{center}

\begin{table}[ht]
\centering
\caption{Estimated flood expected losses.}\label{tab:max.val_f1}
\begin{tabular}{rlll}
  \hline
Structure & $\max(l_{j,c}^{f})$(euro) & $\max(L^f_j)$(Mln euro) & tot $L^f_j$(Mln euro)\\ 
     \hline
 1 storey & 19.61 Vigarano Mainarda (Ferrara) & 7.93 S. Michele al T. (Venezia) & 105.75 \\ 
 2 storeys & 15.16 Vigarano Mainarda (Ferrara) & 36.53 Ferrara & 536.14 \\ 
 3 storeys & 11.56 Vigarano Mainarda (Ferrara) & 18.24 Rimini & 234.01 \\ 
 tot       &                                   &               & 875.90 \\ 
   \hline
\end{tabular}
\caption*{\textbf{Note}: the Table shows descriptive statistics of flood expected losses per number of storeys. In order: maximum expected loss per square metre $l_{j,c}^{f}$, maximum expected loss at the municipal level $L^f_{j,c}$ and the total expected loss $L^f_{j}$.}
\end{table}

\begin{table}[ht]
\centering
\small
\caption{Municipalities with higher expected loss per natural risk.}\label{tab:rank}
\begin{tabular}{rlllr}
  \hline
  \multicolumn{5}{c}{Seismic Expected Loss}\\
  \hline
   & Municipality & Province & Region & $L^s_c$(Mln euro) \\ 
  \hline
1 & Roma & Roma & Lazio & 337.46 \\ 
  2 & Napoli & Napoli & Campania & 114.03 \\ 
  3 & Bologna & Bologna & Emilia-Romagna & 105.26 \\ 
  4 & Verona & Verona & Veneto & 59.58 \\ 
  5 & Firenze & Firenze & Toscana & 58.55 \\ 
  6 & Torino & Torino & Piemonte & 45.17 \\ 
  7 & Reggio di Calabria & Reggio di Calabria & Calabria & 43.23 \\ 
  8 & Modena & Modena & Emilia-Romagna & 39.57 \\ 
  9 & Prato & Prato & Toscana & 33.36 \\ 
  10 & Terni & Terni & Umbria & 33.03 \\ 
  11 & Ravenna & Ravenna & Emilia-Romagna & 31.69 \\ 
  12 & Rimini & Rimini & Emilia-Romagna & 30.38 \\ 
  13 & Messina & Messina & Sicilia & 29.92 \\ 
  14 & Pistoia & Pistoia & Toscana & 29.50 \\ 
  15 & Catania & Catania & Sicilia & 29.14 \\ 
  \hline
  \hline
\multicolumn{5}{c}{Flooding Expected Loss}\\
\hline
   & Municipality & Province & Region & $L^f_c$(Mln euro) \\ 
  \hline
1 & Ferrara & Ferrara & Emilia-Romagna & 56.22 \\ 
  2 & Ravenna & Ravenna & Emilia-Romagna & 52.89 \\ 
  3 & Rimini & Rimini & Emilia-Romagna & 45.03 \\ 
  4 & Pisa & Pisa & Toscana & 37.33 \\ 
  5 & San Michele al Tagliamento & Venezia & Veneto & 34.48 \\ 
  6 & Jesolo & Venezia & Veneto & 27.83 \\ 
  7 & Parma & Parma & Emilia-Romagna & 23.51 \\ 
  8 & Bologna & Bologna & Emilia-Romagna & 21.63 \\ 
  9 & San Donà di Piave & Venezia & Veneto & 21.31 \\ 
  10 & Cesenatico & Forlì-Cesena & Emilia-Romagna & 17.02 \\ 
  11 & Piacenza & Piacenza & Emilia-Romagna & 16.05 \\ 
  12 & Cervia & Ravenna & Emilia-Romagna & 15.40 \\ 
  13 & Verbania & Verbano-Cusio-Ossola & Piemonte & 14.58 \\ 
  14 & Forlì & Forlì-Cesena & Emilia-Romagna & 13.33 \\ 
  15 & Abano Terme & Padova & Veneto & 12.94 \\ 
   \hline
\hline
\end{tabular}
\end{table}

\subsection{Results}
\label{sec:expected loss}
Earthquake and flood losses have been estimated per municipality and structural typology. Seismic risk is described in Table \ref{tab:max.val_s1}, where results from \cite{Asprone} are also reported for comparison. We can note that, though the model adopted is the same, huge differences emerge between the two analysis. Several reasons contribute to these discrepancies and should be discussed for a better understanding results.

First of all, (i) estimates are highly sensitive to the probability distribution of hazard intensities, and while $\lambda(PGA)$ has been here fitted from INGV data, \cite{Asprone} rely on some distributional assumption. In addition, (ii) we assumed PGA values ranging in $\left[0,\infty\right[$, while the previous analysis considers $\left[0,2g\right]$ only. (iii) INGV data on PGA fails to represent many smaller municipalities that have here been approximated by means of neighbours' values and this assumption may have further contributed to the differences in results. (iv) Exposure strongly affect results too and while MRCI collects the number of dwellings per structural typology at the municipal level, \cite{Asprone} had information at the provincial level only. Moreover, MRCI refers to the 2011 population census, while the database used by \cite{Asprone} date back to 10 years earlier.

Arguments (i)-(iv) determine the different loss scenario, and, in particular, Table \ref{tab:max.val_s1} shows that estimated loss per square metre obtained by our model are considerably lower than those of \cite{Asprone}. The main reason is the adoption of a power law distribution that concentrates the probability on weaker events. However, our model highlights the gap in expected losses between more and less fragile buildings more than the older version.

Though our losses per square metre are lower than previous findings, the second column of the Table \ref{tab:max.val_s1} ($\max(l^s_{j,c})$) describe similar patterns. By contrast, expected losses per municipality and structural typology $L_{j,c}^s$ in the third column do not even show the same pattern. As argued before, exposure strongly affect results and the detailed information on buildings in MRCI allowed us to better represent real estate assets. In fact, Rome is the biggest municipality in Italy, and therefore its exposure produces expected losses that are extremely higher than those of other municipalities. By contrast, the homogeneous distribution of provincial structures among the municipalities in \cite{Asprone} very likely underestimates the exposure of major areas.

\begin{figure}
\centering
\caption{Seismic expected loss per square metre (masonry buildings).}\label{fig:masonry_loss}
\includegraphics[width=0.65\textwidth, trim=30mm 25mm 10mm 15mm, clip]{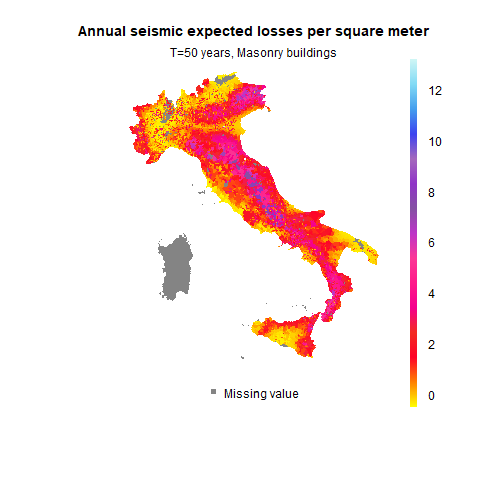}
\caption*{\textbf{Note}: the minimum value is $l^s_{j,c}=0.025$, maximum is $12.69$, and average value is $2.23$ euro per square metre.}\notag
\end{figure}

\begin{figure}
\centering
\caption{Expected seismic annual loss per municipality.}
\includegraphics[width=0.65\textwidth, trim=30mm 35mm 10mm 15mm, clip]{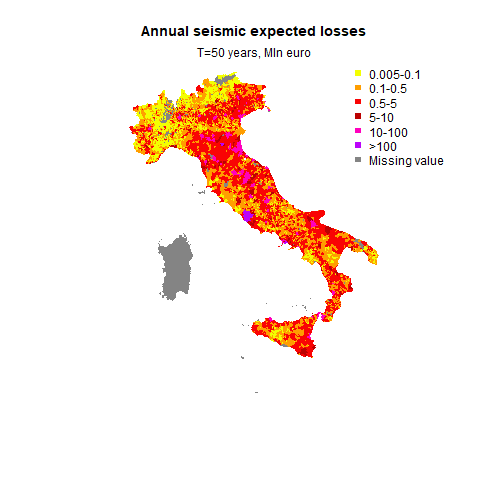}
\label{fig:seismic_loss_map}
\end{figure}

The fundamental role of exposure becomes clear when comparing $l_{j,c}^s$ and $L_{j,c}^s$ geographically. Figure \ref{fig:masonry_loss} represents the expected loss per square metre on the most vulnerable buildings - the masonry structures - in each municipality. The map reflects the hazard component of the risk model and clearly shows the proximity to risk sources. By contrast, this pattern in risk distribution is not evident in Figure \ref{fig:seismic_loss_map} showing annual total expected losses per municipality. In fact, the risky dark area delimited in Figure \ref{fig:masonry_loss} largely corresponds to the Appennino mountain chain, where several municipalities are sparsely inhabited. On the other hand, densely populated municipalities on the coast do not show extremely high level of loss per square metre but reach the highest expected losses at the aggregate level because of large real estates.

In order to appreciate the effect of different hazard and exposure components, one can consider reinforced concrete gravity loaded structures: though the power law distribution gets to a lower $\max (l^s_{j,c})$, the associated estimate of the expected loss $L^s_{j,c}$ in Rome is four times greater than that obtained in the previous paper.

Our analysis of seismic risk led to total expected loss equal to $6234.66$ million, which is almost half the value obtained by \cite{Asprone}. The value is seven times greater than the expected loss estimated for flood risk, equal to $875.90$ million per year, thus indicating that the earthquakes are the natural hazard of main concern in Italy.

\begin{figure}
\centering
\caption{Flood expected loss per square metre (one-storey buildings)}
\includegraphics[width=0.65\textwidth, trim=30mm 25mm 10mm 15mm, clip]{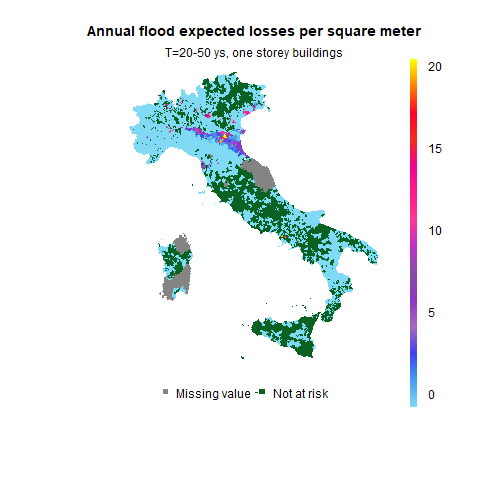}
\caption*{\textbf{Note}: ``Not at risk'' identifies municipalities where $l^f_{j,c}=0$. Among the other municipalities, the minimum loss is $2.24e^{-08}$. Maximum value is $l_{j,c}^f=19.61$. On average, expected loss in risky areas (municipalities ``Not at risk'' not included) is $0.37$ euro per square metre.}
\label{fig:one_storey_loss}
\end{figure}

\begin{figure}
\centering
\caption{Expected flood annual loss per municipality.}
\includegraphics[width=0.65\textwidth, trim=30mm 35mm 10mm 15mm, clip]{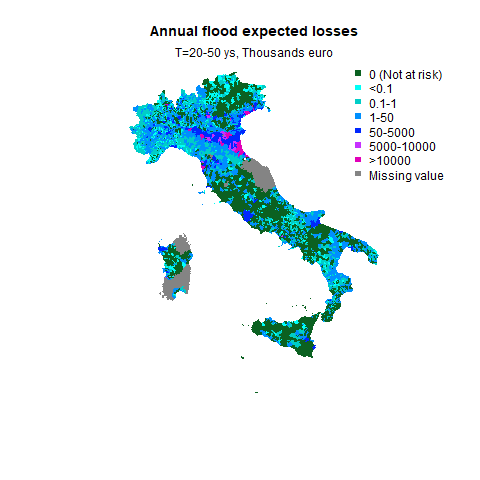}
\label{fig:flood_loss_map}
\end{figure}

As far as flood losses concern, main findings are presented in Table \ref{tab:max.val_f1}. Maximum losses per square metre $l^f_{j,c}$ are higher than the seismic ones, but Figure \ref{fig:one_storey_loss} shows that a great part of the territory does not appear to be affected by hydrological risk and most municipalities are associated to values of $l^f_{j,c}$ close to $0$. The map shows that the risk mostly affects northern Italy, and in particular the Emilia-Romagna, Veneto and Lombardia regions. More or less the same risk distribution is obtained at the aggregate level in Figure \ref{fig:flood_loss_map}, where the effect of exposure highlights additional areas of interest, such as the north-west coast, north Sardinia and Rome.

By comparing Figures \ref{fig:seismic_loss_map} and \ref{fig:flood_loss_map}, we can observe that north-east Italy is highly affected by both the two hazards, though the effect of floods remains consistently limited with respect to that of earthquakes. To conclude, Table \ref{tab:rank} ranks the fifteen largest expected municipal losses per each hazard. One can notice that three cities in Emilia-Romagna are listed for both: Bologna, Ravenna and Rimini.

\section{Public-private insurance model}
\label{Premium}
When constructing an insurance scheme, two fundamental quantities should be carefully evaluated: the premium per policyholder and the amount of reserves needed for the business given the solvency constraint. In the private market, insurers aim at profit and constitutes reserves through premium's collection. Therefore, rates should be sufficiently high to yield profit and avoid unacceptable levels of loss, while at the same time meeting an acceptable level of risk. Moreover, in order for the policy to be purchased, premiums should also meet the demand. Premium rating can hence be represented as a typical decision problem of profit and utility maximizing agents \citep{Mossin,Ehrlich}. However, when the government takes over the market, it radically changes the management objectives of the insurance company, and the traditional model is no longer suitable to capture agent's behavior. In this respect, this section presents a public-private insurance model for natural disasters where homeowners, insurers and government cooperate in risk financing.

Although the traditional private insurance-model has to be suitably modified to describe a public-private partnership, the problem can still be represented by comparing two perspectives: on the one hand, individuals who are willing to spend up to a certain amount on coverage; on the other hand, the insurer who, supported by the government, offers the policy subject to some solvency constraints.
The next two Subsections address the problem on the demand and supply perspectives respectively.

On the demand side, individuals face a standard decision problem and their utility functions can be defined as in the private-market literature. We keep the standard assumption of perfect information and rationality of individuals, though these hypotheses are often criticized as inappropriate to describe real world conditions \citep{Goda}.\footnote{Common shared information between insurer and insured is questionable in real contexts \citep{Cooper,KunreutherPauly1985}. In addition to lack of data for risk assessment, individuals also have to face limited cognitive capacity \citep{Kahneman,Goda} and imperfect rationality: \cite{Kunreuther} asserts that policy adoption conveys individual risk perception; \cite{Palm} observes that appreciation of earthquake policies’ benefits depends on personal attitude, socioeconomic and demographic characteristics, proximity to physical hazards, and past experience.} These criticisms are extremely important for the private market, but less relevant to our analysis since governments have the ability to modify the behavior of individuals by investing in risk education, promoting public awareness or introducing the obligation to purchase the policy.

On the supply side, the goal of insurance business is substantially affected by the partnership with the government. In the free market, insurer's goal is profit maximization subject to survival and/or stability constraints that require low ruin probability and low probability of high operational costs \citep{Goda}. By contrast, when entering the business, the government forces insurers to set the lowest premium possible given both the demand and the solvency constraints. Our model departs from traditional literature by assuming that the business endorses social welfare and therefore rates do not include profit-load. On the other side the government also supports the business by relaxing its financial burden: it partially subsidises reserves through capital injections and contributes to the reimbursement whenever stored funds are not sufficient for claim compensation.

Several additional issues arise in specifying the supply side. In particular, insurers' solvency constraint refer to the aggregate loss distribution, which is difficult to represent due to lack of information. While expected losses can be reconstructed through risk modeling as in Section \ref{Risk}, particular attention should be devoted to the variance as spatial correlations strongly affect insurer's potential of extreme losses. Quantifying correlation is difficult when records of past events are available, and it is practically impossible when they are not. However, it is reasonable to assume that spatial correlation between municipalities depends on their proximity, so that it can be identified a sufficiently large threshold $r$ such that two municipalities that are at least $r$-km far away are independent. We assume $r=50$ km, and include spatial correlation by means of the Hoeffding bound for $r$-dependent variables.

In addition to risk quantification, private insurers are also affected by  state regulations, market competition \citep{Grossi} and social or political decisions that may result in moral hazard and adverse selection \citep{KunreutherPauly2009}. Coordinating government and insurers actions can prevent these drawbacks, that should therefore not be included in the public-private model.

Finally, agents' attitudes toward risk should also be carefully evaluated. While in the literature is widespread agreed that homeowners are risk-averse, some evidence suggests that insurers may also exhibit risk-aversion \citep{Gollier}. Actuarial practice also encourages cautious behaviors, emphasizing the importance of adjusting rates by a risk-load component proportional to aggregate loss variance for extraordinary uncertain events such as natural hazards \citep{Kunreuther,Larsen}. However, by entering the business as a guarantor, the government release the insurer from its strict capital constraint and there is no need for the insurer to over-protect the reserves. We thus assume risk-averse homeowners and a risk-neutral insurer.

The two agent's perspectives are combined in Subsection \ref{ch:national_insurance_scheme} where the insurance scheme is defined. The application to Italian data is discussed in the following Subsection, and results for both the two hazards conclude the Section. Four different policies have been estimated and, thought seismic risk generates highest expected losses, the analysis shows that almost the same amount of public funds is necessary to manage the two risks. This section discusses single hazards policies only, multi-hazard analysis follows in the next Section.

\subsection{Homeowner's purchase decision}
\label{ch:homeowner_convenience}
Since the seminal papers by \cite{Mossin} and \cite{Ehrlich}, several premium setting models on insurance purchase decision have been developed. These models describe policies offered by the private sector and set premiums by comparing the risk-averse individual's willingness to pay with the profit maximization sought by the insurer. Thought we are considering a public-private partnership, individual's willingness to pay is left substantially unchanged. This subsection deals with the demand side, defines the utility function of the owners and gets to the maximum premium that they are willing to pay.

Let us consider a single peril insurance (i.e. earthquakes or floods only) in Italy. Any generic homeowner $i$ has an $m_i$ square metres property. The $N_i$ individuals gather in municipalities, thus any $i$ belongs to a generic Italian municipality $c$. A negative event has an annual probability $1-\pi_c(0)$ to hit the Municipality $c$ and ruin the $i$-th individual property at time $t$ causing a loss $l^a_{i,t}$ per square metre. Consider discrete time period $t$ equal to one year.

Individual $i$ may incur in a loss $l^a_{i,t}$ with probability $1-\pi_{c}(0)$, $i \in c$. This loss affects his wealth $w_{i,t}$, that we assume equal to the house value for simplicity. However, the individual may buy an insurance coverage and pay a premium $p_{i,t}$ per square metre to get a reimbursement $x_{i,t}$ per square metre in case that the event occurs. Let us define $x_{i,t}$ as a function of the actual loss $l^a_{i,t}$:
\begin{equation}\label{eq:rimborso_x}
x_{i,t} = \begin{cases} 0, & \mbox{with probability } \pi_{c}(0), \\ x \left(l^a_{i,t}\right), & \mbox{with probability } 1-\pi_{c}(0), \qquad 0 < x\left(l^a_{i,t}\right) \leq l^a_{i,t}, \end{cases} 
\end{equation}
with $i \in c$ and 
\begin{equation}\label{eq:rimborso_semplice}
x \left(l^a_{i,t}\right) = \begin{cases} 0 & \mbox{if } l^a_{i,t}\leq D,
\\ l^a_{i,t} - D & \mbox{if } D<l^a_{i,t}<E + D,
\\ E & \mbox{if } l^a_{i,t} \geq E + D , \end{cases} 
\end{equation}
where $D$ and $E$ are the deductible and the maximum coverage provided per square metre by the insurer.

The homeowner's utility of not being insured is traditionally expressed as the sum of two components representing the case of no events occurring during the year and a unique loss scenario:
\begin{equation}\label{eq:u_not_insured}
U_{\mbox{not insured}} = \pi_c (0) u(w_{i,t}) + \left( 1-\pi_c(0)\right)u(w_{i,t}-l^a_{i,t}m_{i,t}).
\end{equation}
Similarly, the utility of purchase is defined as:
\begin{equation}\label{eq:u_insured}
U_{\mbox{insured}} = \pi_c (0) u(w_{i,t}-p_{i,t}m_{i,t}) + (1- \pi_c(0) )u(w_{i,t}-p_{i,t}m_{i,t}-l^a_{i,t}m_{i,t}+x\left(l^a_{i,t}\right) m_{i,t}).
\end{equation}
Therefore, assuming rational behaviour, we can assume that the homeowner will buy an insurance coverage for its property if and only if its utility of purchasing is greater than or equal to that of not purchasing the policy: $U_{\mbox{insured}} \geq U_{\mbox{not insured}}$.

Considering any possible loss level, hence any possible phenomena intensity $\zeta$, we can define the probability $\pi_c (\zeta)$ that $c$ will experience a $\zeta$-intensity event in a year and that the homeowner $i$ living in municipality $c$ will suffer a loss $l^a_{i,t}(\zeta)$ expressed as a function of $\zeta$. In case he is owning a residential insurance coverage, its claim value will be then:
\begin{equation}
x_{i,t} = \begin{cases} 0, & \mbox{with probability } \pi_{c}(0), \\ x \left(l^a_{i,t}(\zeta)\right), & \mbox{with probability } \pi_{c}(\zeta), \qquad 0 < x\left(l^a_{i,t}(\zeta)\right) \leq l^a_{i,t}, \end{cases} \quad \mbox{with } i \in c.
\end{equation}
with
\begin{equation}\label{eq:rimborso}
x \left(l^a_{i,t}(\zeta)\right) = \begin{cases} 0 & \mbox{if } l^a_{i,t}(\zeta)\leq D,
\\ l^a_{i,t}(\zeta) - D & \mbox{if } D<l^a_{i,t}(\zeta)<E+D,
\\ E & \mbox{if } l^a_{i,t}(\zeta) \geq E+D \end{cases} 
\end{equation}
and the insured purchase-convenience condition becomes:
\begin{multline}\label{eq:Assicurato_base}
\pi_c (0) \cdot u(w_{i,t}) + \int_{0}^{\infty} \pi_c(\zeta) \cdot u(w_{i,t}-l^a_{i,t}(\zeta)m_{i,t}) \mbox{d}\zeta \\ \leq \pi_c (0) \cdot u(w_{i,t}-p_{i,t}m_{i,t}) + \int_{0}^{\infty} \pi_c(\zeta) \cdot u(w_{i,t}-p_{i,t}m_{i,t}-l^a_{i,t}(\zeta)m_{i,t}+x \left(l^a_{i,t}(\zeta)\right) m_{i,t}) \mbox{d}\zeta.
\end{multline}
According to the traditional literature on insurance purchasing decision, we assume the individual to be risk-averse and we represent its preferences by means of the utility function $u(x) = \log(x +1)$. We set $w_{i,t}$ equal to the house value and assume for simplicity that it corresponds to the reconstruction cost, equal to $RC$ per square metre. The logarithmic specification allows us to simplify the model considering losses per square metre, so we can rewrite eq. (\ref{eq:Assicurato_base}) as:
\begin{multline}\label{eq:Assicurato}
\pi_c (0) \cdot \log(RC+1) + \int_{0}^{\infty} \pi_c(\zeta) \log(RC-l^a_{i,t}(\zeta)+1) \mbox{d}\zeta \\ \leq \pi_c (0) \cdot \log(RC-p_{i,t}+1) + \int_{0}^{\infty} \pi_c(\zeta) \log(RC-p_{i,t}-l^a_{i,t}(\zeta)+x \left(l^a_{i,t}(\zeta)\right)+1) \mbox{d}\zeta.
\end{multline}
We assume that the premium $p_{i,t}$ is fixed at $t=0$ and does not vary with respect to time, $p_{i,t}=p_i$, and neither do inhabited square metres, so $m_{i,t}=m_i$. We can compute the highest premium that homeowners are willing to pay by restricting condition in eq. (\ref{eq:Assicurato}) to the equality, obtaining:
\begin{equation}\label{eq:Assicurato_ug}
\pi_c (0) \cdot \log \frac{(RC+1)}{(RC -p_{i}+1)} + \int \pi_c(\zeta) \log \frac{(RC-l^a_{i,t}(\zeta)+1)}{(RC-p_{i}-l^a_{i,t}(\zeta)+x \left(l^a_{i,t}(\zeta)\right)+1)} \mbox{d}\zeta = 0 .
\end{equation}
This equality states that the individual is indifferent to the decision to purchase the policy or not, and allows us to derive the risk-based maximum premium $p^H_{i}$ that the individual is willing to pay per structural typology and municipality.

\subsection{Public-private partnership}
\label{ch:Governmental_issues_SH}
We now consider the supply side, where the insurer and the government cooperate in risk management. As previously discussed, the goal of the business is maximizing social well-being, while financially protecting the insurer. Therefore, the government forces insurers to apply the lowest possible premiums, given both the demand and the solvency constraints, and offers its support to the business by partially subsidising reserves and committing to pay reimbursements whenever the reserve is not sufficient for claim compensation.

As the demand can be represented through the maximum premium that individuals are willing to pay, supply is concerned about the constitution of reserves in order to cope with possible future claims. At the beginning of the activity, say $t=0$, the insurer should create a reserve $W$, that will be increased every year by annual premiums $p_{i}$ collected from the $N_i$ individuals. Since the government supports the insurers, the reserve is partially subsidises by public capitals. Assume for simplicity that all the premiums are paid at the beginning of the year, while claims are paid when experienced. Hence, a minimum capital requirement $W_d$ should be fixed, so that the government will have to pay $W_d$ in $t=0$ and to refill the fund at the end of the year $t$ if $W_t$ goes below this threshold. So, at the beginning ($b$) of the year $t=0$ the initial reserve $W^{b}_{0}$ is created:
\begin{equation}
W_0^b = W_d + \sum^{N_i}_{i=1} p_{i} m_{i},
\end{equation}
and at the end ($e$) of the year it will be decreased of the total amount of reimbursement paid during the year:
\begin{equation}
W_0^e = W_0^b - \sum_{i=1}^{N_i} x_{i,0} m_{i}.
\end{equation}
Since claims $x \left(l^a_{i,t}(\zeta)\right)$ may incur at any random time $t$ and more events may happen close in time, the minimum capital requirement $W_d$ is necessary to guarantee money availability for reimbursement with a sufficiently high probability. Thus, if $W_0^e < W_d$ the government will refill it with an additional amount $W_r = W_d - W_{0}^e$.

At any subsequent time $t$, the fund value at the beginning of the year is:
\begin{equation}
W_t^b = W_{t-1} + \sum^{N_i}_{i=1} p_{i} m_{i} \qquad \mbox{with} \qquad W_{t-1} = \max(W^e_{t-1}; W_d),
\end{equation}
while at the end it will be:
\begin{equation}
W_t^e = W_t^b - \sum_{i=1}^{N_i} x_{i,t} m_{i}.
\end{equation}
However, the insurer is legally asked to meet some solvency constraint and hence need the government to set $W_d$ such that the probability of not being able to promptly pay the claims (``insolvency'' probability) below a certain low value $\epsilon_1$.

Let us assume that a negative event hits any building within a municipality. We assume that every policy can generate at most one claim per year and per individual; since reconstructing or restoring a building requires long time, this hypothesis is reasonable. Moreover, assume that actual square metre losses $l^a_{i,t}$ are equal for all the individuals within the same municipality and so does $x_{i,t}$. Consider the $N_c$ municipalities in Italy and indicate the total number of inhabited squared metres in the municipality $c$ as $M_{c}$, we have:
\begin{equation}
M_{c}= \sum_{i \in c} m_{i}, \qquad  \sum_{i \in c} x_{i,t} m_{i} = X_{c,t} M_{c}, \qquad \mbox{hence} \qquad X_{c,t}= \frac{\sum_{i \in c} x_{i,t} m_{i}}{M_{c}},
\end{equation}
so we can compute the total amount of claims as:
\begin{equation}\label{eq:total_claims_1}
Y_t = \sum_{i=1}^{N_i} x_{i,t} m_{i} = \sum_{c=1}^{N_c} \sum_{i \in c} x_{i,t} m_{i} = \sum_{c=1}^{N_c} X_{c,t} M_{c}.
\end{equation}
Since our policy covers at most one claim per year and per individual, claim occurrence per year and per municipality can be modelled as a Bernoulli random variable $\bar{X}_{c,t} \sim Ber(q_c)$ \cite{OlivieriPitacco} 
\begin{equation}
\bar{X}_{c,t} = \begin{cases} 1 & \mbox{with probability } q_c,
\\ 0 &  \mbox{with probability } 1-q_c . \end{cases} 
\end{equation}
with $q_c = \pi_c \left( \zeta > \zeta_D \right)$ and $\zeta_D$ such that $l^a_{i,t}(\zeta_D) = D$.

We can rewrite $Y_t$ as:
\begin{equation}\label{eq:total_claims}
Y_t = \sum_{c=1}^{N_c} X_{c,t} M_{c} = \sum_{c=1}^{N_c} M_{c} \bar{X}_{c,t} x \left( l^a_{c,t,j} \right) =  \sum_{c=1}^{N_c} \bar{X}_{c,t} \sum_j  M_{j,c} x \left( l^a_{c,t,j} \right)   = \sum_{c=1}^{N_c} \bar{X}_{c,t} a_{c,t},
\end{equation}
where $j$ indicates the structural typology and $M_{j,c}$ is the number of squared metres of properties of type $j$ in municipality $c$.

A main issue related to covering natural disasters is the high level of correlation between individual risks, which makes the description of the probability distribution of $Y_t$ non-trivial. There is no physical bound for energy propagation and this means that we cannot consider municipalities as perfectly independent among each other, especially in the earthquakes' case. By the way, natural phenomena hit neighbour cities, but far enough municipalities fairly never experience the same event. Therefore, it could be found a certain distance $r$ such that municipalities whose centroids are at least $r$ km far are independent. This assumption is similar to the \cite{Hoeffding}'s definition of $(r-1)$-dependence, and allows us to follow his work to model the national claim amount $Y_t$.

We sample municipalities in $N_g$ groups $Y^g$ of independent units, namely we create the groups in such a way that all the municipalities within a group are at least $r$ km apart from each other. The number $n_g$ of municipalities in group $g$ varies.

The total amount of claims in Italy can thus be obtained as:
\begin{equation}
Y_t = Y^1_t + Y^2_t + Y^3_t + \dots + Y^{N_g}_t,
\end{equation}
with
\begin{equation}
Y^g_t = \sum_{c\in g} \bar{X}_{c,t} a_{c,t}, \qquad c=1, \dots ,n_g.
\end{equation}
Each group claim amount $Y^g_t$ is the sum of $n_g$ independent and bounded random variables.

Assuming that the hazard distribution does not vary with respect to time too, expected losses do not depend on $t$, and neither do $E \left[ Y_t \right]$ and $E \left[ Y^g_{t} \right]$. Considering that $\int_{0}^{\zeta_D} \pi_c (\zeta)  x \left[ l^a_{j,c} \left( \zeta \right) \right]  \mbox{d} \zeta = 0$, the $g$-th group expected value:
\begin{equation}\label{eq:rimborso_gruppo_1}
E \left[ Y^g_{t} \right] = E \left[ Y^g \right] = \sum_{c\in g} \sum_j M_{j,c} \int_{0}^{\infty} \pi_c (\zeta)  x \left[ l^a_{j,c} \left( \zeta \right) \right]  \mbox{d}\zeta = \sum_{c\in g} \sum_j M_{j,c} \cdot E \left[ x \left( l^a_{j,c} \right) \right].
\end{equation}
The expected total amount of claims in Italy is:
\begin{equation}\label{eq:expected_Y}
    E \left[ Y_{t} \right] = E \left[ Y \right] = \sum_{g=1}^{N_g} E \left[ Y^g \right].
\end{equation}
Now we can define the insolvency probability at time $t$, and impose an upper bound $\epsilon_1$ on it:
\begin{equation}
Prob \Biggl\{ W_{t-1} + \sum_{i=1}^{N_i} p_{i}m_{i} - Y_t < 0 \Biggr\} < \epsilon_1 ,
\end{equation}
or equivalently:
\begin{equation}
Prob \Biggl\{ Y_t > W_{t-1} + \sum_{i=1}^{N_i} p_{i}m_{i} \Biggr\} < \epsilon_1 .
\end{equation}
We consider the worst case scenario $W_{t-1} = W_d$:
\begin{equation}\label{eq:prob}
Prob \Biggl\{ Y_t > W_d + \sum_{i=1}^{N_i} p_{i}m_{i} \Biggr\} < \epsilon_1 .
\end{equation}
The minimum capital requirement $W_d$ that the government should guarantee is then obtained by applying the \cite{Hoeffding} bound to our weighted sum of independent and bounded random variables:
\begin{equation}\label{eq:eq_Hoeff_epsilon1}
Prob \Bigg\{ Y_t > N_c \phi + E \left[ Y \right] \Bigg\} < \sum_{g=1}^{N_g} w_g e^{-h_1 \phi} E \left[ e^{\frac{h_1}{n_g}  \left(Y^g_t -E \left[ Y^g \right] \right) } \right], \qquad h_1>0,
\end{equation} 
with $w_g = \frac{n_g}{N_c}$.

Set
\begin{equation}\label{eq:fondo_base}
W_d + \sum_{i=1}^{N_i} p_{i}m_{i} = N_c \phi + E \left[ Y \right],
\end{equation}
and fix the right hand side of eq. (\ref{eq:eq_Hoeff_epsilon1}) equal to $\epsilon_1$:
\begin{equation}\label{epsilon_1}
\begin{split}
\epsilon_1= \sum_{g=1}^{N_g} w_g e^{-h_1 \phi} E \left[ e^{\frac{h_1}{n_g}  \left(Y^g_t -E \left[ Y^g \right] \right) } \right] = e^{-h_1 \phi} \sum_{g=1}^{N_g} w_g E \left[ e^{\frac{h_1}{n_g}Y_t^g}  e^{-\frac{h_1}{n_g}E \left[ Y^g \right] } \right]= \\ = e^{-h_1 \phi} \sum_{g=1}^{N_g} w_g e^{-\frac{h_1}{n_g}E \left[ Y^g \right] } E \left[ e^{\frac{h_1}{n_g}Y_t^g} \right] = e^{-h_1 \phi} \sum_{g=1}^{N_g} w_g e^{-\frac{h_1}{n_g}E \left[ Y^g \right] } E \left[ e^{\frac{h_1}{n_g}\sum_{c\in g} \bar{X}_{c,t} a_{c,t}} \right].
\end{split}
\end{equation}
The last expected value in eq. (\ref{epsilon_1}) is the moment generating function of the sum of random variables $\mathcal{M}_{Y^g_t}\left(\frac{h_1}{n_g}\right)$:
\begin{equation}
\begin{split}
E \left[ e^{\frac{h_1}{n_g}\sum_{c\in g} \bar{X}_{c,t} a_{c,t}} \right] = \prod_{c \in g} \mathcal{M}_{Y^g_t}\left(\frac{h_1}{n_g}\right)= \prod_{c \in g} \mathcal{M}_{\bar{X}_{c,t} a_{c,t}} \left( \frac{h_1}{n_g}\right),
\end{split}
\end{equation}
hence eq. (\ref{epsilon_1}) can be rewritten as:
\begin{equation}\label{eq:epsilon:1}
\epsilon_1 =e^{-h_1 \phi} \sum_{g=1}^{N_g} w_g e^{-\frac{h_1}{n_g}E \left[ Y^g \right] } \prod_{c \in g} \mathcal{M}_{\bar{X}_{c,t} a_{c,t}} \left(\frac{h_1}{n_g}\right) .
\end{equation}
Solving eq. (\ref{eq:epsilon:1}) we obtain $\phi$ as:
\begin{equation}
 \phi = \frac{1}{h_1} \log \left( \frac{\sum_{g=1}^{N_g} w_g e^{-\frac{h_1}{n_g}E \left[ Y^g \right] } \prod_{c \in g}  \mathcal{M}_{\bar{X}_{c,t} a_{c,t}} \left(\frac{h_1}{n_g}\right)}{\epsilon_1} \right)
\end{equation}
and estimate $W_d$ from eq. (\ref{eq:fondo_base}):
\begin{equation}\label{eq:fondo_Gov}
W_d = N_c \phi + E \left[ Y \right] -\sum_{i=1}^{N_i} p_{i}m_{i}. 
\end{equation}
Eq. (\ref{eq:fondo_Gov}) may result in a negative value of $W_d$, but we bind possible solutions to
\begin{equation}\label{eq:cond_Wd_positivo}
    W^*_d \geq 0.
\end{equation}
In case of $W_d < 0$, we assume that the government will decide to set it equal to $0$ and keep an insolvency probability even lower than the desired level: $\epsilon^*_1 \leq \epsilon_1$.

Moreover, it is reasonable to suppose that the government aims to minimize the probability to refill the fund with additional capital $W_r = W_d - W^e_{t}$, so it will need to set a premium sufficiently high to guarantee a low probability bounded from above by $\epsilon_2$ to pay that quantity at any time $t$:
\begin{equation}\label{eq:prob_2}
Prob \Biggl\{ W_d - W^e_{t} > 0 \Biggr\} = Prob \Biggl\{ W_d - W_{t-1} - \sum^{N_i}_{i=1} p_{i} m_{i} + Y_t > 0 \Biggr\} < \epsilon_2 .
\end{equation}
Once again, consider the worst case scenario $W_{t-1} = W_d$:
\begin{equation}
Prob \Biggl\{ W_d - W_d - \sum^{N_i}_{i=1} p_{i} m_{i} + Y_t > 0 \Biggr\} =  Prob \Biggl\{ Y_t - \sum^{N_i}_{i=1} p_{i} m_{i} > 0 \Biggr\} < \epsilon_2 ,
\end{equation}
or equivalently:
\begin{equation}
Prob \Biggl\{ Y_t > \sum^{N_i}_{i=1} p_{i} m_{i} \Biggr\} < \epsilon_2 .
\end{equation}
Note that this condition applies a new constraint on the premiums' value.

Given a sufficiently low probability $\epsilon_2$, we can define the minimum amount of total premiums by applying again the \cite{Hoeffding} inequality:
\begin{equation}\label{eq:eq_Hoeff_epsilon2}
Prob \Bigg\{ Y_t > N_c \gamma + E \left[ Y \right]  \Bigg\} < e^{-h_2 \gamma} \sum_{g=1}^{N_g} w_g e^{-\frac{h_2}{n_g} E\left[ Y^g \right]} \prod_{c \in g}  \mathcal{M}_{\bar{X}_{c,t} a_{c,t}} \left(\frac{h_2}{n_g}\right)
 , \qquad h>0.
\end{equation}
Set
\begin{equation}
\epsilon_2 = e^{-h_2 \gamma} \sum_{g=1}^{N_g} w_g e^{-\frac{h_2}{n_g} E\left[ Y^g \right]} \prod_{c \in g}  \mathcal{M}_{\bar{X}_{c,t} a_{c,t}} \left(\frac{h_2}{n_g}\right)
\end{equation}
and get
\begin{equation}
 \gamma = \frac{1}{h_2} \log \left( \frac{\sum_{g=1}^{N_g} w_g e^{-\frac{h_2}{n_g}E\left[ Y^g \right]} \prod_{c \in g}  \mathcal{M}_{\bar{X}_{c,t} a_{c,t}} \left(\frac{h_2}{n_g}\right)}{\epsilon_2} \right)
\end{equation}
which in turn allows us to estimate the minimum allowable value of the sum of premiums $\sum_{i=1}^{N_i} p^G_{i} m_i$:
\begin{equation}\label{eq:min_tot_premium}
\sum_{i=1}^{N_i} p^G_{i} m_{i} =  N_c \gamma + E \left[ Y \right].
\end{equation}

\subsection{Insurance model}
\label{ch:national_insurance_scheme}
The maximum value $p^H_{i}$ that each individual is willing to pay in eq. (\ref{eq:Assicurato_ug}) and the minimum amount of total premium necessary to avoid excessive government risk-exposure $\sum_{i=1}^{N_i} p^G_{i} m_{i}$ in eq. (\ref{eq:min_tot_premium}) are the two constraints that the supply faces when defining a national insurance scheme. The two equations pose conditions on rates and they may either identify a range of possible values or fail to find a unique solution. However, since we are focused on a publicly supported insurance scheme, it is reasonable to assume that the government will keep the premium as low as possible in order not to financially over-stress homeowners, though this may imply a higher probability of found refill at each $t$, thus a greater risk for public resources. Therefore, given the desired probability $\epsilon_2$ of government non-financial over-stress we define the optimal premium level $p^{*}_{i}$ as:

\begin{equation}\label{eq:optimum_premium}
p^{*}_{i}= \min (c, 1) \cdot p^H_{i} \qquad \mbox{with} \qquad c= \frac{\sum_{i=1}^{N_i} p^G_{i} m_{i}}{\sum_{i=1}^{N_i} p^H_{i} m_{i}}.
\end{equation}
Premiums as defined in eq. (\ref{eq:optimum_premium}) are risk-based on municipality hazard and individual structural typology, thus guaranteeing social fairness. Moreover, the equation implies that:
\begin{equation}
\sum_{i=1}^{N_i} p^{*}_{i} m_{i} = \min (c, 1) \sum_{i=1}^{N_i} p^{H}_{i} m_{i} = \min \left( 1, \frac{1}{c} \right) \sum_{i=1}^{N_i} p^{G}_{i} m_{i} = \min \left( 1, \frac{1}{c} \right) \left( N_c \gamma + E \left[ Y \right] \right) = N_c \gamma^{*} +  E \left[ Y \right],
\end{equation}
thus $\gamma^{*}$ is
\begin{equation}
    \gamma^{*} = \frac{ \min \left( 1, \frac{1}{c} \right) \left( E \left[ Y \right] + N_c \gamma \right) E \left[ Y \right] }{N_c} \leq \gamma
\end{equation}
and the insurer is thus able to guarantee an upper bound $\epsilon^{*}_2$ on the probability to refill the fund equal to: 
\begin{equation}\label{epsilon_2_optimum}
\epsilon^*_2 = \frac{\sum_{g=1}^{N_g} w_g e^{-\frac{h_2}{n_g}E\left[ Y^g \right]} \prod_{c \in g}  \mathcal{M}_{\bar{X}_{c,t} a_{c,t}} \left(\frac{h_2}{n_g}\right)}{e^{h_2 \gamma^{*}}} \geq \epsilon_2.
\end{equation}
Given the desired upper bound on insolvency probability $\epsilon_1$, the optimal capital minimum requirement $W^{*}_d$ is then obtained from condition (\ref{eq:fondo_Gov}):
\begin{equation}\label{eq:optimum_fondo}
W^{*}_d = \max \left\{ N_c \phi + E \left[ Y \right] -\sum_{i=1}^{N_i} p^{*}_{i}m_{i}  ; 0  \right\} = N_c \phi^{*} + E \left[ Y \right] -\sum_{i=1}^{N_i} p^{*}_{i}m_{i},
\end{equation}
with
\begin{equation}
    \phi^{*} = \frac{ W^{*}_d + \sum_{i=1}^{N_i} p^{*}_{i}m_{i} - E \left[ Y \right]}{N_c} \geq \phi.
\end{equation}
Thus, the optimal value $\epsilon^{*}_1$ is:
\begin{equation}\label{epsilon_1_optimum}
\epsilon^{*}_{1}= \frac{\sum_{g=1}^{N_g} w_g e^{-\frac{h_1}{n_g}E\left[ Y^g \right]} \prod_{c \in g}  \mathcal{M}_{\bar{X}_{c,t} a_{c,t}} \left(\frac{h_1}{n_g}\right)}{e^{h_1 \phi^{*}}}.
\end{equation}
Since $\epsilon_1$ decreases as $\phi$ increases, the optimal insolvency probability will be at most equal to the level desired by the insurer: $\epsilon^*_1 \leq \epsilon_1$.

Moreover, note that:
\begin{multline}\label{eq:condition_fondo}
W^{*}_d = N_c \phi^{*} + E \left[ Y \right] -\sum_{i=1}^{N_i} p^{*}_{i}m_{i} = N_c \phi^{*} + E \left[ Y \right] - E \left[ Y \right] - N_c \gamma^{*} = N_c \left( \phi^{*} - \gamma^{*} \right).
\end{multline}
From eq. (\ref{epsilon_2_optimum}) and (\ref{epsilon_1_optimum}), $\gamma^{*}$ and $\phi^{*}$ can be defined as:
\begin{equation}\label{eq:gamma_star}
    \gamma^{*} = \frac{1}{h_2} \log \left( \frac{\sum_{g=1}^{N_g} w_g e^{-\frac{h_2}{n_g}E\left[ Y^g \right]} \prod_{c \in g}  \mathcal{M}_{\bar{X}_{c,t} a_{c,t}} \left(\frac{h_2}{n_g}\right)}{\epsilon_2^{*}} \right)
\end{equation}
and
\begin{equation}\label{eq:phi_star}
    \phi^{*} = \frac{1}{h_1} \log \left( \frac{\sum_{g=1}^{N_g} w_g e^{-\frac{h_1}{n_g}E\left[ Y^g \right]} \prod_{c \in g}  \mathcal{M}_{\bar{X}_{c,t} a_{c,t}} \left(\frac{h_1}{n_g}\right)}{\epsilon_1^{*}} \right).
\end{equation}
Given the condition in eq. (\ref{eq:cond_Wd_positivo}), eq. (\ref{eq:condition_fondo}) implies
\begin{equation}\label{eq:fondo_sviluppo}
         \frac{\left( \sum_{g=1}^{N_g} w_g e^{-\frac{h_1}{n_g}E\left[ Y^g \right]} \prod_{c \in g}  \mathcal{M}_{\bar{X}_{c,t} a_{c,t}} \left(\frac{h_1}{n_g}\right) \right)^{\frac{1}{h_1}}}{\left( \sum_{g=1}^{N_g} w_g e^{-\frac{h_2}{n_g}E\left[ Y^g \right]} \prod_{c \in g}  \mathcal{M}_{\bar{X}_{c,t} a_{c,t}} \left(\frac{h_2}{n_g}\right) \right) ^{ \frac{1}{h_2}} } \cdot \frac{ \left(\epsilon_2^{*}\right)^{\frac{1}{h_2}} }{ \left( \epsilon_1^{*}\right)^{\frac{1}{h_1}}} \geq 1.
\end{equation}
In particular, if a parameter $h=h_1=h_2$ is chosen, eq. (\ref{eq:condition_fondo}) becomes
\begin{equation}\label{eq:condition_fondo_finale}
    W^{*}_d = \frac{N_c}{h} \log \left( \frac{\epsilon_2^{*}}{\epsilon_1^{*}} \right).
\end{equation}
Eq. (\ref{eq:condition_fondo_finale}) shows that the amount of public resources needed increases with the ratio $\epsilon_2^{*} / \epsilon_1^{*}$, and more importantly, eq. (\ref{eq:fondo_sviluppo}) collapses to:
\begin{equation}
\epsilon_2^{*} \geq \epsilon_1^{*},
\end{equation}
indicating that insolvency should never be preferred to the disbursement of public funds, thus enforcing the Government role of social guarantor. The minimum $W^{*}_d$ value corresponds to $\epsilon_1^{*}=\epsilon_2^{*}$ and is equal to 0.

However, $\epsilon_2^{*} / \epsilon_1^{*}$ affects $W^{*}_d$ logarithmically, while the capital requirement is largely determined by $N_c / h$. Therefore, $W^{*}_d$ is directly proportional to the number of municipalities, and inversely related to the parameter $h$, whose value is determined by the Government's initial preferences $\epsilon_1$ and $\epsilon_2$ and the overall risk distribution.

\subsection{Application to Italy}\label{sec:Application}
\subsubsection{Individuals' willingness to pay for seismic policies.}
\label{sec:will_seismic}
Premium model application to the seismic case requires particular attention due to the hazard component $\zeta = PGA$. We can estimate $\pi(\zeta)$ as $\pi(PGA) = \lvert \frac{ \mbox{d}\lambda(PGA)}{\mbox{d}(GPA)} \rvert$. The absence of seismic movements $\zeta = 0$ corresponds to the case of no seismic event happening in the year, thus we have $l_{i,t}(0)=0$ and $x \left(l_{i,t}(\zeta)\right)=0$. This allows us to include the case of no seismic event in the integral term of condition (\ref{eq:Assicurato_ug}):
\begin{equation}\label{eq:ass_def_sismico}
\int_0^{\infty} \pi_c(PGA) \log \frac{(RC-l^a_{i,t}(PGA)+1)}{(RC-p_{i,t}-l^a_{i,t}(PGA)+x \left(l^a_{i,t}(PGA)\right)+1)} \mbox{d}(PGA) = 0 .
\end{equation}
In Section \ref{seismic_hazard} we have shown that $\lambda_c(PGA)$ approximately behaves as a Power Law distribution and therefore we have:
\begin{equation}
\pi_c(PGA) = \myabs{\frac{d(\lambda(PGA))}{d(PGA)}}  = \alpha_c  PGA^{-\beta_c},
\end{equation}
whose domain does not include values in $[0,PGA_{min_c}[$, with
\begin{equation}
PGA_{min_c} = e^{\frac{\log\left( \frac{\alpha_c}{\beta_c-1} \right)}{\beta_c-1}}.
\end{equation}
This implies that, in this case, the integral in condition (\ref{eq:ass_def_sismico}) cannot be evaluated in $[0,+\infty[$ but in $[ PGA_{min_c}, +\infty[$ only. However, $PGA_{min_c}$ take values ranging from $ 7.92 e^{-09}$ to $0.002$, and are small enough to include the case of no seismic loss.

The loss function per structural typology $l^a_{j,t}(PGA)$ is derived from eq. (\ref{eq_model}):
\begin{equation}
l^a_{j,t}(PGA) = \frac{1}{K_j} \sum_{k=1}^{K_j} \sum_{LS=1}^{N_{LS_k}} RC(LS) \cdot \left[ P_k \left(LS \vert PGA \right)- P_k \left(LS+1 \vert PGA \right)\right],
\end{equation}
with $P_k \left(N_{LS_k} +1 \vert PGA \right) = 0$.

Condition (\ref{eq:Assicurato_ug}) for seismic risk becomes:
\begin{equation}
    \int_{PGA_{min_c}}^{\infty} \alpha_c PGA^{-\beta_c} \log \left( \frac{RC-l^a_{i,t}(PGA)+1}{RC-p_{i,t}-l^a_{i,t}(PGA)+x \left(l^a_{i,t}(PGA)\right)+1} \right) \mbox{d}(PGA) = 0.
\end{equation}

\subsubsection{Individuals' willingness to pay for flood policies.}\label{sec:will_flood}
The premium model application to flood is simpler with respect to the seismic. Here, hazard is represented by depth $\zeta = \delta$ and $l_{i,t}(\delta)$ is obtained by the depth-percent damage curve $g_j(\delta)$ for the number of storeys $j$. The probabilistic component $\pi_c(\zeta)$ is given by $f(N_F)$ defined in equation (\ref{eq:f_prob}), whose estimation has been discussed in section \ref{ch:flood_hazard}. We define the individual flooding probability from equation (\ref{eq:flood_prob}) as:
\begin{equation}
P(N_F \geq 1) = (1 - f_{N_F}^{A_P}(0)) \cdot ext_{c,P3} \cdot \frac{\bar{c}^{f}}{N_c^{AP}}  .
\end{equation}
The probability of no flood events in a year $\pi_c(0)$ is then defined as:
\begin{equation}
f_{N_F}(0) = \left[ 1 - (1 - f_{N_F}^{A_P}(0)) \cdot ext_{c,P3} \cdot \frac{\bar{c}^{f}}{N_c^{AP}} \right] := \pi_c(0) ;
\end{equation}
while $\pi_c(\delta)$ corresponds to:
\begin{equation}
\pi_c(\delta)= (1 - f_{N_F}^{A_P}(0)) \cdot ext_{c,P3} \cdot \frac{\bar{c}^{f}}{N_c^{AP}} \cdot f_{\delta \lvert N_F} (\delta \lvert N_F\geq 1) 
\end{equation}
So condition (\ref{eq:Assicurato_ug}) becomes:
\begin{multline}
\left[ 1 - (1 - f_{N_F}^{A_P}(0)) \cdot ext_{c,P3} \cdot \frac{\bar{c}^{f}}{N_c^{AP}} \right] \cdot \log \left( \frac{RC+1}{RC-p_{i,t}+1} \right) + (1 - f_{N_F}^{A_P}(0)) \cdot  ext_{c,P3} \cdot \frac{\bar{c}^{f}}{N_c^{AP}} \cdot \\ \cdot \int_{0}^{\infty} f_{\delta \lvert N_F}(\delta \lvert N_F \geq 1) \cdot \log \left( \frac{RC-\frac{RC}{100} g_j(\delta)+1}{RC-p_{i,t}-\frac{RC}{100} \cdot g_j(\delta) + x \left[ \frac{RC}{100} \cdot g_j(\delta) \right] +1} \right) \mbox{d}\delta  = 0 .
\end{multline} 
We focus on the integral in the second term, and split it into two components:
\begin{multline}
\int_{0}^{\infty} f_{\delta \lvert N_F}(\delta \lvert N_F \geq 1) \cdot \log \left( \frac{RC- \frac{RC}{100} \cdot g_j(\delta)+1}{RC-p_{i,t}- \frac{RC}{100} \cdot g_j(\delta) + x \left[ \frac{RC}{100} \cdot g_j(\delta) \right] +1} \right) \mbox{d}\delta = \\ = \int_{0}^{\infty} f_{\delta \lvert N_F}(\delta \lvert N_F \geq 1) \cdot \log \left( RC- \frac{RC}{100} \cdot g_j(\delta)+1 \right) \mbox{d}\delta +  
\\ - \int_{0}^{\infty} f_{\delta \lvert N_F}(\delta \lvert N_F \geq 1 ) \cdot \log \left(RC-p_{i,t}- \frac{RC}{100} \cdot g_j(\delta) + x \left[ \frac{RC}{100} \cdot g_j(\delta) \right] +1 \right) \mbox{d}\delta,
\end{multline}
then, we consider them separately.

Since $g_j$ is a positive non-decreasing function that becomes constant at level $100\%$ corresponding to a certain depth $\delta_{max}$, the first integral can be simplified to:
\begin{multline}
\int_{0}^{\infty} f_{\delta \lvert N_F}(\delta \lvert N_F \geq 1) \cdot \log \left( RC-\frac{RC}{100} \cdot g_j(\delta)+1 \right) \mbox{d}\delta = \\
= \int_{0}^{\delta_{max}} f_{\delta \lvert N_F}(\delta \lvert N_F \geq 1) \cdot \log \left( RC- \frac{RC}{100} \cdot g_j(\delta)+1 \right) \mbox{d}\delta
+ \int_{\delta_{max}}^{\infty} f_{\delta \lvert N_F}(\delta \lvert N_F \geq 1) \cdot \log \left( 1 \right) \mbox{d}\delta = \\ = \int_{0}^{\delta_{max}} f_{\delta \lvert N_F}(\delta \lvert N_F \geq 1) \cdot \log \left( RC- \frac{RC}{100} \cdot g_j(\delta)+1 \right) \mbox{d}\delta.
\end{multline}
The second integral involves two piecewise functions: $g_j(\delta)$  and $x \left( g_j(\delta) \right)$. Defining $\delta_D$ such that $g_j({\delta_D}) \cdot \frac{RC}{100} = D $ and $\delta_E$ such that $g_j({\delta_E})  \cdot \frac{RC}{100} = E + D$ and considering $\delta_{max}$, we can split it into 4 components:
\begin{multline}
\int_{0}^{\infty} f_{\delta \lvert N_F}(\delta \lvert N_F \geq 1) \cdot \log \left(RC-p_{i,t}- \frac{RC}{100} \cdot g_j(\delta) + x \left[ \frac{RC}{100} \cdot g_j(\delta) \right] +1 \right) \mbox{d} \delta = \\ =
\int_{0}^{\delta_D} f_{\delta \lvert N_F}(\delta \lvert N_F \geq 1) \cdot \log \left(RC-p_{i,t}- \frac{RC}{100} \cdot g_j(\delta) +1 \right) \mbox{d} \delta + \\ 
+ \log \left( RC - p_{i,t} - D +1 \right) \cdot \left[ F_{\delta \lvert N_F} (\delta_E \lvert N_F \geq 1) - F_{\delta \lvert N_F} (\delta_D \lvert N_F \geq 1) \right] + \\
+ \int_{\delta_E}^{\delta_{max}} f_{\delta \lvert N_F}(\delta \lvert N_F \geq 1) \cdot \log \left(RC-p_{i,t}- \frac{RC}{100} \cdot g_j(\delta) + E  +1 \right) \mbox{d} \delta + \\ + \log \left( E  - p_{i,t} + 1 \right) \left[ 1 - F_{\delta \lvert N_F} (\delta_{max} \lvert N_F \geq 1) \right] .
\end{multline}
Summing up, insured purchasing indifference condition (\ref{eq:Assicurato_ug}) for the flood case study is:
\begin{multline}\label{eq:ass_flood_def}
\left[ 1 - (1 - f_{N_F}^{A_P}(0)) \cdot ext_{c,P3} \cdot \bar{c}^{f} \right] \cdot \log \left( \frac{RC+1}{RC-p_{i,t}+1} \right) + (1 - f_{N_F}^{A_P}(0)) \cdot ext_{c,P3} \cdot \bar{c}^{f} \cdot \\
\cdot \Bigg\lbrace \int_{0}^{\delta_{max}} f_{\delta \lvert N_F}(\delta \lvert N_F \geq 1) \cdot \log \left( RC- \frac{RC}{100} \cdot g_j(\delta)+1 \right) \mbox{d}\delta + \\ - \int_{0}^{\delta_D} f_{\delta \lvert N_F}(\delta \lvert N_F \geq 1) \cdot \log \left(RC-p_{i,t}- \frac{RC}{100} \cdot g_j(\delta) +1 \right) \mbox{d} \delta + \\ 
- \log \left( RC - p_{i,t} - D +1 \right) \cdot \left[ F_{\delta \lvert N_F} (\delta_E \lvert N_F \geq 1) - F_{\delta \lvert N_F} (\delta_D \lvert N_F \geq 1) \right] + \\
- \int_{\delta_E}^{\delta_{max}} f_{\delta \lvert N_F}(\delta \lvert N_F \geq 1) \cdot \log \left(RC-p_{i,t}- \frac{RC}{100} \cdot g_j(\delta) + E  +1 \right) \mbox{d} \delta + \\ - \log \left( E - p_{i,t} + 1 \right) \left[ 1 - F_{\delta \lvert N_F} (\delta_{max} \lvert N_F \geq 1) \right]
 \Bigg\rbrace = 0 .
\end{multline}

\subsubsection{Aggregate claim distribution}
\label{sec:Application_aggregate_claim}
In order to apply the model, $\mathcal{M}_{\bar{X}_{c,t} a_{c,t}}\left(\frac{h}{n_g}\right)$ should be defined and perhaps some distributional assumption should be introduced. The best distributional form depends on the scope of the coverage, and the analysis might rather compare multiple significant scenarios represented by alternative distributional hypotheses.

An informative choice is focusing on the expected value of claims, and thus assuming that $Y_t$ is a weighted sum of Bernoulli random variables. Assuming that the properties within a municipality are perfectly correlated, $Y_t$ equal to:
\begin{multline}\label{eq:assumption_Y_t}
        Y_t = \sum_{c=1}^{N_c} X_{c,t} M_{c} = \sum_{c=1}^{N_c} M_{c} \bar{X}_{c,t} \int_{\zeta_D}^{\infty} \pi_c \left( \zeta | \zeta > \zeta_D \right) x \left[ l^a_{c,t} \left( \zeta \right) \right] \mbox{d}\zeta = \\ =  \sum_{c=1}^{N_c} \bar{X}_{c,t} \sum_j  M_{j,c} \int_{\zeta_D}^{\infty} \pi_c \left( \zeta | \zeta > \zeta_D \right) x \left[ l^a_{j,c,t} \left( \zeta \right) \right] \mbox{d}\zeta  = \sum_{c=1}^{N_c} \bar{X}_{c,t} a_{c}.
\end{multline}
Note that now parameters $a_{c}$ do not depend on time $t$ and are constants. The expected reimbursement of the $g$-th group $E \left[ Y^g_t \right]$ in eq. (\ref{eq:rimborso_gruppo_1}) therefore becomes
\begin{multline}
        E \left[ Y^g_t \right] = \sum_{c \in g} q_c a_{c} = \sum_{c \in g} \pi \left( \zeta > \zeta_D \right) \sum_j  M_{j,c} \int_{\zeta_D}^{\infty} \pi_c \left( \zeta | \zeta > \zeta_D \right) x \left[ l^a_{j,c,t} \left( \zeta \right) \right] \mbox{d}\zeta = \\ = \sum_{c \in g} \pi \left( \zeta > \zeta_D \right) \sum_j  M_{j,c} \int_{\zeta_D}^{\infty} \frac{\pi_c \left( \zeta \right)}{\pi_c \left( \zeta > \zeta_D \right)}  x \left[ l^a_{j,c,t} \left( \zeta \right) \right] \mbox{d}\zeta = \\ = \sum_{c \in g} \sum_j  M_{j,c} \int_{\zeta_D}^{\infty} \pi_c \left( \zeta \right) x \left[ l^a_{j,c,t} \left( \zeta \right) \right] \mbox{d}\zeta = \sum_{c \in g} \sum_j  M_{j,c} E \left[ x\left( l^a_{j,c,t} \right)\right],
\end{multline}
and the moment generating function of $\bar{X}_{c,t} a_{c}$ can be written through the probability generating function of a weighted sum of Bernoulli variables:
\begin{equation}
        \mathcal{M}_{\bar{X}_{c,t} a_{c}}\left(\frac{h}{n_g}\right) = \mathcal{G}_{\bar{X}_{c,t} a_{c}}\left(e^{\frac{h}{n_g}}\right) = \left[ 1+ \left( e^{\frac{h}{n_g}a_{c}} - 1 \right) q_c \right].
\end{equation}
However, since Bernoulli variables are bounded in $[0,1]$, each $Y_t^g$ is bounded between $0 \leq Y_t^g \leq \sum_{c \in g} a_c = b_g $. In the seismic model,
\begin{equation}
       a_c^s = \sum_j x_{j,c}^{s} 
\end{equation}
with
\begin{multline}
x_{j,c}^{s} = \int_{PGA_{min_c}}^{\infty} x \Bigg(  \frac{1}{K_j} \sum_{k=1}^{K_j} \sum_{LS=1}^{N_{LS_k}} RC(LS) \cdot  \left[ P_k \left(LS \vert PGA \right) - P_k \left(LS+1 \vert PGA \right)\right] \cdot \\ \cdot \myabs{\frac{ \mbox{d} \lambda_c \left( PGA \right) }{\mbox{d}(PGA)}}  \Bigg) \mbox{d}(PGA),
\end{multline}
while expected claims for flood policies are
\begin{equation}
    a_c^f = \sum_j x_{j,c}^{f}
\end{equation}
with
\begin{multline}
x^{f}_{j,c} = x \left( \int_0^{\infty} \frac{RC}{100} \cdot P(N_F \geq 1) \cdot  g_j(\delta) f_{\delta\vert N_F}(\delta \vert N_F \geq 1) \mbox{d}\delta \right) = \\
= P(N_F \geq 1) \cdot  x \left( \int_0^{\infty} \frac{RC}{100} \cdot  g_j(\delta) f_{\delta\vert N_F}(\delta \vert N_F \geq 1) \mbox{d}\delta \right) = \\
= P(N_F \geq 1) \cdot \Bigg( \int_{\delta_D}^{\delta_E} \frac{RC}{100} \cdot  g_j(\delta) f_{\delta\vert N_F}(\delta \vert N_F \geq 1) \mbox{d}\delta + \int_{\delta_E}^{\infty} E \cdot f_{\delta\vert N_F}(\delta \vert N_F \geq 1) \mbox{d}\delta \Bigg) = \\
= P(N_F \geq 1) \cdot \Bigg( \int_{\delta_D}^{\delta_E} \frac{RC}{100} \cdot  g_j(\delta) f_{\delta\vert N_F}(\delta \vert N_F \geq 1) \mbox{d}\delta +  E \cdot \lambda_{\delta \vert N_F }(\delta_E \vert N_F \geq 1) \Bigg).
\end{multline}
According to \cite{Hoeffding}, the bounds in eq. (\ref{eq:eq_Hoeff_epsilon1}) and (\ref{eq:eq_Hoeff_epsilon2}) simplify for the case of bounded weighted random variables. Consider, for instance, the bound as in eq. (\ref{eq:eq_Hoeff_epsilon1})
\begin{equation}\notag
Prob \Bigg\{ Y_t > N_c \phi + E \left[ Y \right] \Bigg\} < \sum_{g=1}^{N_g} w_g e^{-h_1 \phi} E \left[ e^{\frac{h_1}{n_g}  \left(Y^g_t -E \left[ Y^g \right] \right) } \right], \qquad h_1>0.
\end{equation} 
According to Lemma 1 in \cite{Hoeffding}, since the final term in the right-hand side of the inequality is convex, we know that: 
\begin{multline}
 E \left[ e^{\frac{h_1}{n_g}  \left(Y^g_t -E \left[ Y^g \right] \right) } \right] \leq e^{\frac{h_1}{n_g} E \left[ Y^g_t \right]} \left[  \frac{b_g - E \left[ Y^g \right]}{b_g} + \frac{E \left[ Y^g \right]}{b_g} e^{\frac{h_1}{n_g} b_g} \right] = \\ = e^{\frac{h_1}{n_g} E \left[ Y^g \right]} \left[  1 +  \frac{E \left[ Y^g \right]}{b_g} \left( e^{\frac{h_1}{n_g} b_g} -1 \right)  \right] = e^{L(h_g)}.
\end{multline}
$L(h_g)$ can be specified as
\begin{equation}
    L(h_g) = - h_g p_g + \ln{  \left( 1 + p_g \left( e^{h_g} -1 \right) \right) }
\end{equation}
with
\begin{equation}\notag
 h_g = \frac{h_1}{n_g} b_g \qquad \mbox{and} \qquad p_g = \frac{E \left[ Y^g \right]}{b_g}.
\end{equation}
According to the proof of Theorem 2 in \cite{Hoeffding}, 
\begin{equation}
    L(h_g) \leq \frac{1}{8} h_g^2 = \frac{1}{8} \left( \frac{h_1 b_g}{n_g} \right)^2,
\end{equation}
hence the bound can be rewritten as 
\begin{equation}\label{eq:bound_semplificato}
    Prob \Bigg\{ Y_t > N_c \phi + E \left[ Y \right] \Bigg\} < \sum_{g=1}^{N_g} w_g e^{-h_1 \phi} \left( e^{\frac{1}{8} \left( \frac{h_1 b_g}{n_g} \right)^2} \right) = \sum_{g=1}^{N_g} w_g e^{-h_1 \phi + \frac{1}{8} \left( \frac{h_1 b_g}{n_g} \right)^2 }.  
\end{equation}
In order to get the best possible upper bound, we find the minimum of the right-hand side of the inequality as a function of $\phi$, thus obtaining
\begin{equation}\label{eq:h_par}
    h_1 = \frac{4 \phi n_g^2}{b_g^2}.
\end{equation}
Substituting the parameter $h_1$ as defined in eq.(\ref{eq:h_par}) into eq.(\ref{eq:bound_semplificato}), the Hoeffding's bound simplifies to
\begin{equation}\label{eq:new_Hoeff_phi}
    Prob \Bigg\{ Y_t > N_c \phi + E \left[ Y \right] \Bigg\} < \sum_{g=1}^{N_g} w_g e^{- \frac{2 \phi^2 n_g^2}{b_g^2}}.
\end{equation}
Similarly, the bound in eq.(\ref{eq:eq_Hoeff_epsilon2}) can be rewritten as
\begin{equation}\label{eq:new_Hoeff_gamma}
Prob \Bigg\{ Y_t > N_c \gamma + E \left[ Y \right]  \Bigg\} < \sum_{g=1}^{N_g} w_g e^{- \frac{2 \gamma^2 n_g^2}{b_g^2}}.
\end{equation}

\subsubsection{Insurance model}
We now revise the insurance model by applying the distributional assumptions in section \ref{sec:Application_aggregate_claim}.

Once again, parameters $\phi$ and $\gamma$ are obtained by fixing the desired probabilities $\epsilon_1$ and $\epsilon_2$ equal to the right-hand side of inequalities (\ref{eq:new_Hoeff_phi}) and (\ref{eq:new_Hoeff_gamma}) respectively.

Premiums $p_i^H$ are obtained as in section \ref{sec:will_seismic} and \ref{sec:will_flood}, and $p_i^G$ are computed as in eq. (\ref{eq:min_tot_premium}). The optimal premium amount $\sum_{i=1}^{N_i} p_i^* m_i$ is again computed according to eq. (\ref{eq:optimum_premium}).

While $\phi^*$ and $\gamma^*$ remain unchanged as in eq. (\ref{eq:phi_star}) and (\ref{eq:gamma_star}),
\begin{equation}\label{eq:neq_epsilon1_star}
    \epsilon_1^* = \sum_{g=1}^{N_g} w_g e^{- \frac{2 \phi^{*^2} n_g^2}{b_g^2}}
\end{equation}
and
\begin{equation}\label{eq:neq_epsilon2_star}
    \epsilon_2^* = \sum_{g=1}^{N_g} w_g e^{- \frac{2 \gamma^{*^2} n_g^2}{b_g^2}}.
\end{equation}
Optimal values $\phi^*$ and $\gamma^*$ here cannot be expressed as explicit functions of $\epsilon_1^*$ and $\epsilon_2^*$ respectively, hence eq. (\ref{eq:condition_fondo}) 
\begin{equation}\notag
    W_d^{*} = N_c \left( \phi^{*} - \gamma^{*} \right) \geq 0
\end{equation}
cannot be directly related to the two probabilities. However, the equation implies $\phi^{*} \geq \gamma^{*}$ and since $\epsilon^*_1$ and $\epsilon^*_2$ are inversely related to $\phi^*$ and $\gamma^*$ respectively, the condition is satisfied if and only if 
\begin{equation}
    \epsilon_2^{*} \geq \epsilon_1^{*}.
\end{equation}

Similarly to the special case $h=h_1=h_2$, the model indicates that providing additional public resources should always be preferred to being insolvent. Coherently, $W^*_d=0$ is obtained if and only if $\epsilon^*_1 = \epsilon^*_2$. Of course, the fund is again directly proportional to the number of municipalities $N_c$. Initial preferences $\epsilon_1$ and $\epsilon_2$ and claim distribution are now reflected by $\phi^{*}$ and $\gamma^{*}$ instead.

\begin{table}[t!]
\centering
\caption{Public-private insurance scheme for earthquake risk management.}\label{tab:seismic_policies}
\begin{tabular}{|cc|ccccc|}
  \cline{3-7}
    \multicolumn{2}{c|}{ } & \multicolumn{5}{c|}{$\epsilon_1=0.01$,  $\epsilon_2=0.02$} \\
  \hline
  \multirow{2}{*}{Deductible}  & Maximum coverage & $\sum_{i=1}^{N_i} p_i^{*}$ & \multirow{2}{*}{$c$} & $W_d^{*}$  & \multirow{2}{*}{$\epsilon_1^{*}$} & \multirow{2}{*}{$\epsilon_2^{*}$} \\ 
   &  (per square metre) & (Mln) & & (Mln) & & \\
  \hline
 \multirow{2}{*}{0} & \multirow{2}{*}{1500} & 10735.784 & 1.394 & 7970.726 & 0.010 & 0.061 \\
    &  & (0.000) & (0.021) & (0.080) & (0.000) & (0.035) \\ 
     \hline
   \multirow{2}{*}{0} & \multirow{2}{*}{1200} & 9725.082 & 1.505 & 8563.810 & 0.010 & 0.080 \\ 
   &  & (0.000) & (0.021) & (0.073) & (0.000) & (0.030) \\ 
    \hline
\multirow{2}{*}{200} & \multirow{2}{*}{1500} & 8837.312 & 1.576 & 8582.130 & 0.010 & 0.095 \\
    &  & (0.000) & (0.021) & (0.068) & (0.000) & (0.027) \\
     \hline
 \multirow{2}{*}{200} & \multirow{2}{*}{1200} & 8221.215 & 1.652 & 8771.088 & 0.010 & 0.112 \\  
  &  & (0.000) & (0.021) & (0.065) & (0.000) & (0.024) \\ 
   \hline
\end{tabular}
\caption*{\textbf{Note}: results are based on $100$ samplings over the $N_c=6404$ municipalities for which data were fully available. Policies are defined on deductible and maximum coverage and listed by row, while columns represent the model's relevant variables. Reported values are mean and coefficient of variation.}
\end{table}

\begin{table}[t!]
\centering
\caption{Optimal seismic premiums per square metre.}\label{tab:optimal_seismic_premiums}
\begin{tabular}{cc|cccc}
  \hline
 \multicolumn{2}{c|}{Deductible} & 0 & 0 & 200 & 200 \\ 
 \multicolumn{2}{c|}{Maximum coverage (per square metre)} & 1500 & 1200 & 1500 & 1200 \\ 
 \hline
\multirow{3}{*}{RC.gl} & min & 0.460 & 0.460 & 0.034 & 0.034 \\ 
 & mean & 6.620 & 6.582 & 4.413 & 4.106 \\ 
 &  max & 32.261 & 32.261 & 30.471 & 30.471 \\ 
 \hline
 \multirow{3}{*}{RC.sl} & min & 0.034 & 0.034 & 0.007 & 0.007 \\ 
 & mean & 2.005 & 1.676 & 1.351 & 1.350 \\ 
 & max & 10.226 & 10.226 & 8.922 & 8.683 \\ 
   \hline
  \multirow{3}{*}{A.gl} & min  & 0.027 & 0.027 & 0.008 & 0.008 \\ 
 & mean  & 1.902 & 1.712 & 1.536 & 1.424 \\ 
 & max & 10.200 & 10.197 & 9.269 & 9.124 \\ 
  \hline
 \multirow{3}{*}{A.sl} & min & 0.012 & 0.012 & 0.011 & 0.011 \\ 
 & mean  & 1.745 & 1.535 & 1.278 & 1.205 \\ 
 & max & 10.153 & 10.153 & 7.810 & 7.696 \\ 
 \hline
  \multirow{3}{*}{M} & min & 0.075 & 0.062 & 0.041 & 0.041 \\ 
 & mean & 4.461 & 3.910 & 3.975 & 3.544 \\ 
 & max & 50.182 & 40.042 & 31.387 & 30.926 \\ 
   \hline
\end{tabular}
\caption*{\textbf{Note}: the Table shows the minimum, average and maximum premium at the municipal level per each combination of structural typology (row) and coverage limits (columns).}
\end{table}

\begin{table}[ht]
\centering
\caption{Public-private insurance scheme for flood risk management.}\label{tab:flood_policies}
\begin{tabular}{|cc|ccccc|}
  \cline{3-7}
        \multicolumn{2}{c|}{ }  & \multicolumn{5}{c|}{$\epsilon_1=0.01$,  $\epsilon_2=0.02$} \\
  \hline
  \multirow{2}{*}{Deductible}  & Maximum coverage & $\sum_{i=1}^{N_i} p_i^{*}$ & \multirow{2}{*}{$c$} & $W_d^{*}$  & \multirow{2}{*}{$\epsilon_1^{*}$} & \multirow{2}{*}{$\epsilon_2^{*}$} \\ 
   &  (per square metre) & (Mln) & & (Mln) & & \\
  \hline
 \multirow{2}{*}{0} & \multirow{2}{*}{1500} & 1021.072 & 5.644 & 7422.276 & 0.010 & 0.408 \\  
    &  & (0.000) & (0.035) & (0.041) & (0.000) & (0.029) \\ 
     \hline
   \multirow{2}{*}{0} & \multirow{2}{*}{1200} & 823.444 & 6.957 & 7567.733 & 0.010 & 0.534 \\ 
   &  &  (0.000) & (0.035) & (0.040) & (0.000) & (0.028) \\ 
    \hline
\multirow{2}{*}{200} & \multirow{2}{*}{1500} & 1020.885 & 5.607 & 7366.177 & 0.010 & 0.402  \\ 
    &  &  (0.000) & (0.035) & (0.041) & (0.000) & (0.029)\\
     \hline
 \multirow{2}{*}{200} & \multirow{2}{*}{1200} &   823.257 & 6.898 & 7496.555 & 0.010 & 0.547 \\ 
  &  &  (0.000) & (0.035) & (0.040) & (0.000) & (0.027)   \\ 
   \hline
\end{tabular}
\caption*{\textbf{Note}: results are based on $100$ samplings on $N_c=7772$ municipalities for which data were fully available. Policies are defined on deductible and maximum coverage and listed by row, while columns represent the model's relevant variables. Reported values are mean and coefficient of variation.}
\end{table}

\begin{table}[ht]
\centering
\caption{Optimal seismic premiums per square metre.}\label{tab:optimal_flood_premiums}
\begin{tabular}{cc|cccc}
  \hline
\multicolumn{2}{c|}{Deductible} & 0 & 0 & 200 & 200 \\ 
\multicolumn{2}{c|}{Maximum coverage (per square metre)} & 1500 & 1200 & 1500 & 1200 \\ 
 \hline
\multirow{3}{*}{1 storey} &  min & 2e-06 & 1e-06 & 2e-06 & 1e-06 \\ 
&  mean & 1.355 & 1.088 & 1.355 & 1.088 \\ 
&  max & 16.140 & 12.962 & 16.139 & 12.961 \\ 
 \hline
\multirow{3}{*}{2 storeys} &  min & 1e-16 & 1e-16 & 1e-16 & 1e-16 \\ 
&  mean & 0.198 & 0.161 & 0.198 & 0.161 \\ 
&  max & 2.382 & 1.933 & 2.381 & 1.933 \\ 
 \hline
\multirow{3}{*}{3 or more storeys} &  min & 1e-16 & 1e-16 & 1e-16 & 1e-16 \\ 
&  mean & 0.182 & 0.147 & 0.182 & 0.147 \\ 
&  max & 2.188 & 1.771 & 2.187 & 1.770 \\ 
   \hline
\end{tabular}
\caption*{\textbf{Note}: the Table shows the minimum, average and maximum premium at the municipal level per each combination of structural typology (row) and coverage limits (columns).}
\end{table}

\subsection{Results}\label{sec: res_SH}
The insurance model has been applied to the Italian residential building stock according to the assumption discussed in section \ref{sec:Application}. Results here presented refer to initial preferences $\epsilon_1=0.01$ and $\epsilon_2=0.02$. The former value is representative of solvency requirement in Solvency II, the latter has been fixed slightly greater than $\epsilon_1$ according to the model description. In addition, we assumed $r=50$ km, thus adopting a precautionary hypothesis on spatial correlation. This criteria allows for multiple sampling solutions, each resulting in different optimal values of the relevant variables. Therefore, final results have been averaged over 100 samplings.

Four policies have been considered, differing on the level of deductible (none or 200) and maximum coverage (none or 1200 per square metre). Note that deductible equal to 0 corresponds to the absence of it, while maximum coverage equal to $1500$ per square metre indicates that no maximum coverage applies.

Results for seismic policies are reported in Table \ref{tab:seismic_policies}, where relevant variables are presented in terms of their mean and coefficient of variation ($CoV$). It can be noticed that the optimal premiums always corresponds to the maximum price that individuals are willing to pay, $p_i^H$, as shown by (i) $c \geq 1$; (ii) the coefficient of variation of $\sum_{i=1}^{N_i} p_i^*$ equal to 0; and (iii) $\epsilon^*_1 = \epsilon_1$. When interpreting these findings, there are two elements that should be carefully evaluated: individuals' risk aversion and spatial correlation.

On one hand, because of risk aversion, individuals are keen to buy policies at a premium greater than their expected loss; the more individuals are risk averse, the higher is the amount of premium that the insurer is able to collect and, in turn, the lower is the additional capital needed to satisfy the solvency constraint $\epsilon_1$. On the other hand, spatial correlation between individual risks inflate loss volatility and bumps the tail of the aggregate loss distribution, thereby increasing the amount of capital corresponding to $\epsilon_1$. Parameter $c>1$ indicates that individual's risk aversion is not sufficient to tackle the risk-enhancing effect of spatial correlation at the aggregate level.

As a consequence of $c>1$, the premium $p_i^G$ that would satisfy the desired solvency constraint $\epsilon_1$ and capital re-investment probability $\epsilon_2$ does not meet market demand, and would generate a market failure. This result suggests a potential weakness of the free market: since the government has easier access to capital markets than private companies, it is reasonable to assume that a private insurer will require a probability of capital re-investment at most equal to the one desired by the government; under this condition, $p_i^G$ would be the minimum pure\footnote{Without profit load and expenses.} premium that traditional insurers would be able to charge to the homeowner, and the policy would not be purchased.

Limiting coverage might help the insurer controlling risk's volatility, thus allowing for lower premiums. In particular, being earthquakes low frequency-high intensity perils, the aggregate loss distribution is strongly affected by rare events causing severe damages and therefore we expect maximum coverage to reduce the insurer's financial exposure more than deductibles. In Table \ref{tab:seismic_policies} we can clearly notice that increasing coverage limits reduces the overall minimum amount of reserves that should be guaranteed at the beginning of each year $W_{min} =\sum_{i=1}^{N_i} p^{*}_i + W^{*}_d$, but the minimum capital requirement $W^*_d$ increases and $\epsilon^*_2$ deviates more and more from the desired level. As confirmed by the greater values of $c$, individuals are in fact reluctant to deductibles and maximum coverage due to increasing risk aversion\footnote{Risk aversion has been here represented by means of the utility function $u(x)=\log(x)+1$, whose relative risk aversion coefficient is increasing in $x$.}. Coverage limits negatively affect individual willingness to pay, that in turn lower their contribution to reserves and the insurer is left with an enhanced financial pressure. As said, limits-reluctance is here generated by risk-aversion, but unfortunately $c>1$ even for full-coverage policies, thus suggesting the need for a government intervention on the private market.

Results for floods are collected in Table \ref{tab:flood_policies}. Once again $c>1$ and the need for a government intervention in the insurance private market is even more strongly suggested (higher value of $c$). However, deductibles are here beneficial to the insurer and, in fact, both $W^*_d$ and $\epsilon^*_2$ are lower when $D=200$ applies. Though findings are completely different from the seismic case study, this evidence still generates due to a combination of risk aversion and loss distribution. Being high frequency-low intensity perils, floods mostly cause small claims on relatively low return periods and the aggregate loss distribution therefore concentrates on low values. On the other side, increasing risk aversion makes individuals extremely averse to high losses and less concerned about low damages that can afford by them own: in Table \ref{tab:optimal_flood_premiums}, when applying the deductible $D=200$, $p_i^H$ remains substantially unchanged. Combining the two effects, deductibles relieve the insurer commitment while not substantially modifying individual's willingness to pay.

By contrast, introducing an maximum coverage worsens the insurer condition by increasing both $\epsilon^*_2$ and $W_d^*$. This effect is clear when comparing the policy $(D=0,E=1500)$ with the $(D=0,E=1200)$ or $(D=200,E=1500)$ with $(D=200,E=1200)$. This limit in fact diminishes the risk of the insurer by lowering the tail of its aggregate loss distribution, but leaves highest level of individual risk to property-owners. Because of increasing risk aversion, the premium individuals are willing to pay is therefore much lowered, and the amount of public funds needed much increased.

The most interesting result is obtained when comparing policies with estimated losses in Section \ref{sec:expected loss}: though earthquakes produce expected losses that are more than seven times greater than those from floods, the minimum capital requirement $W_d^{*}$ for the two hazards almost coincide. Once again, the shape of the aggregate loss distribution and individual's increasing risk aversion jointly determine this surprising result. As low frequency-high intensity perils, earthquakes sometimes produce enormous damages that individuals are extremely concerned about. Therefore, owners are willing to pay a premium consistently higher than their expected loss. On the other side, floods happen quite more often but their damages are usually minor and can mostly be afforded by homeowners themselves. People are risk averse, and hence keen to buy a policy and get rid of their flood risk, but the amount they are willing to pay for the insurance service is lower. In other words, both the two premiums are higher than the corresponding expected loss but the difference between the premium that the homeowner pays for the earthquake policy and its expected seismic loss is greater than that of floods:
\begin{equation}
p_i^{H,s} - E(L^s_i) > p_i^{H,f} - E(L^f_i).
\end{equation}
The higher is the difference between premiums and expected losses, the lower is the additional capital needed by the insurance in order to manage the risk, and hence the lower is the capital requirement $W^*_d$. This becomes clear when considering the ratio $\frac{\sum_{i=1}^{N_i} p_i^*}{W^*_d}$. Ratios for seismic policies span between $0.93$ and $1.34$ and indicate that the government and the homeowners almost equally contribute to the constitution of reserves. On the other side, flood risk is much unfairly distributed between the two agents with ratios $\left[0.11, 0.14 \right]$.

Evidence suggests that spatial correlation strongly affects the development of the private insurance market for both the two perils, but larger values of $\epsilon^*_2$ indicate that flood risk is even more difficult to insure. A second level risk transfer (such as a reinsurance contract or a catastrophe linked securities) might help reducing $\epsilon_2$ by lowering the aggregate loss tail.

\begin{figure}[t]
   \centering
   \caption{Optimal premiums per square metre for earthquake policies on masonry buildings.}\label{fig:Italy_SH_M}
       \includegraphics[width=0.65\textwidth, trim=30mm 25mm 10mm 15mm, clip]{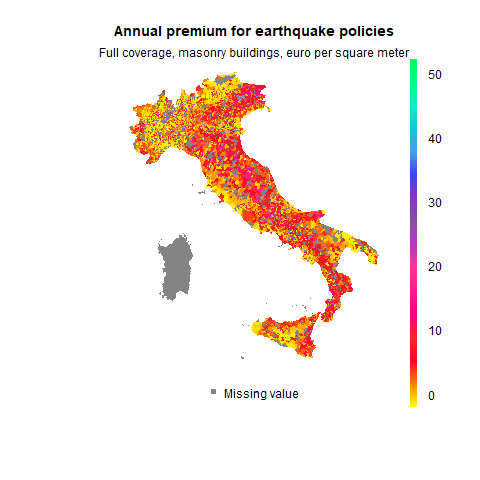}
   \caption*{\textbf{Note}: the map refers to the full coverage ($D=0$,$E=1500$ per square metre) policy, that has been estimated over $N_c=6404$. The minimum value reported is $0.075$ and therefore yellow municipalities should be interpreted as approximately $0$. The maximum premium is $50.182$.}
\end{figure}

\begin{figure}[t]
   \centering
   \caption{Optimal premiums per square metre for flood policies on one-storey buildings.}\label{fig:Italy_SH_1s}
       \includegraphics[width=0.65\textwidth, trim=30mm 25mm 10mm 15mm, clip]{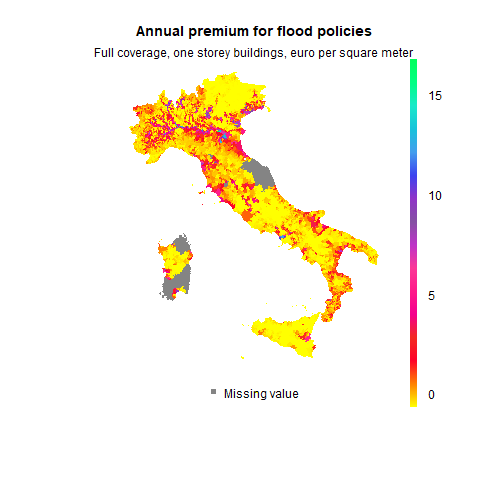}
   \caption*{\textbf{Note}: the map refers to the full coverage policy ($D=0$,$E=1500$ per square metre), that has been estimated over $N_c=7772$.}
\end{figure}

To conclude, Figures \ref{fig:Italy_SH_M} and \ref{fig:Italy_SH_1s} show annual optimal premiums per square metre $p_i^{H}$ for the most vulnerable structural typology per each municipality for the two perils respectively. Since premiums are risk-based, the two maps reflect the hazard component of risk modeling and hence report a pattern similar to the maps on loss per square metre in Figure \ref{fig:masonry_loss} and \ref{fig:one_storey_loss}.

\section{Managing multiple hazards}\label{ch:MH}
In the previous section a public-private insurance model has been created for single peril policies. As well known in finance, merging portfolios of different risks is beneficial if risks are uncorrelated, as floods and earthquakes are likely to be. It remains to be seen whether benefits from risk diversification counteract the negative impact of spatial correlation emerged in previous results. This section goes through what has been discussed so far and extends the analysis to multi-hazard.

The first subsection is devoted to risk assessment, and therefore comes back to the risk-modeling introduced in Section \ref{Risk}. While assessing single hazard risk is challenging, studying possible consequences from several perils is even more complicated. \cite{Kappes} identifies two major issues raising in a multi-hazard context: finding a common measure suitable to describing all the hazards considered, and understanding the relationship linking them. Regarding the former issue, since it is impossible to find a geological or atmospheric indicator describing both flood and earthquakes, the two risk have been here assessed separately and compared in monetary terms only. The second point refers to the correlation between the two phenomena. Based on some empirical evidence in the literature, we argue that the two risks are uncorrelated.

To our knowledge, \cite{MarzocchiGarcia} is the only work addressing multi-hazard risk assessment in Italy by studying seismic, volcanic, hydrogeological, flooding, landslide and industrial risk in the municipality of Casalnuovo. However, this analysis is restricted to a municipality and has been applied therein to human and societal risk only, and does not pursue any insurance decision. A different framework is therefore needed for our case study.

The following subsection extends the insurance model in Section \ref{Premium} to multi-hazard. The model is shaped by redefining supply and demand. In particular, we show that the maximum premium that individuals are willing to pay is equal to the sum of the premiums for the two single hazard policies, while the required amount of public capital is less than or equal to the sum necessary when managing the risk separately.

The last subsection presents results, which clearly show that benefit from risk diversification are not sufficient to override the effect of spatial correlation. However, additional positive externalities emerge: for example, premiums for multi-hazard policies are geographically more homogeneous with respect to the single hazard's, thus favoring the perception of fair-treatment among the population.

\subsection{Risk assessment}
\label{ch:multi_risk_assessment}
As discussed in Section \ref{Risk}, four elements determine losses from natural events : $H$, $E$, $V$ and $L$. A mathematical model for loss estimation should therefore be able to capture the relevant characteristics of each component and describe those process that link them. Since every natural phenomenon has specific characteristics, studying the effects of multiple hazards furthermore complicates risk assessment.

In particular, in our two-peril framework, $H$ should encompass both floods and earthquakes and therefore a common physical measure able to describe both the two perils should be identified. However, given the different characteristics of the two perils, this is impossible, and we are able to compare the two risks in money-value only. As a consequence, we are forced to assess the two risks separately, though this approach cannot capture the potential dependence among them.

In our multi-hazard risk assessment, we refer to two hazard-indicators $\zeta^s$ and $\zeta^f$, where apexes $s$ and $f$ indicate seismic and flood risks respectively. Each indicator is associated to a certain probability of occurrence $F^s (\zeta^s)$ and $F^f (\zeta^f)$. Hence, hazard is described by both $\zeta^h$ and $F^h (\zeta^h)$. As we have previously shown, vulnerability functions are defined over a specific hazard-indicator, and their output can be easily converted into monetary terms. Referring to the definition of loss in Sections \ref{ch: seismic_loss} and \ref{ch:flood_loss}, for simplicity we convey $L$ and $V$ in a unique function $v^h(\zeta^h)$, with $h=s,f$. Expected losses per square metre generated by a peril can hence be estimated as
\begin{equation}
    l^h = F^h(\zeta^h) v^h(\zeta^h) E^h, \qquad h=s,f.
\end{equation}
As anticipated, multi-hazard expected loss might be affected by potential interactions of the two perils, and therefore we need some assumption on the degree of dependence between floods and earthquakes. Unfortunately, our database do not offer any information about if and how the two perils interact, but some empirical analysis in the literature \citep{Tavarnien,IVASS} support the hypothesis of independence between seismic and flood risks. However, various degrees of independence are also possible. Following the work of \cite{Brunette}, we now discuss two possible independence scenarios. First, we consider the hazards to be mutually exclusive, thus assuming that floods and earthquakes cannot happen simultaneously and the structure can get damaged by one peril only; we will refer to this case as ``mutual exclusion scenario''. As an alternative, we consider perils to be ``mutually independent'', allowing them to potentially happen together. In this case, the property may be damaged by at least one of the two events.

\begin{itemize}
    \item{\textbf{Mutual exclusion scenario}}\\
    If the two hazards are mutually exclusive the joint probability of an event $F^{MH}_{dep}$ is obtained by simply summing the single hazard probabilities
    \begin{equation}\label{eq:mutually_dep_prob}
    F^{MH}_{exc} = F^s(\zeta^s) + F^f(\zeta^f),     
    \end{equation}
    and we can compute expected losses per square metre as:\\
    \begin{equation}
    l^{MH}_{exc} = F^s(\zeta^s) v^s(\zeta^s) + F^f(\zeta^f) v^f(\zeta^f).
    \end{equation}
    We can notice that in case of mutually exclusion the multi-hazard loss per square metre coincides with the sum of the single hazards expected losses:
    \begin{equation}\label{eq:multi_losses}
    l^{MH}_{exc} = l^s +l^f.
    \end{equation}
    \item{\textbf{Mutual independence scenario}}\\
    Avoid now any dependence and allow the hazards to happen simultaneously. The joint occurrence probability $F^{MH}_{ind}$ becomes:
    \begin{equation}
    F^{MH}_{ind} = F^s (\zeta^s) + F^f (\zeta^f) - F^s(\zeta^s) F^f(\zeta^f).
    \end{equation}
    Expected losses now arise from flood, earthquakes or a combination of the two. When the two events happen together the damages suffered by the property are defined by a new vulnerability function $v^{s+f} (\zeta^s, \zeta^f)$, therefore expected losses per square metre are obtained as:
    \begin{equation}
    \begin{split}
    l^{MH}_{ind} = \left[ F^s (\zeta^s) - F^s(\zeta^s) F^f(\zeta^f) \right] v^s(\zeta^s) + \left[ F^f (\zeta^f) - F^s(\zeta^s) F^f(\zeta^f) \right]
    v^f(\zeta^f) + \\ + F^s(\zeta^s) F^f(\zeta^f)v^{s+f} (\zeta^s, \zeta^f) = \\ = F^s (\zeta^s) v^s(\zeta^s) + F^f (\zeta^f) v^f(\zeta^f) + F^s(\zeta^s) F^f(\zeta^f) \left[ v^{s+f} (\zeta^s, \zeta^f) - v^s(\zeta^s) - v^f(\zeta^f)  \right] = \\ = l^s + l^f + F^s(\zeta^s) F^f(\zeta^f) \left[ v^{s+f} (\zeta^s, \zeta^f) - v^s(\zeta^s) - v^f(\zeta^f)  \right].
    \end{split}
    \end{equation}
\end{itemize}

We can notice that $l^{MH}_{ind} > l^{MH}_{exc}$ if $v^{s+f} (\zeta^s, \zeta^f) > v^s(\zeta^s) + v^f(\zeta^f)$ and this happens when the interaction of the two events amplifies the damages they cause on the property. We are unable to define the function $v^{s+f} (\zeta^s, \zeta^f)$ or to state whether it is smaller or greater than the sum of the two single hazard vulnerability functions. However, the low number of reported events suggests that the associated probability $F^s(\zeta^s) F^f(\zeta^f)$ is reasonably close to 0. Moreover, assuming the expected multi-hazard loss $l^{MH}_{ind}$ equal to $l^{MH}_{exc}$ is a prudential assumption if $v^{s+f} (\zeta^s, \zeta^f) < v^s(\zeta^s) + v^f(\zeta^f)$ because it requires the insurer to create slightly greater funds, thus effectively getting the probability of insolvency and fund-refill lower than the required level. For these reasons, we estimate expected losses as:
\begin{equation}
    l_{ind}^{MH} = l^s +l^f.
\end{equation}

\subsection{Homeowner's purchase decision}
In Section \ref{Premium} we have argued that premiums should meet the demand and that maximum rates that individuals are willing to pay pose a constraint to an insurance model. Similarly to the single-hazard policy, the demand constraint in a multi-hazard framework is therefore given by the equality:
\begin{equation}
    u^{MH}_{\mbox{not insured}} = u^{MH}_{\mbox{insured}}.
\end{equation}
Given the assumption of independence between floods and earthquakes and the individual utility functions defined in Section \ref{Premium}, we can now address the multi-hazard purchase decision problem. We refer to seismic events by means of the apex $s$ and flood by $f$, and for simplicity individual loss $l_{i,t}$ are indicated by $l_{i,t}^s$ for earthquakes and $l_{i,t}^f$ for floods. In addition, single hazard and multi-hazard policies are specified by means of apexes as $SH$ and $MH$.

Given the probability of multiple events' probabilities as defined in eq. (\ref{eq:mutually_dep_prob}) and losses computed as in eq. (\ref{eq:multi_losses}), the reimbursement function in eq. (\ref{eq:rimborso_x}) becomes:
    \begin{equation}\label{eq:MH_reimb_x}
    x^{MH} = \begin{cases} 0, & \mbox{with probability } \pi^s_{c}(0) + \pi^f_{c}(0), \\ x^s = x \left(l^s_{i,t}\right), & \mbox{with probability } 1-\pi^s_{c}(0), \qquad 0 < x\left(l^s_{i,t}\right) \leq l^s_{i,t}, \\ x^f = x \left(l^f_{i,t}\right), & \mbox{with probability } 1-\pi^f_{c}(0), \qquad 0 < x\left(l^f_{i,t}\right) \leq l^f_{i,t}, \end{cases} \quad \mbox{with } i \in c,
    \end{equation}
    with
    \begin{equation}\label{eq:MH_reimb_x^h}
    x^h = x \left(l^h_{i,t}\right) = \begin{cases} 0 & \mbox{if } l^h_{i,t}\leq D,
    \\ l^h_{i,t} - D & \mbox{if } D<l^h_{i,t}<E,
    \\ E-D & \mbox{if } l^h_{i,t} \geq E, \end{cases} \qquad h= s,f.
    \end{equation}
Hence, individual utilities of being and not being insured in eq. (\ref{eq:u_not_insured})-(\ref{eq:u_insured}) for multi-hazard policies are:
    \begin{equation}\label{eq:u_multi_notbuy}
        u_{\mbox{not insured}}^{MH} = \left[ \pi^s_c(0) + \pi^f_c(0) \right] u(RC) + \left[1-\pi^s_c(0)\right] u(RC - l^s) + \left[1-\pi^f_c(0)\right] u(RC - l^f)
    \end{equation}
    and
    \begin{equation}\label{eq:u_multi_buy}
    \begin{split}
        u_{\mbox{insured}}^{MH} = \left[ \pi^s_c(0) + \pi^f_c(0) \right] u(RC - p^{MH}) + \left[1-\pi^s_c(0)\right] u(RC - p^{MH} - l^s + x^s) + \\ + \left[1-\pi^f_c(0)\right] u(RC - p^{MH} - l^f + x^f).
    \end{split}
    \end{equation}
In Section \ref{ch:homeowner_convenience} the maximum premium that an homeowner is willing to pay for a single hazard policy is the quantity $p^{SH}$ solving the equality:
    \begin{equation}\label{eq:convenience_SH}
        u^{SH}_{\mbox{not insured}} = u^{SH}_{\mbox{insured}}.
    \end{equation}
    with
    \begin{equation}\label{eq:multi_u_SH_not}
        u^{SH}_{\mbox{not insured}} = \pi^{SH}_c(0)u(RC) + \left[1-\pi^{SH}_c(0)\right] u(RC - l^{SH})
    \end{equation}
    and
    \begin{equation}\label{eq:multi_u_SH_yes}
        u_{\mbox{insured}}^{SH} =  \pi^{SH}_c(0) u(RC - p^{SH}) + \left[1-\pi^{SH}_c(0)\right] u(RC - p^{SH} - l^{SH} + x^{SH}),
    \end{equation}
    for $SH=f,s$.\\
Comparing $MH$ and $SH$ utilities, we get:
\begin{equation}\label{eq:multi_single_condition}
    u^{MH}_{\mbox{insured}} = u^{MH}_{\mbox{not insured}} = u^s_{\mbox{not insured}} + u^f_{\mbox{not insured}} = u^s_{\mbox{insured}} + u^f_{\mbox{insured}}.
\end{equation}
This equality states that the homeowner utility of buying both the two single hazard policies is equal to the utility of buying a multi-hazard one. However, when evaluating one peril per time, policies prices are fixed by solving a consume decision with two options - to buy or not to buy the policy-, but a multi-hazard framework extends the range of possible choice: the individual may decide to buy a $MH$ policy, both the $SH$ policies, one out of the two $SH$, or neither of them. We know that if the policy is priced at $p^SH$ the individual is indifferent between buying or not the single-hazard policy, and eq. (\ref{eq:multi_single_condition}) states that the sum of the two utilities equals the utility of buying a $MH$ one. We should then investigate the option of buying both the two single hazard policies ($s+f$):
\begin{equation}
u^{s+f}_{\mbox{insured}} \geq u^{s+f}_{\mbox{not insured}}
\end{equation}
\begin{equation}
\begin{split}
    u^{s+f}_{\mbox{insured}} = \left[ \pi^s_c(0) + \pi^f_c(0) \right] u(RC-p^s-p^f) + \left[1-\pi^s_c(0)\right] u(RC-p^s-p^f-l^s+x^s) + \\ + \left[1-\pi^f_c(0)\right] u(RC-p^s-p^f-l^f+x^f)
\end{split}
\end{equation}
while
\begin{equation}
    u^{s+f}_{\mbox{not insured}} = \left[ \pi^s_c(0) + \pi^f_c(0) \right] u(RC) + \left[1-\pi^s_c(0)\right] u(RC-l^s+x^s) + \left[1-\pi^f_c(0)\right] u(RC-l^f+x^f).
\end{equation}
Note that the premium paid by the owner in this scenario is $p^{s+f}=p^s + p^f$. Assuming consumer's perfect rationality and neglecting any operational cost that a policy may generate, the individual prefers a multi-hazard policy to two single-hazard ones if $p^{MH} < p^s + p^f$ because it implies that $u^{MH}_{\mbox{insured}} > u^{s+f}_{\mbox{insured}}$. Therefore
\begin{equation}
    u^{s+f}_{\mbox{not insured}} = u^{MH}_{\mbox{not insured}} = u^{s}_{\mbox{not insured}} + u^{f}_{\mbox{not insured}}.
\end{equation}
which in turn implies:
\begin{equation}
    u^{s+f}_{\mbox{insured}} = u^{MH}_{\mbox{insured}} = u^{s}_{\mbox{insured}} + u^{f}_{\mbox{insured}}.
\end{equation}
Thus, the maximum premium that an individual is willing to pay for a multi-hazard policy makes him indifferent between any purchase choice and is equal to
\begin{equation}\label{eq:multi_premiums}
    p^{MH,H} = p^s +p^f.
\end{equation}

\subsection{Public-private partnership}
Main differences in risk-pooling single- or multi- hazard policies are determined by the different loss, reimbursement and premium functions, that are now described by eq. (\ref{eq:multi_losses}), (\ref{eq:MH_reimb_x}) - (\ref{eq:MH_reimb_x^h}) and (\ref{eq:multi_premiums}).

We now construct the fund $W^{MH}$ for multi-hazard policies by extending the single-hazard model. The reader can find the extended description of the procedure in Subsection \ref{ch:Governmental_issues_SH}.

The multi-hazard fund at the beginning $W_t^{MH,b}$ and at the end $W_t^{MH,e}$ of the year $t$, are now:
\begin{equation}
W_t^{MH,b} = W^{MH}_{t-1} + \sum^{N_i}_{i=1} p^{MH}_{i} m_{i} = W^{MH}_{t-1} + \sum^{N_i}_{i=1} \left( p^{s}_{i} + p^{f}_{i} \right) m_{i} \qquad \mbox{with} \qquad W^{MH}_{t-1} = \max(W^{MH,e}_{t-1}; W^{MH}_d),
\end{equation}
and
\begin{equation}
W_t^{MH,e} = W_t^{MH,b} - \sum_{i=1}^{N_i} x^{MH}_{i,t} m_{i} = W_t^{MH,b} - \sum_{i=1}^{N_i} \left( x^{s}_{i,t} + x^{f}_{i,t} \right) m_{i} .
\end{equation}
Assume that an earthquakes or a flood hits any building within a municipality and that every policy can generate at most one claim per hazard per year. Square metre expected losses $l^{MH}_{i,t}$ are equal for all the individuals within the same municipality and so does $x^{MH}_{i,t}$. Given the number of inhabited squared metres $M_c = \sum_{i \in c} m_{i}$, the total claims value per municipality is:
\begin{equation}
\sum_{i \in c} x^{MH}_{i,t} m_{i} = \sum_{i \in c} \left( x^{s}_{i,t} + x^{f}_{i,t} \right) m_{i} = \left( X^{s}_{c,t} + X^{f}_{c,t} \right) M_{c},
\end{equation}
and the total national amount is
\begin{equation}
Y^{MH}_t = \sum_{i=1}^{N_i} x^{MH}_{i,t} m_{i} = \sum_{c=1}^{N_c} \sum_{i \in c} x^{MH}_{i,t} m_{i} = \sum_{c=1}^{N_c} \left( X^{s}_{c,t} + X^{f}_{c,t} \right) M_{c} = Y^{s}_t + Y^{f}_t,
\end{equation}
and therefore is equal to the sum of the national claims for earthquakes $Y^s_t$ and floods $Y^f_t$ computed by means of eq. (\ref{eq:total_claims_1}). We model claim probabilities by means of Bernoulli variables $\bar{X}^{s}_{c,t} \sim Ber(q^s_c)$ and $\bar{X}^{f}_{c,t} \sim Ber(q^f_c)$ with $q^s_c = \pi^s_c \left( \zeta^s > \zeta_D \right)$ and $q^f_c = \pi^f_c \left( \zeta^f > \zeta_D \right)$ and apply equation (\ref{eq:total_claims}):
\begin{equation}
\begin{split}
Y^{MH}_t  = \sum_{c=1}^{N_c} \left( \bar{X}^s_{c,t} a^s_{c,t} + \bar{X}^f_{c,t} a^f_{c,t} \right) .
\end{split}
\end{equation}
Assuming municipalities that are at least $50$ km far each other to be independent, we can recall the sample that have been created for single hazard policies. Considering the two hazard separately, we will have $N_g$ groups of municipalities' seismic risks and other $N_g$ groups for floods. Each group will contain $n_g$ municipalities:
\begin{equation}
    Y^{s,g}_t = \sum_{c \in g} \bar{X}^s_{c,t} a^s_{c,t} \qquad c=1,\dots,n_g
\end{equation}
\begin{equation}
    Y^{f,g}_t = \sum_{c \in g} \bar{X}^f_{c,t} a^f_{c,t} \qquad c=1,\dots,n_g
\end{equation}
such that
\begin{equation}
    Y^{MH}_t = Y^{s,1}_t + Y^{s,2}_t + \dots + Y^{s,N_g}_t + Y^{f,1}_t + Y^{f,2}_t + \dots + Y^{f,N_g}_t.
\end{equation}
Defining $w_g = \frac{n_g}{N_c}$, the expected total amount of claims in Italy is:
\begin{equation}
    E \left[ Y^{MH}_{t} \right] = E \left[ Y^{MH} \right] = \sum_{g=1}^{N_g} w_g \left( E \left[ Y^{s,g}_{t}\right] + E \left[ Y^{f,g}_{t}\right] \right),
\end{equation}
with $E \left[ Y^{s,g}_{t}\right]$ and $E \left[ Y^{f,g}_{t}\right]$ computed as in (\ref{eq:rimborso_gruppo_1}).\\
Applying the \cite{Hoeffding} bound as in eq.(\ref{eq:prob})-(\ref{eq:fondo_Gov}), we get to the definition of both insolvency probability and $W_d$. We fix the insolvency probability $\epsilon_1$: 
\begin{multline}
    \epsilon_1 =  e^{-h_1 \phi} \sum_{g=1}^{N_g} w_g e^{-\frac{h_1}{n_g}E \left( Y^{s,g} + Y^{f,g} \right) } E \left[ e^{\frac{h_1}{n_g}\left( Y^{s,g}_{t} + Y^{f,g}_{t} \right)} \right] = \\ = e^{-h_1 \phi} \sum_{g=1}^{N_g} w_g e^{-\frac{h_1}{n_g}E \left( Y^{s,g} + Y^{f,g} \right) } E \left[ e^{\frac{h_1}{n_g}\left( Y^{s,g}_{t}\right)} e^{\frac{h_1}{n_g}\left( Y^{f,g}_{t} \right)} \right] = \\ = e^{-h_1 \phi} \sum_{g=1}^{N_g} w_g e^{-\frac{h_1}{n_g}E \left( Y^{s,g} + Y^{f,g} \right) } E \left[ e^{\frac{h_1}{n_g}\left( \bar{X}^s_{c,t} a^s_{c,t} \right)} e^{\frac{h_1}{n_g}\left( \bar{X}^f_{c,t} a^f_{c,t} \right)} \right],
\end{multline}
and since seismic and flood risk are independent: 
\begin{multline}
 \epsilon_1 = e^{-h_1 \phi} \sum_{g=1}^{N_g} w_g e^{-\frac{h_1}{n_g}E \left( Y^{s,g} + Y^{f,g} \right) } E \left[ e^{\frac{h_1}{n_g}\left( \bar{X}^s_{c,t} a^s_{c,t} \right)} \right] E \left [e^{\frac{h_1}{n_g}\left( \bar{X}^f_{c,t} a^f_{c,t} \right)} \right] = \\ = e^{-h_1 \phi} \left( \sum_{g=1}^{N_g} w_g e^{- \frac{h_1}{n_g} \left( E \left( Y^{s,g} \right) + E \left( Y^{f,g} \right) \right) }  \prod_{c \in g}  \mathcal{M}_{\bar{X}^s_{c,t} a^s_{c,t}} \left(\frac{h_1}{n_g}\right) \prod_{c \in g}  \mathcal{M}_{\bar{X}^f_{c,t} a^f_{c,t}} \left(\frac{h_1}{n_g}\right) \right).
\end{multline}
The minimum capital requirement for a multi-hazard public insurance is
\begin{equation}
    W_d^{MH} = N_c \phi + E \left[ Y \right] - \sum_{i=1}^{N_i} \left( p^s_i + p^f_i \right)m_i
\end{equation}
with
\begin{equation}
    \phi = \frac{1}{h_1} \log \left( \frac{\sum_{g=1}^{N_g} w_g e^{- \frac{h_1}{n_g} \left( E \left( Y^{s,g} \right) + E \left( Y^{f,g} \right) \right) }  \prod_{c \in g}  \mathcal{M}_{\bar{X}^s_{c,t} a^s_{c,t}} \left(\frac{h_1}{n_g}\right) \prod_{c \in g}  \mathcal{M}_{\bar{X}^f_{c,t} a^f_{c,t}} \left(\frac{h_1}{n_g}\right) }{ \epsilon_1} \right)
\end{equation}
The probability of fund-refill $\epsilon_2$ and the minimum amount of premiums $\sum_{i=1}^{N_i} p^{MH,G}_{i} m_{i}$ that the insurer needs given a certain $W_d$ are obtained by applying the \cite{Hoeffding} bound as in (\ref{eq:prob_2})-(\ref{eq:min_tot_premium}). Hence, fixing
\begin{equation}
    \epsilon_2 = e^{-h_2 \gamma} \left( \sum_{g=1}^{N_g} w_g e^{- \frac{h_2}{n_g} \left( E \left( Y^{s,g} \right) + E \left( Y^{f,g} \right) \right) }  \prod_{c \in g}  \mathcal{M}_{\bar{X}^s_{c,t} a^s_{c,t}} \left(\frac{h_2}{n_g}\right) \prod_{c \in g}  \mathcal{M}_{\bar{X}^f_{c,t} a^f_{c,t}} \left(\frac{h_2}{n_g}\right) \right) 
\end{equation}
we get
\begin{equation}\label{eq:multi_gov_premium}
\sum_{i=1}^{N_i} p^{MH,G}_{i} m_{i} = N_c \gamma + E \left[ Y \right]
\end{equation}
where $\gamma$ is computed as:
\begin{equation}
    \gamma = \frac{1}{h_2} \log \left( \frac{ \sum_{g=1}^{N_g} w_g e^{- \frac{h_2}{n_g} \left( E \left( Y^{s,g} \right) + E \left( Y^{f,g} \right) \right) }  \prod_{c \in g}  \mathcal{M}_{\bar{X}^s_{c,t} a^s_{c,t}} \left(\frac{h_2}{n_g}\right) \prod_{c \in g}  \mathcal{M}_{\bar{X}^f_{c,t} a^f_{c,t}} \left(\frac{h_2}{n_g}\right) }{ \epsilon_2} \right)
\end{equation}

\subsection{Insurance model}
The model for the definition of a public-private insurance scheme with multi-hazard policies can be defined as in Section \ref{ch:national_insurance_scheme}, therefore here we briefly extend the model to the multi-hazard scenario, but the reader can refer to the previous Section for technical details.

The two fundamental conditions are now defined by equations (\ref{eq:multi_premiums}) and (\ref{eq:multi_gov_premium}). The optimal premium $p^{*MH}_i$ is estimated as:
\begin{equation}\label{eq:multi_optimum_premium}
p^{MH^{*}}_{i}= \min (c, 1) \cdot p^{MH,H}_{i} \qquad \mbox{with} \qquad c= \frac{\sum_{i=1}^{N_i} p^{MH,G}_{i} m_{i}}{\sum_{i=1}^{N_i} p^{MH,H} m_i},
\end{equation}
from which we obtain
\begin{equation}
\sum_{i=1}^{N_i} p^{MH^{*}}_{i} m_i = \min \left(c, \frac{1}{c} \right) N_c \gamma + E \left[ Y \right]=  N_c \gamma^{*} + E \left[ Y \right],
\end{equation}
and the optimal probability of fund-refill $\epsilon_2^{*}$:
\begin{equation}\label{eq:multi_epsilon_2_optimum}
\epsilon^*_2 = \frac{\sum_{g=1}^{N_g} w_g e^{- \frac{h_2}{n_g} \left( E \left( Y^{s,g} \right) + E \left( Y^{f,g} \right) \right) }  \prod_{c \in g}  \mathcal{M}_{\bar{X}^s_{c,t} a^s_{c,t}} \left(\frac{h_2}{n_g}\right) \prod_{c \in g}  \mathcal{M}_{\bar{X}^f_{c,t} a^f_{c,t}} \left(\frac{h_2}{n_g}\right) }{e^{h_2 \gamma^*}},
\end{equation}
where $\gamma^{*}$ is 
\begin{equation}
    \gamma^{*} = \frac{\min \left( 1, \frac{1}{c} \right) \left( E \left[ Y \right] - N_c \gamma \right) - E \left[ Y \right] }{N_c}.
\end{equation}
The optimal $W^{MH^*}_d $ is estimated as in equation (\ref{eq:optimum_fondo}):
\begin{equation}
\begin{split}
W^{MH^*}_d  = \max \left\{ N_c \phi + E \left[ Y \right] -\sum_{i=1}^{N_i} p^{MH^{*}}_{i}m_{i}  ; 0  \right\} = N_c \phi^{*} + E \left[ Y \right] -\sum_{i=1}^{N_i} p^{MH^{*}}_{i}m_{i} = \\ = N_c \left( \phi^{*} - \gamma^{*} \right),
\end{split}
\end{equation}
with
\begin{equation}
    \phi^{*} = \frac{ W^{*}_d + \sum_{i=1}^{N_i} p^{MH^{*}}_{i}m_{i} - E \left[ Y \right]}{N_c},
\end{equation}
and the optimal value $\epsilon^{*}_1$ is:
\begin{equation}\label{eq:MH_epsilon_1}
\epsilon^*_1 = \frac{\sum_{g=1}^{N_g} w_g e^{- \frac{h_1}{n_g} \left( E \left( Y^{s,g} \right) + E \left( Y^{f,g} \right) \right) }  \prod_{c \in g}  \mathcal{M}_{\bar{X}^s_{c,t} a^s_{c,t}} \left(\frac{h_1}{n_g}\right) \prod_{c \in g}  \mathcal{M}_{\bar{X}^f_{c,t} a^f_{c,t}} \left(\frac{h_1}{n_g}\right) }{e^{h_1 \gamma^*}}.
\end{equation}
As in the single-hazard scenario, some distributional assumptions are needed in order to solve the model. We keep the assumptions as in Section \ref{sec:Application_aggregate_claim}, and therefore we represent $Y_t$ as a weighted sum of Bernoulli random variables. We assume that the properties within a municipality are perfectly correlated. Hence, the \cite{Hoeffding} bound simplifies and the probabilities $\epsilon^*_1$ and $\epsilon^*_2$ become:
\begin{equation}
        \epsilon_1^{*} = \sum_{g=1}^{N_g} w_g e^{- \frac{2 \phi^{*^2} n^2_g}{b^2_g}}
\end{equation}
and
\begin{equation}
    \epsilon_2^{*} = \sum_{g=1}^{N_g} w_g e^{- \frac{2 \gamma^{*^2} n^2_g}{b^2_g}},
\end{equation}
where
\begin{equation}
    b^2_g = \sum_{c \in g} a^s_c + a^f_c.
\end{equation}

\begin{table}[t]
\centering
\caption{Multi-hazard public-private insurance scheme.}\label{tab:MH_policies}
\begin{tabular}{|cc|c|ccccc|}
  \cline{4-8}
    \multicolumn{3}{c|}{} & \multicolumn{5}{c|}{$\epsilon_1=0.01$,  $\epsilon_2=0.02$} \\
  \hline
  \multirow{2}{*}{Deductible}  & Maximum coverage & & $\sum_{i=1}^{N_i} p_i^{*}$ & \multirow{2}{*}{$c$} & $W_d^{*}$  & \multirow{2}{*}{$\epsilon_1^{*}$} & \multirow{2}{*}{$\epsilon_2^{*}$} \\ 
   &  (per square metre) & & (Mln) & & (Mln) & & \\
  \hline
\multirow{6}{*}{0} & \multirow{6}{*}{1500} & \multirow{2}{*}{MH} & 11185.123 & 1.694 & 13281.008 & 0.010 & 0.091 \\ 
                     &                       &                     & (0.000) & (0.024) & (0.068) & (0.000) & (0.029)  \\
                     \cline{3-8}
                     &                       & \multirow{2}{*}{S}  & 10356.859 & 1.424 & 8194.243 & 0.010 & 0.063 \\ 
                     &                       &                     & (0.000) & (0.022) & (0.080) & (0.000) & (0.035) \\
                     \cline{3-8}
                     &                       & \multirow{2}{*}{F}  & 828.264 & 6.346 & 7359.961 & 0.010 & 0.403 \\  
                     &                       &                    & (0.000) & (0.036) & (0.044) & (0.000) & (0.029)   \\
   \hline
\multirow{6}{*}{0} & \multirow{6}{*}{1200} & \multirow{2}{*}{MH} & 10046.430 & 1.852 & 13983.445 & 0.010 & 0.120 \\ 
                     &                       &                      & (0.000) & (0.024) & (0.063) & (0.000) & (0.025)   \\ \cline{3-8}
                     &                       & \multirow{2}{*}{S}  & 9378.433 & 1.537 & 8760.141 & 0.010 & 0.082 \\ 
                     &                       &                     & (0.000) & (0.022) & (0.074) & (0.000) & (0.030) \\
                    \cline{3-8}
                     &                       & \multirow{2}{*}{F}   & 667.997 & 7.820 & 7469.278 & 0.010 & 0.427 \\ 
                     &                       &                    & (0.000) & (0.036) & (0.043) & (0.000) & (0.028)  \\
   \hline
 \multirow{6}{*}{200} & \multirow{6}{*}{1500} & \multirow{2}{*}{MH} & 9335.925 & 1.919 & 13879.044 & 0.010 & 0.132 \\ 
                     &                       &                     & (0.000) & (0.024) & (0.061) & (0.000) & (0.024)   \\ \cline{3-8}
                     &                       & \multirow{2}{*}{S}  & 8507.814 & 1.612 & 8766.645 & 0.010 & 0.097 \\ 
                     &                       &                     & (0.000) & (0.022) & (0.069) & (0.000) & (0.027) \\ \cline{3-8}
                     &                       & \multirow{2}{*}{F}  & 828.110 & 6.305 & 7305.976 & 0.010 & 0.396 \\  
                     &                       &                    & (0.000) & (0.036) & (0.044) & (0.000) & (0.029)  \\
   \hline
 \multirow{6}{*}{200} & \multirow{6}{*}{1200} & \multirow{2}{*}{MH} & 8575.042 & 2.047 & 14183.694 & 0.010 & 0.163 \\  
                     &                       &                     & (0.000) & (0.024) & (0.058) & (0.000) & (0.022)   \\ \cline{3-8}
                     &                       & \multirow{2}{*}{S}  & 7907.199 & 1.692 & 8945.297 & 0.010 & 0.114 \\ 
                     &                       &                     & (0.000) & (0.022) & (0.066) & (0.000) & (0.024) \\ \cline{3-8}
                     &                       & \multirow{2}{*}{F}   & 667.843 & 7.756 & 7401.415 & 0.010 & 0.435 \\ 
                     &                       &                     & (0.000) & (0.036) & (0.043) & (0.000) & (0.028) \\
   \hline
\end{tabular}
\caption*{\textbf{Note}: the Table shows multi-hazard (MH), seismic (S) and flood (F) insurance for the Italian residential building stock. Results have been estimated over $100$ samplings on $N_c=6217$ municipalities for which data were fully available for both flood and earthquakes. Policies are defined on deductible and maximum coverage and listed by row, while columns represent the model's relevant variables. Reported values are mean and coefficient of variation.}
\end{table}

\begin{table}[t!]
\centering
\caption{Optimal multi-hazard premiums per square metre.}\label{tab:optimal_MH_premiums}
\begin{tabular}{|rr|rr|ccc|ccc|}
  \cline{5-10}
\multicolumn{4}{c|}{ } & \multicolumn{6}{c|}{Deductible} \\
\multicolumn{4}{c|}{ } & \multicolumn{3}{c|}{0} & \multicolumn{3}{c|}{200} \\ \cline{5-10}
\multicolumn{4}{c|}{ } &  1s & 2s & 3s & 1s & 2s & 3s \\ \cline{5-10}
  \hline
\parbox[t]{2mm}{\multirow{30}{*}{\rotatebox[origin=c]{90}{Maximum coverage (per square metre)}}} & \parbox[t]{2mm}{\multirow{15}{*}{\rotatebox[origin=c]{90}{1500}}} & \multirow{3}{*}{ RC.gl} & min & 0.644 & 0.592 & 0.591 & 0.051 & 0.051 & 0.051 \\ 
& & & mean & 7.541 & 6.730 & 6.719 & 5.322 & 4.512 & 4.500 \\ 
& & & max & 32.261 & 32.261 & 32.261 & 30.471 & 30.471 & 30.471 \\  \cline{3-10}
& & \multirow{3}{*}{ RC.sl} & min & 0.038 & 0.034 & 0.034 & 0.012 & 0.008 & 0.008 \\
& & & mean & 2.932 & 2.122 & 2.110 & 2.195 & 1.384 & 1.373 \\ 
& & &  max & 16.951 & 10.679 & 10.638 & 16.935 & 9.277 & 9.248 \\ \cline{3-10}
 & & \multirow{3}{*}{ A.gl} & min & 0.036 & 0.036 & 0.036 & 0.021 & 0.021 & 0.021 \\   
 & & & mean & 2.849 & 2.038 & 2.027 & 2.478 & 1.667 & 1.656 \\
& & &  max & 19.136 & 10.793 & 10.730 & 19.111 & 9.625 & 9.595 \\  \cline{3-10}
& & \multirow{3}{*}{ A.sl} &  min & 0.021 & 0.021 & 0.021 & 0.015 & 0.015 & 0.015 \\
& & & mean & 2.684 & 1.873 & 1.862 & 2.221 & 1.410 & 1.399 \\ 
& & & max & 18.860 & 10.283 & 10.272 & 18.614 & 8.165 & 8.136 \\   \cline{3-10}
& & \multirow{3}{*}{ M} & min & 0.092 & 0.092 & 0.092 & 0.058 & 0.058 & 0.058 \\ 
& & &  mean & 5.383 & 4.573 & 4.561 & 4.898 & 4.088 & 4.077 \\  
& & &  max & 50.428 & 50.218 & 50.215 & 31.387 & 31.387 & 31.387 \\ 
\cline{2-10}
\cline{2-10}
& \parbox[t]{2mm}{\multirow{15}{*}{\rotatebox[origin=c]{90}{1200}}} & \multirow{3}{*}{ RC.gl} & min & 0.644 & 0.589 & 0.588 & 0.038 & 0.035 & 0.035 \\ 
& & &  mean & 7.315 & 6.666 & 6.656 & 4.818 & 4.169 & 4.159 \\ 
& & &  max & 32.261 & 32.261 & 32.261 & 30.471 & 30.471 & 30.471 \\   \cline{3-10}
& & \multirow{3}{*}{ RC.sl} & min & 0.037 & 0.034 & 0.034 & 0.011 & 0.008 & 0.008 \\ 
& & &  mean & 2.428 & 1.779 & 1.770 & 2.133 & 1.484 & 1.474 \\ 
& & &  max & 13.964 & 10.583 & 10.549 & 13.910 & 8.962 & 8.938 \\  \cline{3-10}
& & \multirow{3}{*}{ A.gl} & min & 0.031 & 0.028 & 0.028 & 0.011 & 0.008 & 0.008 \\ 
& & &  mean & 2.469 & 1.820 & 1.811 & 2.175 & 1.526 & 1.517 \\ 
& & &  max & 15.046 & 10.648 & 10.596 & 14.109 & 9.339 & 9.315 \\    \cline{3-10}
& & \multirow{3}{*}{ A.sl} & min & 0.015 & 0.012 & 0.012 & 0.015 & 0.015 & 0.015 \\
& & &  mean & 2.286 & 1.637 & 1.627 & 1.958 & 1.308 & 1.299 \\  
& & &  max & 13.952 & 10.258 & 10.250 & 13.608 & 7.985 & 7.960 \\   \cline{3-10}
& & \multirow{3}{*}{ M} & min & 0.068 & 0.063 & 0.062 & 0.045 & 0.042 & 0.041 \\ 
& & &  mean & 4.641 & 3.992 & 3.983 & 4.275 & 3.626 & 3.616 \\  
& & &  max & 40.891 & 40.168 & 40.157 & 30.926 & 30.926 & 30.926 \\
   \hline
\end{tabular}
\caption*{\textbf{Note}: the table is divided in four sub-tables, each representing a specific combination of deductible (by column) and maximum coverage (by row). Each sub-table presents minimum, average and maximum premium at the municipal level per each combination of structural typology (row) and number of storeys (column).}
\end{table}

\begin{figure}[h]
   \centering
      \caption{Optimal multi-hazard premium per square metre for one storey masonry buildings.}\label{fig:Italy_MH_M1}
       \includegraphics[width=0.65\textwidth, trim=30mm 25mm 10mm 15mm, clip]{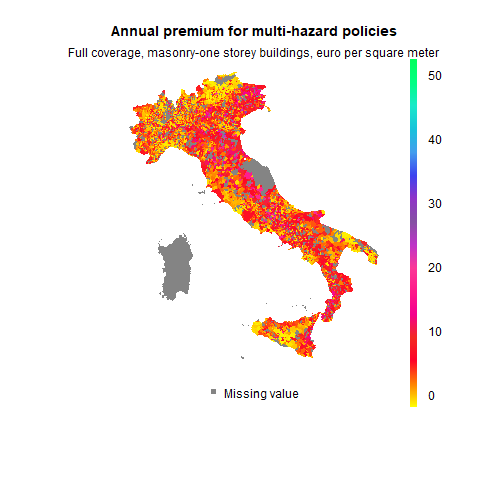}
    \caption*{\textbf{Note}: the map represents a full coverage ($D=0$, $E=1500$ per square metre) policy, that has been estimated on $N_c$= 6217. The minimum value represented is $0.092$ and therefore yellow municipalities should be interpreted as approximately $0$. The maximum premium reported is $50.428$.}
\end{figure}

\subsection{Results}\label{ch:results}
For multi-hazard analysis, only municipalities where both seismic and flood data are available have been considered, thereby restricting the database to $N_c=6217$.

As in Section \ref{sec: res_SH}, municipalities have been assumed independent if centroids are at least 50 km far and 100 samplings have been considered for final results. The four policies considered for single hazard policies have also been estimated for multi-hazard: (i) a full coverage policy ($D=0$, $E=1500$); (ii) one with a maximum coverage equal to 1200 per square metre ($D=0$, $E=1200$); (iii) one with a deductible equal to 200 ($D=200$, $E=1500$); (iv) a policy with both the maximum coverage and the deductible ($D=200$, $E=1200$). Initial preferences have been again fixed to $\epsilon_1=0.01$ and $\epsilon_2=0.02$.

Results are presented in Table \ref{tab:MH_policies} together with the corresponding single hazard policies, that have been re-estimated on the restricted number of municipalities for the sake of comparability. As expected, seismic risk dominates the multi-hazard scenario because of the consistently higher impact on the national territory. In particular, since 
\begin{equation}
    \sum_{i=1}^{N_i}p_i^{MH*} = \sum_{i=1}^{N_i}p_i^{s*} + \sum_{i=1}^{N_i}p_i^{f*},
\end{equation}
we can notice that multi-hazard premiums amount $\sum_{i=1}^{N_i}p_i^*$ is mostly determined by seismic risk and just a small portion of it is due to floods. Though premiums for the two single hazard policies are extremely different, the corresponding minimum capital requirement $W^*_d$ is similar (see Section \ref{sec: res_SH}), and for the multi-hazard policy 
\begin{equation}
W_d^{MH*} \leq W_d^{s*} + W_d^{f*}.
\end{equation} 
Therefore, multi-hazard policies need for less public capital than managing the two hazards separately, and this finding is attributable to risk differentiation.

However, advantages from multi-hazard are evident with respect to flood risk, but a bit controversial when we look at the seismic risk. The multi-hazard parameter $c$ is a bit greater than that of the seismic case and much smaller than in the flood case but, unfortunately, is always $c \leq 1$. Our analysis suggests that benefits from risk differentiation are not sufficient for the natural risks to be entirely managed by the private market and once again, a government intervention is highly recommended. This evidence is confirmed by the probability $\epsilon_2^*$, that shows a behaviour similar to $c$ and is always greater than the desired level, and $\epsilon_1^*=\epsilon_1$.

As far as coverage limits concern, the minimum amount of public funds $W^*_d$ and the minimum probability $\epsilon^*_2$ are obtained with the full coverage policy, while applying a deductible $D=200$ or a maximum coverage $E=1200$ lead to similar results. In any case, the worst solution would be applying the two limits together, since both the greatest $W^*_d$ and the highest $\epsilon_2^*$ are here obtained.

In addition to benefits from risk differentiation, a government may prefer multi-hazard policies for another interesting feature: risk-based premiums are much more geographically uniform than those of single hazards. In fact, Figure \ref{fig:Italy_MH_M1} mapping premiums for the most vulnerable buildings (masonry-one storey) shows a quite homogeneous price at the municipal level, while differences are a bit more emphasised in the corresponding single hazard (see Figures \ref{fig:Italy_SH_M} and \ref{fig:Italy_SH_1s}). From the public sector perspective, a uniform rating system is desirable because it weakens the perception of unequal treatment between the property-owners from different areas and therefore allows easier acceptance by the population. On the other hand, different risk-based premiums signal the riskiness of the area to its inhabitants and is therefore important to discourage the construction of most vulnerable housing structures and to encourage preventive behavior. The current rating also preserves this desirable feature since premiums are defined on structural typologies among which rates substantially vary (see Table \ref{tab:optimal_MH_premiums}). 

\section{Conclusion}\label{ch:conclusion}
Seismic and flood risks in Italy have been analysed. Given the limited amount of data available on natural risks, an alternative approach based on risk-modeling has been applied to estimate expected monetary losses. We found that seismic risk results in the highest expected losses at national level, but floods may generate the highest losses per square metre. The two perils differ in geographic extent: while the seismic risk is relevant for almost all the national territory, floods affect a limited area.

In order to cope with the effects of natural risks, a public-private insurance scheme has been proposed. Our insurance model is intended to alleviate the financial burden that natural events place on governments, while at the same time assisting individuals and protecting the insurance business. Therefore, in our model, propertyowners, an insurer and the government co-operate in risk financing. Though expected losses generated by floods and earthquakes are considerably different, we found that the amount of public funds needed to manage the two perils is almost the same. We argue that this evidence is generated by a combination of individuals' increasing risk aversion and hazard loss distributions. Our analysis also shows that the amount of public capital necessary for risk financing can be reduced by jointly managing the two risks. Along with this benefit from risk differentiation, multi-hazard policies allow the insurer to apply rates that are more geographically homogeneous, therefore favoring the perception of fair treatment among the population.

Unfortunately, our results show that neither single- or multi-hazard policies are sustainable by the private market alone: due to spatial correlation among insured assets, the maximum premiums that individuals are willing to pay do not meet the insurer's solvency or capital constraints for any policy considered. Without the government, a private insurer would be forced to drive up premiums, which would not meet the demand and would therefore not be purchased. This evidence suggests the need for the government to intervene in the insurance market for natural disasters.

To conclude, our results show that the probability of the government having to inject further capital may be moderate. Though the insurance scheme reduces the government's financial burden due to natural perils with respect to the current state, adding some layer of risk transfer might be beneficial. For example, CatBonds or some level of reinsurance may reduce losses suffered by the government and their volatility.

\bibliography{main}

@inbook{AgenziaEntrate,
  author       = {{Agenzia delle Entrate}}, 
  title        = {Gli immobili in Italia},
  chapter      = {2},
  publisher    = {Agenzia delle Entrate},
  year         = {2015}
}

@article{Ahmad,
title = "Analytical fragility functions for reinforced concrete and masonry buildings aggregates of Euro-Mediterranean regions – UPAV methodology",
journal = "Internal report, Syner-GProject, 2009/2012",
year = "2011",
author = "Ahmad, Naveed and Crowley, Helen and Pinho, Rui"
}

@Article{Apel,
  author={Apel, Heiko and Thieken, Annegret and Merz, Bruno and Blöschl, Günter},
  title={{A Probabilistic Modelling System for Assessing Flood Risks}},
  journal={Natural Hazards: Journal of the International Society for the Prevention and Mitigation of Natural Hazards},
  year=2006,
  volume={38},
  number={1},
  pages={79-100},
  month={May},
  doi={10.1007/s11069-005-8603-7},
  url={https://ideas.repec.org/a/spr/nathaz/v38y2006i1p79-100.html}
}

@Article{Appelbaum,
  author={Appelbaum, Stuart J.},
  title={Determination of urban flood damage},
  journal={Journal of Water Resources Planning and Management},
  year=1985,
  volume={111},
  number={3},
  pages={269-283},
  month={July},
  doi={https://doi.org/10.1061/(ASCE)0733-9496(1985)111:3(269)}
}

@Article{Arrighi,
AUTHOR = {Arrighi, Chiara and Brugioni, Marcello and Castelli, Fabio and Franceschini, Serena and Mazzanti, Bernardo},
TITLE = {Urban micro-scale flood risk estimation with parsimonious hydraulic modelling and census data},
JOURNAL = {Natural Hazards and Earth System Sciences},
VOLUME = {13},
YEAR = {2013},
NUMBER = {5},
PAGES = {1375--1391},
URL = {https://www.nat-hazards-earth-syst-sci.net/13/1375/2013/},
DOI = {10.5194/nhess-13-1375-2013}
}

@article{Asprone,
title = "Seismic insurance model for the Italian residential building stock",
journal = "Structural Safety",
volume = "44",
pages = "70 - 79",
year = "2013",
issn = "0167-4730",
doi = "https://doi.org/10.1016/j.strusafe.2013.06.001",
url = "http://www.sciencedirect.com/science/article/pii/S0167473013000416",
author = "Domenico Asprone and Fatemeh Jalayer and Saverio Simonelli and Antonio Acconcia and Andrea Prota and Gaetano Manfredi"
}

@article{Borzi1,
  author = {Borzi, Barbara and Crowley, Helen and Pinho, Rui},
  title = {Un metodo meccanico per la definizione della vulnerabilità basato su analisi pushover semplificate},
  booktitle = {Proceedings of XII Convegno L’Ingegneria Sismica in Italia ANIDIS, Paper No. 160},
  address = {Pisa, Italy},
  year = {2007}
}

@article{Borzi2,
  title = {The influence of infill panels on vulnerability curves for RC buildings},
  booktitle = {Proceedings of the 14th world conference on earthquake engineering},
  address = {Beijing, China},
  year = {2008},
  author = {Borzi, Barbara and Crowley, Helen and Pinho, Rui}
}

@article{Brunette,
title = "An actuarial model of forest insurance against multiple natural hazards in fir (Abies Alba Mill.) stands in Slovakia",
journal = "Forest Policy and Economics",
volume = "55",
pages = "46 - 57",
year = "2015",
issn = "1389-9341",
doi = "https://doi.org/10.1016/j.forpol.2015.03.001",
url = "http://www.sciencedirect.com/science/article/pii/S1389934115000398",
author = "Brunette, Marielle and Holecy, Jan and Maroc, Sedliak and Jan, Tucek and Marc, Hanewinkel"
}

@article{Cascini,
author = {Cascini, Leonardo and Ferlisi, Settimio and Vitolo, E.},
title = {Individual and societal risk owing to landslides in the Campania region (southern Italy)},
journal = {Georisk: Assessment and Management of Risk for Engineered Systems and Geohazards},
volume = {2},
number = {3},
pages = {125-140},
year  = {2008},
publisher = {"Taylor and Francis"},
doi = {10.1080/17499510802291310},

URL = { 
        https://doi.org/10.1080/17499510802291310
    
},
eprint = { 
        https://doi.org/10.1080/17499510802291310
    
}

}

@article{Charpentier,
title = "Natural catastrophe insurance: How should the government intervene?",
journal = "Journal of Public Economics",
volume = "115",
pages = "1 - 17",
year = "2014",
issn = "0047-2727",
doi = "https://doi.org/10.1016/j.jpubeco.2014.03.004",
url = "http://www.sciencedirect.com/science/article/pii/S004727271400053X",
author = "Charpentier, Arthur and {Le Maux}, Benoit, "
}

@misc{Colombi,
	title = {Mappe di rischio sismico a scala nazionale con dati aggiornati sulla pericolosità sismica di base e locale},
	url = {http://hdl.handle.net/2122/6524},
	journal = {Progettazione Sismica 2010},
	year= "2010",
	author = {Colombi, Miriam and Crowley, Helen and {Di Capua}, Giuseppe and Peppoloni, Silvia and Borzi, Barbara and Pinho, Rui  and Calvi, {Gian M.}}
}

@book{Consorcio,
  author       = {{Consorcio de Compensación de Seguros}}, 
  title        = {Natural catastrophe risk management and modelling: A practitioner's guide},
  publisher    = {Consorcio de Compensación de Seguros},
  year         = {2008}
}

@ARTICLE{Cooper,
title = {Multi-period insurance contracts},
author = {Cooper, Russell and Hayes, Beth},
year = {1987},
journal = {International Journal of Industrial Organization},
volume = {5},
number = {2},
pages = {211-231},
url = {https://EconPapers.repec.org/RePEc:eee:indorg:v:5:y:1987:i:2:p:211-231}
}

@article{Crowley,
author = {Crowley, Helen and Borzi, Barbara and Pinho, Rui and Colombi, M. and Onida, Mauro},
year = {2008},
month = {06},
pages = {},
title = {Comparison of Two Mechanics-Based Methods for Simplified Structural Analysis in Vulnerability Assessment},
volume = {2008},
journal = {Advances in Civil Engineering},
doi = {10.1155/2008/438379}
}

@misc{DM14012008,
  author={{D.M. 14/01}},
  title={Approvazione delle nuove norme tecniche per le costruzioni},
  journal = {Ministero delle Infrastrutture, Gazzetta Ufficiale 4 febbraio 2008 n.29, S.O.},
  year={2008}
}

@article{Debo,
author = {Thomas N. Debo},
year = {1982},
pages = {1059-1069},
title = {Urban flood damage estimation curves},
volume = {108},
issue = {10},
journal = {Journal of the Hydraulics Division}
}

@misc{DecretoLegislativo,
  author={{Decreto legislativo 23 febbraio 2010 n.4}},
  title={Attuazione della direttiva 2007/60/CE relativa alla valutazione e alla gestione dei rischi di alluvioni},
  year={2010}
}

@article{Degiorgis,
author = {Degiorgis, Massimiliano and Gnecco, Giorgio and Gorni, Silvia and Roth, Giorgio and Sanguineti, Marcello and Taramasso, Angelaceleste},
year = {2012},
month = {11},
pages = {302–315},
title = {Classifiers for the detection of flood prone areas from remote sensed elevation data},
volume = {s 470–471},
journal = {Journal of Hydrology},
doi = {10.1016/j.jhydrol.2012.09.006}
}

@article{Ehrlich,
author = {Ehrlich, Isaac and Becker,Gary S.},
year = {1972},
month = {July-August},
pages = {623-648},
title = {Market Insurance, Self-Insurance, and Self-Protection},
volume = {80 (4)},
journal = {Journal of Political Economy},
doi = {https://ssrn.com/abstract=961496}
}

@article{Erberik,
author = {Erberik, Murat Altug},
year = {2008},
month = {03},
pages = {387 - 405},
title = {Generation of fragility curves for Turkish masonry buildings considering in‐plane failure modes},
volume = {37},
journal = {Earthquake Engineering and Structural Dynamics},
doi = {10.1002/eqe.760}
}

@article{Genovese,
author = {Genovese, Elisabetta},
year = {2006},
month = {01},
pages = {},
title = {A methodological approach to land use-based flood damage assessment in urban areas: Prague case study},
journal = {JRC Report - EUR 22497}
}

@article{Gnecco,
author = {Gnecco, Giorgio and Morisi, Rita and Roth, Giorgio and Sanguineti, Marcello and Taramasso, Angelaceleste},
year = {2015},
month = {11},
pages = {},
title = {Supervised and Semi-Supervised Classifiers for the Detection of Flood-Prone Areas},
journal = {Soft Computing},
doi = {10.1007/s00500-015-1983-z}
}

@inbook{Goda,
author = {Goda, Katsuichiro and Wenzel, Friedemann and Daniell, James},
year = {2015},
month = {01},
pages = {1184-1206},
title = {Insurance and Reinsurance Models for Earthquake},
doi = {10.1007/978-3-642-35344-4_261}
}

@inbook{Gollier,
author = {Gollier, C.},
year = {2013},
pages="107-122",
title = {Dionne G. (eds) Handbook of Insurance},
chapter = {The Economics of Optimal Insurance Design},
editor = { Springer, New York, NY}
}

@inbook{Grossi,
  author={Grossi, Patricia and Kunreuther, Howard and Windeler, Don},
  year= 2005, 
  chapter={An Introduction to Catastrophe Models and Insurance}, 
  editor = {Grossi, Patricia and Kunreuther, Howard}, 
  title= {Catastrophe Modeling: A New Approach to Managing Risk}, 
  publisher= {Springer}
}

@misc{GruppoMPS,
author = {{Gruppo  di  Lavoro  MPS}},
year = {2004},
month = {04},
title = {Redazione  della  mappa  di pericolosità sismica  prevista dall’Ordinanza PCM 3274 del 20 marzo 2003. “Rapporto Conclusivo per il Dipartimento della Pro-tezione Civile},
journal = {INGV}
}

@article{Guzzetti,
author = {Guzzetti, Fausto and Tonelli, G.},
year = {2004},
pages = {212-232},
volume = {4(2)},
title = {Information system on hydrological and geomorphological catastrophes in Italy (SICI): a tool for managing landslide and flood hazards},
journal = {Natural hazards and Earth System Science, Copernicus Publications on behalf of the European Geosciences Union}
}

@article{Hoeffding,
author = {Hoeffding, Wassily},
year = {1963},
month = {03},
volume = {58 (301)},
pages={13-30},
title = {Probability Inequalities for sums of Bounded Random Variables. In: . Springer},
journal = {Journal of the American Statistical Association},
doi = {10.1007/978-1-4612-0865-5_26}
}

@article{Hufschmidt,
author = {Hufschmidt, Gabriele and Glade, Thomas},
year = {2010},
month = {01},
pages = {233-243},
title = {Vulnerability analysis in geomorphic risk assessment},
journal = {Geomorphological Hazards and Disaster Prevention},
doi = {10.1017/CBO9780511807527.019}
}

@article{Kahneman,
author = {Kahneman, Daniel},
year = {2003},
month = {02},
pages = {1449-1475},
title = {Maps of Bounded Rationality: Psychology for Behavioral Economics},
volume = {93},
journal = {American Economic Review},
doi = {10.1257/000282803322655392}
}

@article{Kappes,
author = {Kappes, Melanie S. and Keiler, Margreth and {von Elverfeldt}, Kirsten  and Glade, Thomas},
year = {2012},
pages = {1925–1958},
title = {Challenges of analyzing multi-hazard risk: a review},
volume = {64},
journal = {Natural Hazards}
}

@article{Kappos2003,
author = {Kappos, Andreas J. and Panagopoulos, Georgios and Panagiotopoulos, Christos and Penelis, Gregorios},
year = {2006},
pages = {391–413},
title = {A hybrid method for the vulnerability assessment of R/C and URM buildings},
volume = {4},
journal = {Bulletin of Earthquake Engineering}
}

@article{Kappos2006,
author = {Kappos, Andreas J. and Panagiotopoulos, Christos and Panagopoulos, Georgios and Papadopoulos, E.},
year = {2003},
title = {RISK-UE WP4 – reinforce concrete buildings (Level 1 and Level 2 analysis)},
volume = {4}
}

@article{Kostov,
author = {Kostov, M. and Vaseva, Elena and Kaneva, Antoaneta and Koleva, N.  and Varbanov, G. and Stefanov, D.},
year = {2004},
title = {RISK-UE WP13 – Application to Sofia}
}

@article{Kousky,
author="Kousky, Carolyn and Cooke, Roger",
title="Explaining the Failure to Insure Catastrophic Risks",
journal="The Geneva Papers on Risk and Insurance - Issues and Practice",
year="2012",
month="Apr",
day="01",
volume="37",
number="2",
pages="206--227",
issn="1468-0440",
doi="10.1057/gpp.2012.14",
url="https://doi.org/10.1057/gpp.2012.14"
}

@article{Kunreuther,
author = {Kunreuther, Howard},
year = {1996},
volume = {12 (2/3)},
pages = {171-187},
title = {Mitigating Disaster Losses through Insurance},
journal = {Journal of Risk and Uncertainty, Special Issue: The Stanford University Conference on Social Treatment of Catastrophic Risk (1996)
}
}

@inbook{KunreutherPauly1985,
author = {Kunreuther, Howard and Pauly, Mark},
year = {1985},
month = {01},
pages = {424-443},
title = {Market Equilibrium with Private Knowledge},
volume = {14},
doi = {10.1007/978-94-015-7957-5_22}
}

@inbook{KunreutherPauly2009,
author = {Kunreuther, Howard and Pauly, Mark},
year = {2009},
title = {Insuring against catastrophes},
journal = {The known, the unknown and the unknowable in Financial Risk Management},
editor = {United States: Princeton University Press}
}

@inbook{Larsen,
author = {Larsen, Tom and Kuzak, Dennis},
year = {2005},
title = {Use of Catastrophe Models in Insurance Rate Making},
journal = {Grossi P., Kunreuther H. (eds) Catastrophe Modeling: A New Approach to Managing Risk. Catastrophe Modeling},
volume = {25},
editor = {Springer, Boston, MA},
doi = {10.1007/0-387-23129-3_5}
}

@article{Kwon,
author = {Kwon, Oh-Sung and Elnashai, Amr},
year = {2006},
month = {01},
pages = {289-303},
title = {The effect of material and ground motion uncertainty on the seismic vulnerability curves of RC structure},
volume = {28},
journal = {Engineering Structures},
doi = {10.1016/j.engstruct.2005.07.010}
}

@article{Lagomarsino,
author = {Lagomarsino, Sergio and Giovinazzi, Sonia},
year = {2006},
pages = {415-443},
title = {Macroseismic and mechanical models for the vulnerability and damage assessment of current buildings},
volume = {4},
journal = {Bull Earthquake Eng},
doi = {10.1007/s10518-006-9024-z}
}

@misc{Legge64,
  author={{Legge n. 64, 2 feb}},
  title={Provvedimenti per le costruzioni con particolari prescrizioni per lezone sismiche},
  year={1974}
}

@article{Luino,
author = {Luino, Fabio and Cirio, Chiara G. and Biddoccu, Marcella and Agangi, Andrea and Giulietto, W. and Godone, Franco and Nigrelli, Guido},
year = {2009},
pages = {339-353},
title = {Application of a model to the evaluation of flood damage},
volume = {13},
journal = {Geoinformatica}
}

@article{Maccaferri,
author = "Maccaferri, Sara and Cariboni, Jessica and Campolongo, Francesca",
year = {2012},
title = {Natural Catastrophes: Risk relevance and Insurance Coverage in the EU},
journal = {EUR - Scientific and Technical Research Reports},
editor = {Publications Office of the European Union},
doi = {10.2788/93626}
}

@Article{MarzocchiGarcia,
  author={Marzocchi, Warner and Garcia-Aristizabal, Alexander and Gasparini, Paolo and Mastellone, Maria and {Di Ruocco}, Angela},
  title={{Basic principles of multi-risk assessment: a case study in Italy}},
  journal={Natural Hazards: Journal of the International Society for the Prevention and Mitigation of Natural Hazards},
  year=2012,
  volume={62},
  number={2},
  pages={551-573},
  month={June},
  doi={10.1007/s11069-012-0092-x},
  url={https://ideas.repec.org/a/spr/nathaz/v62y2012i2p551-573.html}
}

@article{Meletti,
title = "Stime di pericolosità sismica per diverse probabilità di superamento in 50 anni: valori di ag",
journal = "Progetto DPC-INGV S1, Deliverable D2",
year = "2007",
author = "Meletti, C. and Montaldo, V.",
url={http://esse1.mi.ingv.it/d2.html}
}

@article{Molinari2012,
  author = {Molinari, Daniela and Aronica, G.T. and Ballio, F. and Berni, Nicola and Pandolfo, C. },
  title = "Le curve di danno quale strumento a supporto della direttiva alluvioni: criticità dei dati italiani",
  journal = {XXXIII Convegno Nazionale diIdraulica e Costruzioni Idrauliche},
  address = {Brescia, Italy},
  day = {10-15},
  month = {09},
  year = {2012}
}

@article{Molinari2014,
author = {Molinari, Daniela and Menoni, Scira and Aronica, G. and Ballio, Francesco and Berni, Nicola and Pandolfo, C. and Stelluti, M. and Minucci, Guido},
year = {2014},
month = {04},
pages = {},
title = {Ex post damage assessment: an Italian experience},
volume = {14},
journal = {Natural Hazards and Earth System Sciences},
doi = {10.5194/nhess-14-901-2014}
}

@TECHREPORT{Mossin,
title = {Aspects of rational insurance purchasing},
author = {Mossin, Jan},
year = {1968},
institution = {Université catholique de Louvain, Center for Operations Research and Econometrics (CORE)},
type = {CORE Discussion Papers RP},
number = {23},
url = {https://EconPapers.repec.org/RePEc:cor:louvrp:23}
}

@book{Palm,
  author       = {Palm, Risa}, 
  title        = {"Earthquake insurance – a longitudinal study of California homeowners"},
  publisher    = {Westview Press, Boulder},
  year         = {1995}
}

@inbook{OlivieriPitacco,
  author       = {Olivieri, Annamaria and Pitacco, Ermanno}, 
  title        = {"Introduction to insurancemathematics. Technical and financial features of risk transfer"},
  chapter      = {"Non-life insurance: pricing and reserving"},
  publisher    = {Springer},
  year         = {2010}
}

@article{OlivieriSantoro,
author = {Oliveri, Elisa and Santoro, Mario},
year = {2000},
month = {09},
pages = {223-234},
title = {Estimation of urban structural flood damages: The case study of Palermo},
volume = {2},
journal = {Urban Water},
doi = {10.1016/S1462-0758(00)00062-5}
}

@misc{OPCM,
  author={{O.P.C.M. 3274}},
  title={ Primi elementi in materia di criteri generali per la classificazione sismica del territorio nazionale e di normative tecniche per le costruzioni in zona sismica},
  journal={Gazzetta Ufficiale 8 maggio 2003 n.108},
  year={2003}
}

@inproceedings{Ozmen,
author = {Ozmen, Hayri and Inel, Mehmet and Meral, Emrah and Bucakli, M},
year = {2010},
month = {09},
pages = {},
title = {Vulnerability of Low and Mid-Rise Reinforced Concrete Buildings In Turkey}
}

@article{CasaItalia,
author = {{Struttura di Missione Casa Italia}},
year = {2017},
month = {06},
journal = {Dipartimento Casa Italia, Presidenza del Consiglio dei Ministri},
title = {Rapporto sulla Promozione della sicurezza dai Rischi naturali del Patrimonio abitativo}
}

@article{Rota1,
author = {Rota, Maria and Penna, Andrea and Magenes, Guido},
year = {2010},
month = {05},
pages = {1312-23},
title = {A methodology for deriving analytical fragility curves for masonry buildings},
journal = {Engineering Structures},
volume = {32}
}

@article{Rota2,
author = {Rota, Maria and Penna, Andrea and Strobbia, Claudio},
year = {2008},
pages = {933-47},
title = {Processing Italian damage data to derive typological fragility curve},
journal = {Soil Dynamics and Earthquake Engineering},
volume = {28},
issue = {10-11}
}

@article{Salvati,
author = {Salvati, Paola and Bianchi, Cinzia and Rossi, Mauro and Guzzetti, Fausto},
year = {2010},
month = {03},
pages = {},
title = {Societal landslide and flood risk in Italy},
volume = {10},
journal = {Natural Hazards and Earth System Sciences},
doi = {10.5194/nhess-10-465-2010}
}

@article{Scorzini,
author = {Scorzini, Anna and Frank, Enrico},
year = {2015},
month = {03},
pages = {},
title = {Flood damage curves: New insights from the 2010 flood in Veneto, Italy},
volume = {10},
journal = {Journal of Flood Risk Management},
doi = {10.1111/jfr3.12163}
}

@article{Spence,
author = {Spence, Robin},
year = {2007},
pages = {165},
title = {Earthquake Disaster Scenario Prediction and Loss Modelling for Urban Areas},
journal = {LESSLOSS Report 7. Pavia, Italy: IUSS Press}
}

@article{Tavarnien,
author = {Tarvainen, Timo and Jarva, Jaana and Greiving, Stefan},
year = {2006},
month = {01},
pages = {83-91},
title = {Spatial pattern of hazards and hazard interactions in Europe},
journal = {Geological Survey of Finland},
volume = {42}
}

@article{Tsionis,
author = {Tsionis, Georgis and Papailia, Alexandra and Fardis, {Michael N.}},
year = {2011},
pages = {83-91},
title = {Analytical fragility functions for reinforced concrete and masonry buildings aggregates of Euro-Mediterranean regions – UPAT methodology},
journal = {Internal report,Syner-G Project,},
volume = {2009/2012}
}

@book{WB,
  author       = {{World Bank}}, 
  title        = {Financial Protection against Natural Disaster: An Operational Framework for Disaster Risk Financing and Insurance scholar},
  publisher    = {World Bank Group},
  year         = {2014}
}

@book{GenevaAssociation,
  author       = {{Geneva Association}}, 
  title        = {Warming of the Oceans and Implications for the (Re)insurance Industry. A Geneva Association Report},
  publisher    = { Geneva: Geneva Association},
  year         = {2013}
}

@article{IVASS,
  author       = {Cesari, Riccardo and {D' Aurizio}, Leandro}, 
  title        = {Natural disasters and insurance cover: risk assessment and policy options for Italy},
  publisher    = {IVASS Working Paper No.12},
  month        = {07},
  year         = {2019}
}

@book{Mitchell-Wallace,
  author       = {K. Mitchell-Wallace and M. Foote and J. Hillier and M. Jones}, 
  title        = {"Natural catastrophe risk management and modelling: A practitioner's guide"},
  publisher    = {Wiley Blackwell},
  year         = {2017}
}

@book{OECD2,
  author       = "OECD", 
  title        = "Disaster Risk Financing: A global survey of practices and challenges",
  publisher    = "OECD Publishing, Paris",
  year         = "2015",
  url           = "http://dx.doi.org/10.1787/9789264234246-en"
}

@book{OECD1,
  author       = "OECD", 
  title        = "Disaster Risk Assessment and Risk Financing. A G20/OECD Methodological Framework",
  publisher    = "G20 meeting in Mexico City",
  day = "4-5",
  month = "11",
  year         = "2012"
}

\end{document}